\documentclass[a4paper,fleqn,usenatbib]{mnras}

\usepackage[T1]{fontenc}
\usepackage{ae,aecompl}

\usepackage{graphicx}	
\usepackage{amsmath}	
\usepackage{amssymb}	
\usepackage{subfigure}
\usepackage{multirow}
\usepackage{xcolor}

\newcommand{\pcc}{\,{\rm cm}^{-3}}
\newcommand{\gcc}{\,{\rm g \, cm}^{-3}}
\newcommand{\pcs}{\,{\rm cm}^{-2}}

\newcommand{\kel}{\, {\rm K}}

\newcommand{\msun}{\, {\rm M}_\odot}
\newcommand{\nh}{n_{\rm H}}
\newcommand{\Nhcol}{N_{\rm H}}
\newcommand{\Nhmol}{N_{\rm H_2}^{\rm shield}}
\newcommand{\Nco}{N_{\rm CO}^{\rm shield}}

\newcommand{\pc}{\, {\rm pc}}

\newcommand{\ssum}{\displaystyle\sum}

\newcommand{\myr}{\, {\rm Myr}}
\newcommand{\kyr}{\, {\rm kyr}}
\newcommand{\ug}{\, {\rm \mu G}}

\newcommand{\kms}{\, {\rm km \, s^{-1}}}
\newcommand{\av}{A_{\rm V}}

\newcommand{\Nhshield}{N_{\rm H}^{\rm shield}}

\title[Non-equilibrium abundances]{Non-Equilibrium Abundances Treated Holistically (NEATH): the molecular composition of star-forming clouds}

\author[Priestley et al.]{
  F. D. Priestley$^1$\thanks{Email: priestleyf@cardiff.ac.uk}, P. C. Clark$^1$, S. C. O. Glover$^2$, S. E. Ragan$^1$, O. Feh\'{e}r$^1$, L. R. Prole$^1$,
  \newauthor R. S. Klessen$^{2,3}$
\\
$^1$School of Physics and Astronomy, Cardiff University, Queen's Buildings, The Parade, Cardiff CF24 3AA, UK \\
$^{2}$Universit\"{a}t Heidelberg, Zentrum f\"{u}r Astronomie, Institut f\"{u}r Theoretische Astrophysik, Albert-Ueberle-Stra{\ss}e 2, D-69120 Heidelberg, Germany\\
$^{3}$Universit\"{a}t Heidelberg, Interdisziplin\"{a}res Zentrum f\"{u}r Wissenschaftliches Rechnen, Im Neuenheimer Feld 205, D-69120 Heidelberg, Germany\\ 
}

\date{Accepted XXX. Received YYY; in original form ZZZ}

\pubyear{2023}

\begin{document}
\label{firstpage}
\pagerange{\pageref{firstpage}--\pageref{lastpage}}
\maketitle

\begin{abstract}

  Much of what we know about molecular clouds, and by extension star formation, comes from molecular line observations. Interpreting these correctly requires knowledge of the underlying molecular abundances. Simulations of molecular clouds typically only model species that are important for the gas thermodynamics, which {tend to be} poor tracers of the denser material where stars form. We construct a framework for post-processing these simulations with a full time-dependent chemical network, {allowing us to model the behaviour of observationally-important species not present in the reduced network used for the thermodynamics.} We use this to investigate the chemical evolution of molecular gas under realistic physical conditions. We find that molecules can be divided into those which reach peak abundances at moderate densities ($10^3 \pcc$) and decline sharply thereafter (such as CO and HCN), and those which peak at higher densities and then remain roughly constant (e.g. NH$_3$, N$_2$H$^+$). {Evolving the chemistry with physical properties held constant at their final values} results in a significant overestimation of gas-phase abundances for all molecules, and {does not capture} the drastic variations {in abundance} caused by different evolutionary histories. The dynamical evolution of molecular gas cannot be neglected when modelling its chemistry.

\end{abstract}

\section{Introduction}

Molecular clouds, as the name suggests, are primarily made up of molecules, the most common being molecular hydrogen (H$_2$) and carbon monoxide (CO). As the birthplaces of stars \citep{bergin2007}, line emission from {molecular rotational transitions with energies low enough to be excited at the $\sim 10 \kel$ temperatures of molecular clouds} contains key information about the process of star formation. Interpreting this data, or producing synthetic observations which can be compared to it, requires knowedge of the abundances of the relevant species. While these can be parametrized as simple functions of the local properties \citep[e.g.][]{tafalla2002,smith2012,smith2013,jones2023}, the comparable chemical and dynamical timescales \citep{banerji2009} mean that molecular abundances are dependent on the prior history of the gas. A detailed treatment of non-equilibrium chemistry in molecular clouds is {therefore} necessary in order to advance our understanding of star formation.

Modern hydrodynamical simulations of molecular clouds \citep[e.g.][]{clark2012b,walch2015,smith2020} typically include time-dependent CO chemistry self-consistently, {albeit in a simplified fashion,} due to its importance as a coolant and its effect on the abundances of other major coolants (C, C$^+$, O). However, its high molecular abundance, and the low critical density of its key rotational lines, make CO a poor tracer of the high-density regions of clouds where star formation actually occurs \citep[e.g.][]{clark2019,priestley2023b}. In cloud-scale simulations, the abundances of molecules tracing denser gas are usually inserted by hand, based on either observational estimates or on smaller-scale simulations of prestellar cores \citep{smith2012,jones2023}. These core-scale simulations \citep[e.g.][]{aikawa2005,sipila2018} make assumptions such as isothermality and/or spherical symmetry which are unlikely to apply in reality, and in addition are highly sensitive to the (unknown) properties of the cloud material exterior to the core \citep{kaminski2014,priestley2023,jensen2023}. It is therefore desirable to follow the chemical evolution of star-forming material from the earliest stages of molecular cloud formation.
 
The abundances of molecular species {other than CO (an important coolant)} typically have little effect\footnote{With the possible exception of their contribution to the ionization state of the gas, when non-ideal magnetohydrodynamic effects are considered \citep{wurster2021}.} on the thermal or dynamical evolution of molecular gas. It is thus possible to decouple the detailed chemical evolution from the hydrodynamics and separately post-process this stage \citep[e.g.][]{panessa2023} with very little loss of accuracy \citep{ferrada2021}, rather than taking the computationally-expensive approach of including hundreds of molecular species directly into the underlying hydrodynamic simulation \citep[e.g.][]{lupi2021}. In this paper, we develop a framework to do this while maintaining consistency between the chemical network incorporated in the hydrodynamics (designed for accuracy and efficiency at low to moderate densities) and the more complex one used in the post-processing stage (designed for completeness, but poorly constrained at temperatures above $\sim 20 \kel$): Non-Equilibrium Abundances Treated Holistically (NEATH).\footnote{Neath is a town in Wales roughly halfway between Cardiff and Swansea.} We use this to investigate the molecular composition of star-forming clouds, and how a full time-dependent treatment of the {physical and chemical evolution} differs from {simplified} approaches common in the literature.

\begin{table}
  \centering
  \caption{{Initial gas-phase elemental abundances, relative to hydrogen nuclei, used in the chemical modelling.}}
  \begin{tabular}{ccccc}
    \hline
    Element & Abundance & & Element & Abundance \\
    \hline
    C & $1.4 \times 10^{-4}$ & & S & $1.2 \times 10^{-5}$ \\
    N & $7.6 \times 10^{-5}$ & & Si & $1.5 \times 10^{-7}$ \\
    O & $3.2 \times 10^{-4}$ & & Mg & $1.4 \times 10^{-7}$ \\
    \hline
  \end{tabular}
  \label{tab:abun}
\end{table}

\begin{figure*}
  \centering
  \includegraphics[width=\columnwidth]{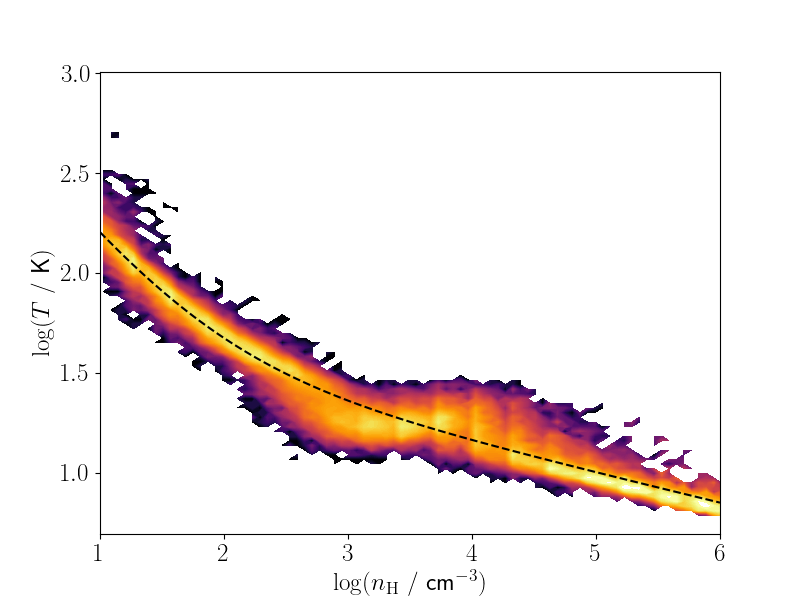}\quad
  \includegraphics[width=\columnwidth]{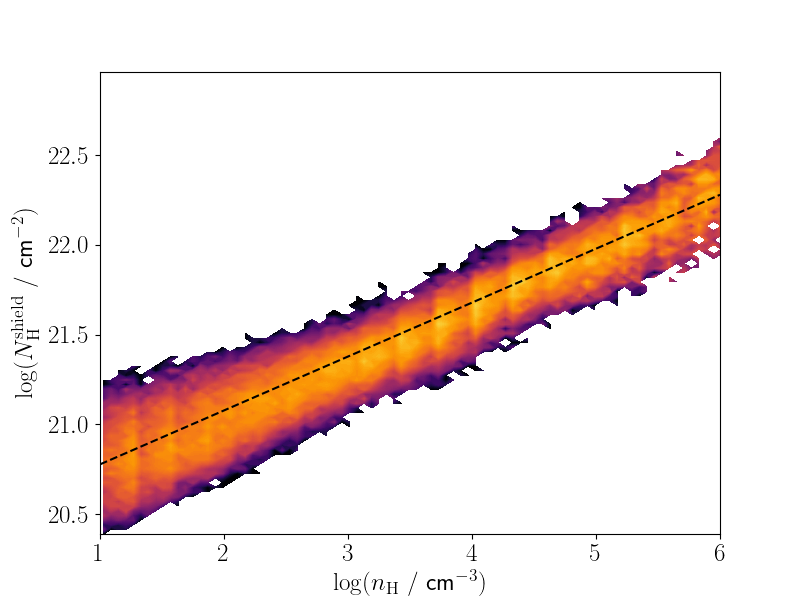}\quad
  \caption{Distribution of gas temperature (left) and shielding column density (right) versus volume density, with the colour scale representing the number of tracer particles. Power law relationships which reproduce the general trend of the data (Equations \ref{eq:temp} and \ref{eq:av}) are shown as dashed black lines.}
  \label{fig:hist}
\end{figure*}

\begin{figure*}
  \centering
  \includegraphics[width=\columnwidth]{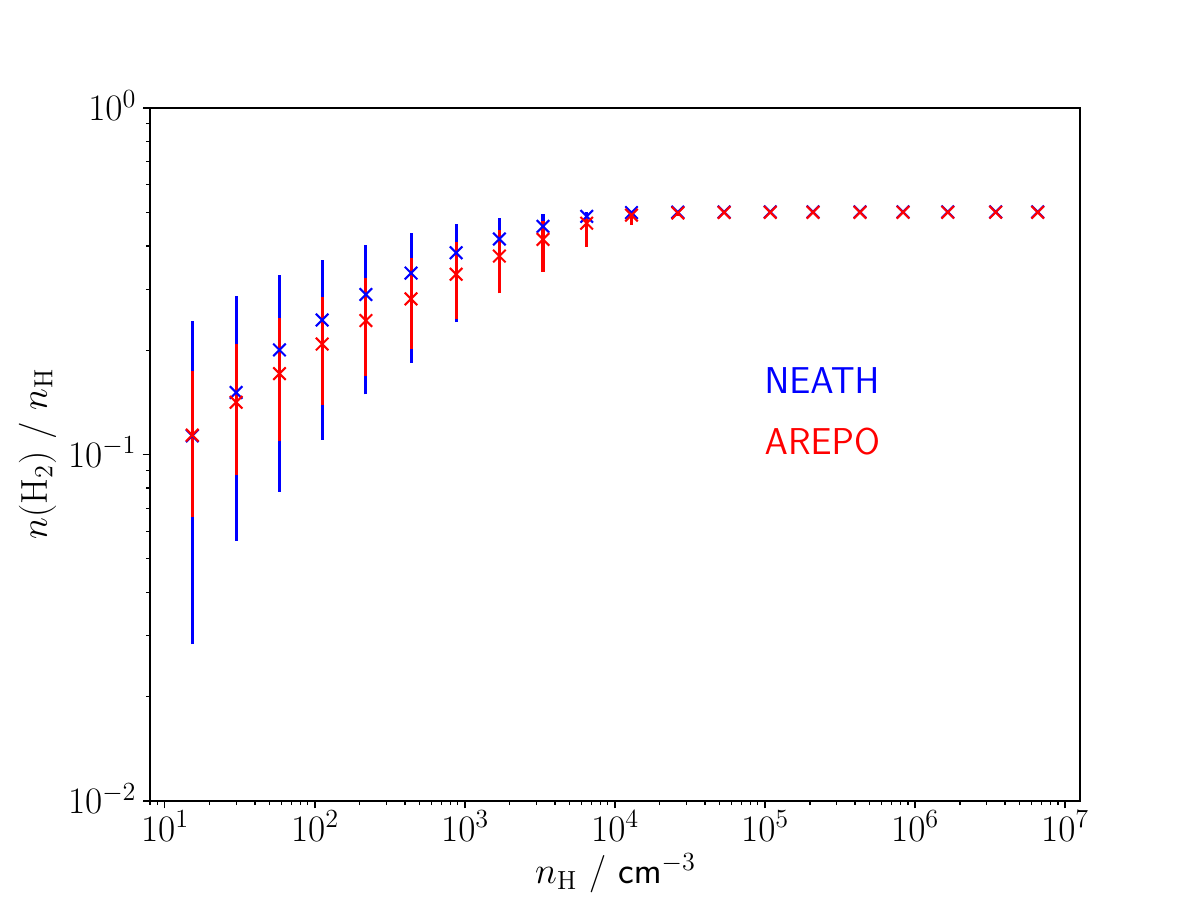}\quad
  \includegraphics[width=\columnwidth]{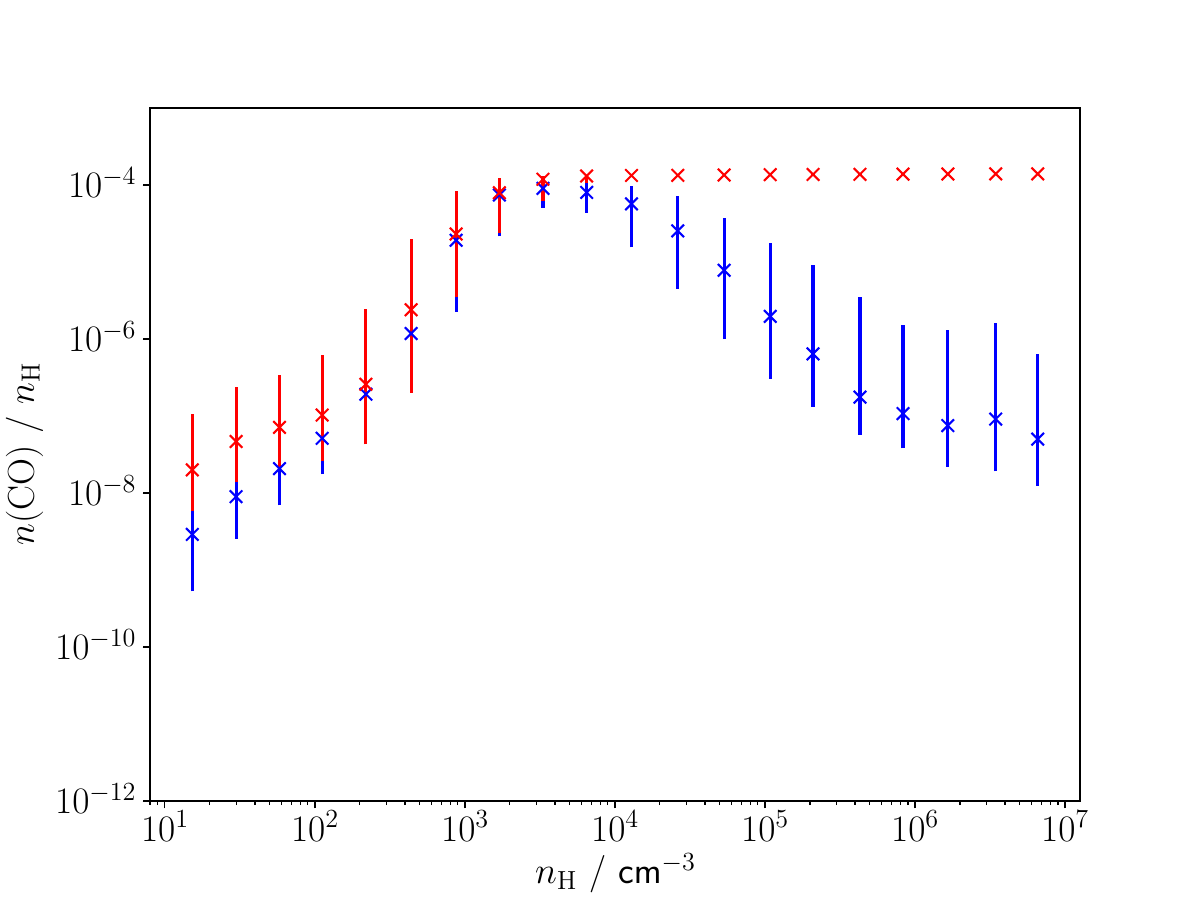}\quad
  \caption{Average abundances versus gas density of H$_2$ (left) and CO (right). Values from {\sc arepo} using the \citet{gong2017} network are shown in red, the NEATH values in blue. Crosses show the median values, with the 16th and 84th percentiles as error bars.}
  \label{fig:chemtest}
\end{figure*}

\begin{figure*}
  \centering
  \includegraphics[width=0.31\textwidth]{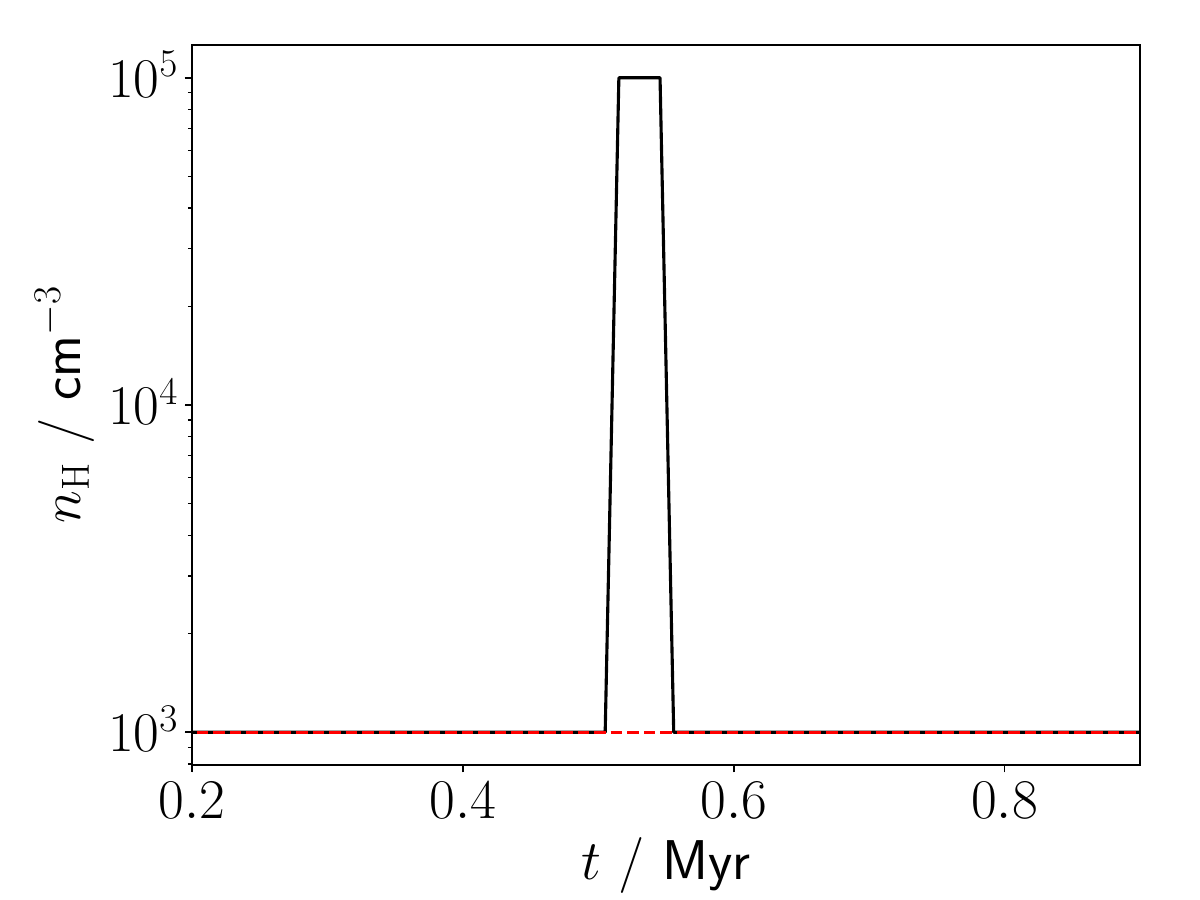}\quad
  \includegraphics[width=0.31\textwidth]{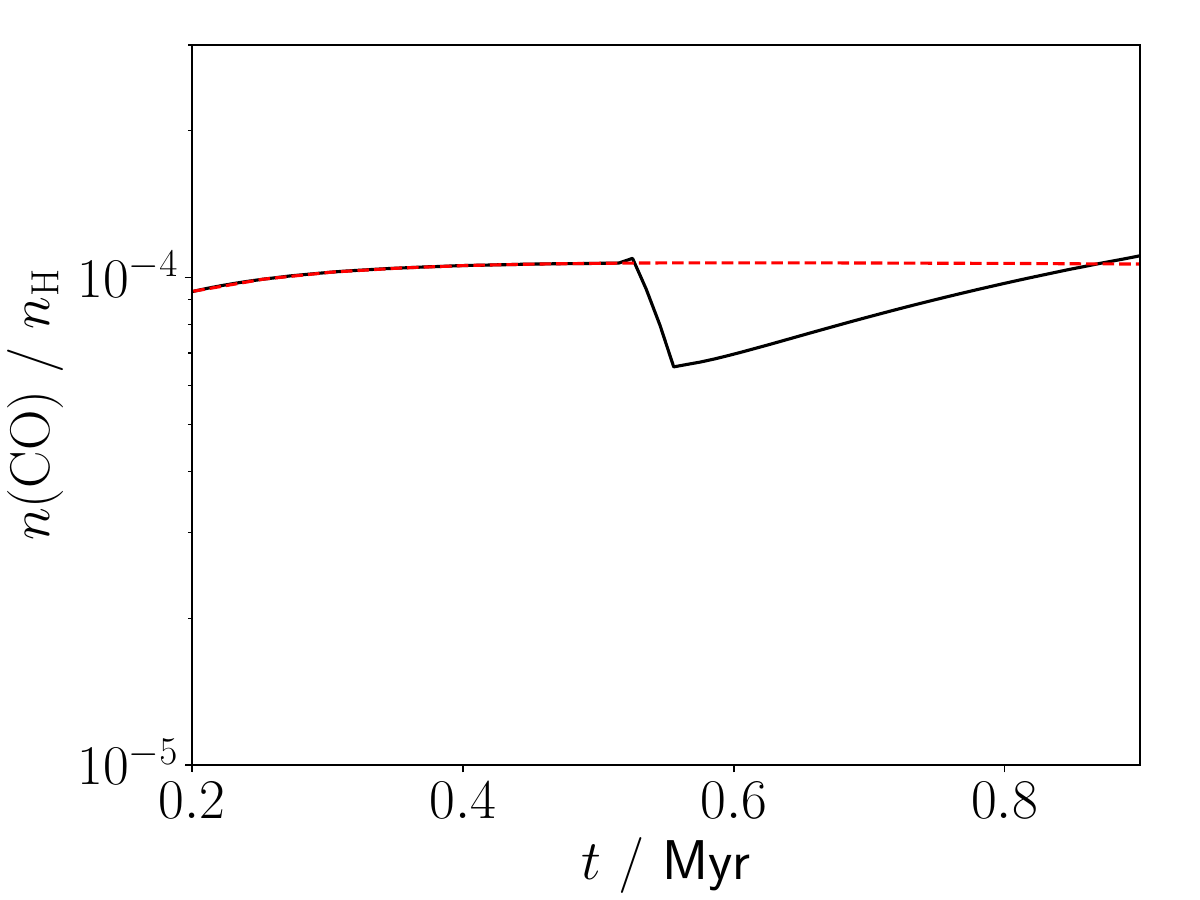}\quad
  \includegraphics[width=0.31\textwidth]{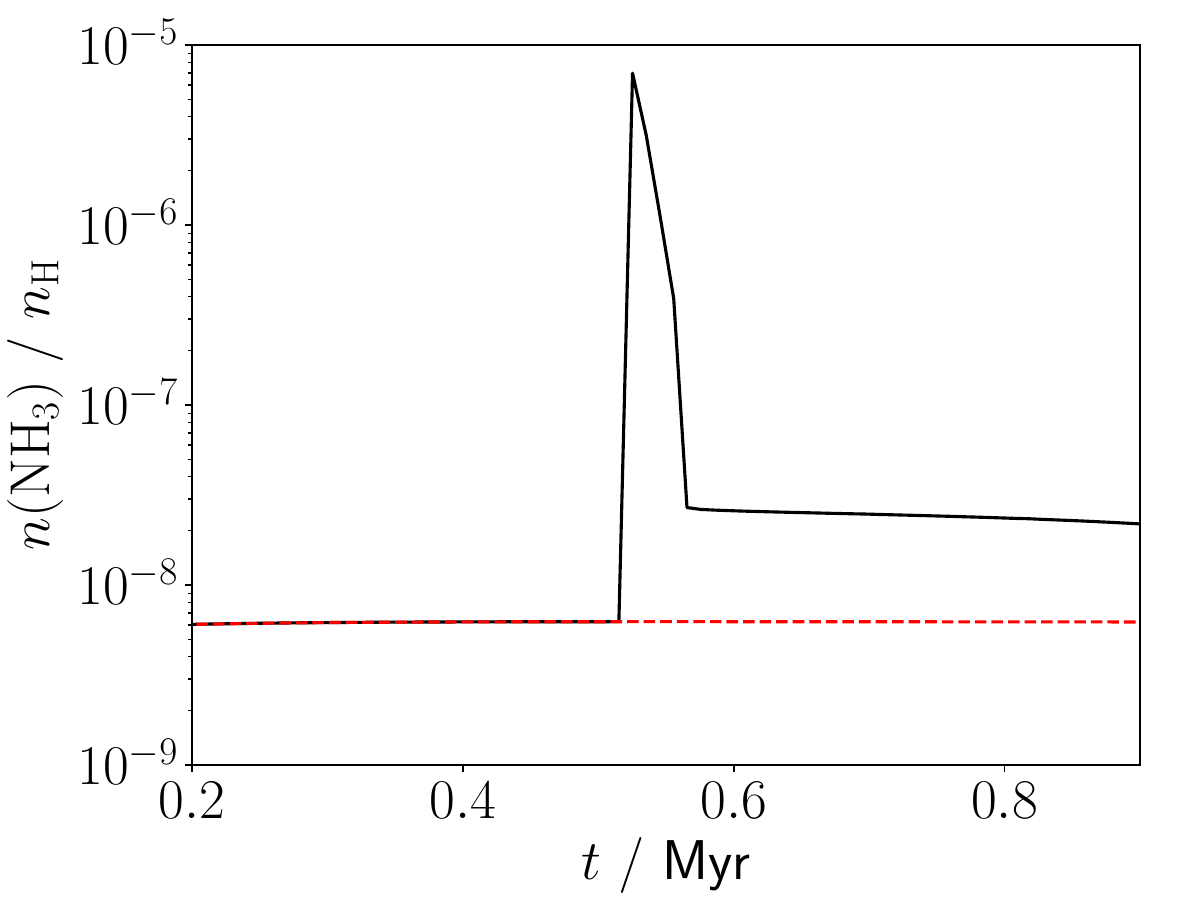}\quad
  \caption{Evolution of the gas density (left) and the CO (middle) and NH$_3$ (right) abundances for the shock test, modelled with a time resolution of $10 \kyr$ (solid black lines) and $50 \kyr$ (dashed red lines).}
  \label{fig:shocktest}
\end{figure*}

\begin{figure}
  \centering
  \includegraphics[width=\columnwidth]{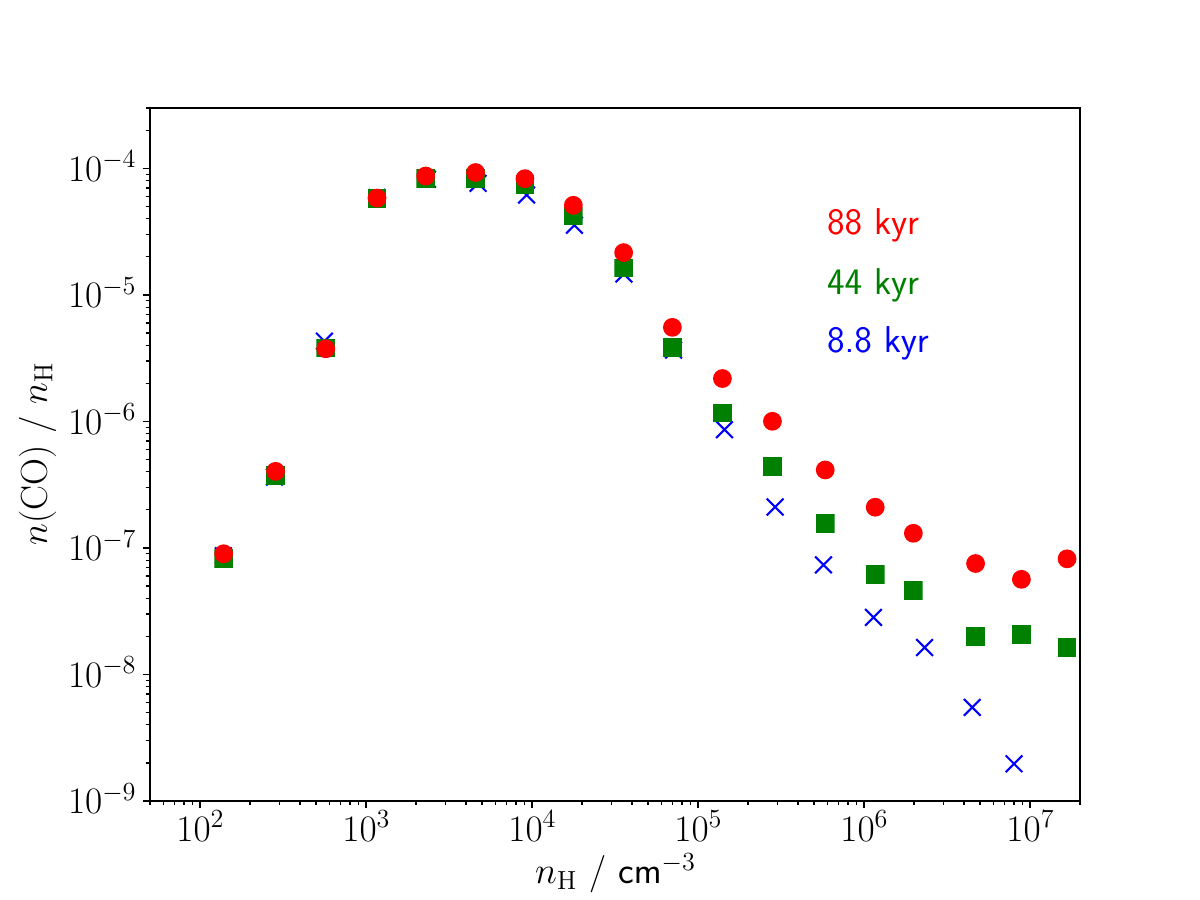}
  \caption{Median CO abundance versus density at a time resolution of $8.8 \kyr$ (blue crosses), $44 \kyr$ (green squares), and $88 \kyr$ (red circles).}
  \label{fig:restest}
\end{figure}

\section{Method}

\subsection{Magnetohydrodynamical model}

We simulate the dynamical evolution of molecular clouds using a modified version of {\sc arepo} \citep{springel2010}, a moving-mesh magnetohydrodynamics (MHD) code. The thermodynamics of the gas and dust are treated according to \citet{glover2007} and \citet{glover2012}; this represents a largely complete account of all important processes, including radiative heating and cooling, cosmic rays, dust-gas thermal coupling, and the energy liberated by exothermic chemical reactions. The abundances of important gas-phase coolants (H$_2$, C, C$^+$, O, CO) are calculated on-the-fly using a time-dependent chemical network based on that of \citet{gong2017}, as described in \citet{hunter2023}. We assume a background ultraviolet (UV) radiation field with the \citet{mathis1983} spectral energy distribution. Shielding from the UV field is also done on-the-fly; column densities of H$_2$, CO, and C for self-shielding, and the total gas column responsible for dust extinction, are calculated using the {\sc treecol} algorithm \citep{clark2012}.

{The on-the-fly chemical network in our {\sc arepo} simulations contains only a limited subsample of the astrophysically-relevant reactions involving key coolant species, and neglects freeze-out of CO onto dust grains entirely. However, benchmarking against more comprehensive networks has shown that the effect of the missing reactions on the abundances of coolants is small \citep{gong2017}. Additionally, at the densities where freeze-out becomes important ($\gtrsim 10^4 \pcc$), the reduced gas-phase CO abundance has little effect on the temperature \citep{goldsmith2001}, due to a combination of highly optically-thick emission and the increasing importance of dust cooling. We therefore expect that the on-the-fly chemical network is sufficient to accurately capture the thermal evolution of the gas.}

The {\sc arepo} grid is refined with a target mass per cell of $0.005 \msun$, with the constraints that the volumes of neighbouring cells do not differ by more than a factor of eight, and enforced minimum (maximum) cell volumes of $5 \times 10^{-11} \pc^3$ ($12 \pc^3$). We also ensure that the local Jeans length is resolved by at least 16 cells.  We include a sink particle algorithm to treat small regions collapsing under self-gravity, as described in \citet{tress2020} and \citet{prole2022}; cells with a density above $2 \times 10^{-16} \gcc$ are converted into sink particles if they are at a local minimum of the gravitational potential, and the gas within a sink formation radius of $9 \times 10^{-4} \pc$ is (by mass-weighted average) both moving and accelerating towards the prospective sink particle.

\subsection{Initial conditions}

Our simulations are of spherical, colliding clouds, each with mass $M = 10^4 \msun$ and radius $R = 19 \pc$, for an initial number density $\nh = 10 \pcc$. The centres of the clouds are initially separated by $2R$, so the clouds are just touching. The initial gas and dust temperatures are set to $300 \kel$ and $15 \kel$ respectively. The clouds are seeded with a virialised turbulent velocity field. The computational domain is a cubic box of side length $96 \pc$ with periodic boundary conditions, filled with a hot, rarefied medium ($0.01 \pcc$, $10^4 \kel$). The initial velocity of each cloud is $7 \kms$, giving a collision velocity of $14 \kms$, and we include a $3 \ug$ magnetic field parallel to the collision axis. Simulations are run for $5.0 \myr$, by which point the clouds have formed $28$ sink particles with a total mass of $24.2 \msun$. Chemical parameters which are shared with the later post-processing stage described in Section \ref{sec:chem} are the elemental carbon ($1.4 \times 10^{-4}$) and oxygen ($3.2 \times 10^{-4}$) abundances from \citet{sembach2000}, the external ultraviolet (UV) radiation field ($1.7$ times the \citet{habing1968} field), and the cosmic ray ionisation rate ($10^{-16} \, {\rm s^{-1}}$ primary ionisations of H$_2$). The `metal' abundance is set to $10^{-7}$, corresponding to the high levels of depletion seen in the dense ISM \citep{jenkins2009}.

\subsection{Chemical post-processing}
\label{sec:chem}

We use tracer particles, {as implemented in {\sc arepo} by \citet{genel2013},} to record the evolving properties of individual parcels of gas, which are used as input for {\sc uclchem} \citep{holdship2017}, a time-dependent gas-grain chemical code.\footnote{{The source code for the version used here is available from {https://fpriestley.github.io/neath/}.}} Each cell is initialised with five tracer particles, which are subsequently propagated in a stochastic manner weighted by the mass fluxes between cells. The properties of the cell the particles find themselves in are recorded {at regular intervals}; any changes on shorter timescales than this {interval} will not be accounted for in the chemical post-processing. {Increasing the time resolution of the model by decreasing the interval between updates will produce more accurate results, at the expense of increasing the memory requirements and so (for a fixed memory allocation) reducing the number of tracers which can be post-processed. We investigate the sensitivity of our model to the chosen time resolution in Section \ref{sec:time}, and for the remainder of the paper use a value of $44 \kyr$, which we find provides an acceptable trade-off between accuracy and computational cost {(the chemical models described below have a runtime of $\sim 200$ CPU-hours, and show excellent parallel scaling, compared to $\sim 6000$ CPU-hours to run the underlying MHD simulation)}.}

The UV field strength and cosmic ray ionisation rate are kept the same as in the {\sc arepo} models. We use the UMIST12 \citep{mcelroy2013} chemical network, which includes additional elements not present in that of \citet{gong2017}. The abundances of these elements, listed in Table \ref{tab:abun}, are also taken from \citet{sembach2000}, but we reduce the values of the refractory elements (Mg and Si) by a factor of 100 so that the level of depletion is consistent with the `metal' abundance in {\sc arepo}. Although astrochemical models often also deplete sulphur in order to match observations of S-bearing species \citep[e.g.][]{navarro2020}, we leave its abundance at the \citet{sembach2000} value; there is no known solid-phase reservoir that can account for any significant level of sulphur depletion \citep{jenkins2009,boogert2015}, and recent work suggests that the gas-phase abundance may in fact be close to its undepleted value \citep{hily2022}.

The quantities required to calculate the chemical evolution of a particle are the number density of hydrogen nuclei ($\nh$), the gas temperature\footnote{{\sc uclchem} assumes that the gas and dust temperatures are equal. This is not necessarily the case, particularly at low densities, but the dust temperature is only used to calculate the rates of diffusion reactions on grain surfaces. These become significant only at relatively high densities ($> 10^4 \pcc$), where dust and gas temperatures are {close to} equal, {so the distinction is unlikely to have much impact on our present results. We plan to extend the model to follow gas and dust temperatures separately in future work.}} ($T$), and the {effective shielding} column densities {(distinct from the line-of-sight columns; \citealt{clark2014})} of hydrogen nuclei ($\Nhshield$), molecular hydrogen ($\Nhmol$), and CO ($\Nco$). For $\Nhshield$, we use the same approach as in {\sc arepo}, and take a weighted average over the $n_{\rm PIX}$ {\sc treecol} rays,
\begin{equation}
  \label{eq:nhcol}
  \exp\left( -2.5 \, \av \right) = \frac{1}{n_{\rm PIX}} \displaystyle\sum_i^{n_{\rm PIX}} \exp\left( -2.5 \, A_{\rm V,i} \right),
\end{equation}
where $\av = 5.3 \times 10^{-22} \left( \Nhshield / \pcs \right) \, {\rm mag}$ \citep{bohlin1978}. The factor of $2.5$ is representative of the key photodissociation reactions in the network \citep{glover2007}.

The self-shielding of H$_2$ and CO is more complicated, as the lines involved may saturate, and broadening and Doppler shifting of the frequencies can be important \citep{glover2007}. While {\sc arepo} calculates the self-shielding factor for each ray individually, we find that the harmonic mean of the column density of the relevant molecule,
\begin{equation}
  \label{eq:sscol}
  (N_X^{\rm shield})^{-1} = \frac{1}{n_{\rm PIX}} \displaystyle\sum_i^{n_{\rm PIX}} (N_{X,i}^{\rm shield})^{-1},
\end{equation}
provides an acceptable approximation. Self-shielding factors are then calculated following \citet{federman1979} for H$_2$ and \citet{vandishoeck1988} for CO. We also replace the default formation rate of H$_2$ on grain surfaces with
\begin{equation}
  \label{eq:h2form}
  R_{\rm H_2} = 3 \times 10^{-18} \, \sqrt{T} \, \exp\left(-\frac{T}{1000 \kel}\right) \, {\rm cm^{3} \, s^{-1}}
\end{equation}
from \citet{dejong1977}, to suppress its formation at high temperatures.

We randomly select $10^5$ tracer particles to post-process, which at the simulation endpoint are located within $16.2 \pc$ of the centre of the computational domain, distributed so that we evenly sample\footnote{{We achieve this by randomly selecting $5000$ particles from each of $20$ logarithmically-spaced density bins. The distribution of particles over these density bins is {\it not} uniform, leading to the `striping' visible in Figure \ref{fig:hist}, where particles are more densely sampled toward one edge of a bin than the other.}} densities in the range $10-10^7 \pcc$. Figure \ref{fig:hist} shows the distributions of $T$ and $\Nhshield$ with $\nh$ for these particles; the chosen density range corresponds to temperatures $\sim 5-200 \kel$ and visual extinctions $\sim 0.1-30$ mag. The general trend of the temperature distribution can be reproduced by the equation of state
\begin{equation}
  \label{eq:temp}
  T = 120 \left(\nh / 10 \pcc\right)^{-0.8} + 10 \left(\nh / 10^5 \pcc\right)^{-0.15} \kel,
\end{equation}
although this fails to capture the plateau at $\sim 20 \kel$ between $10^3-10^4 \pcc$. Similarly, the typical column density is correlated with density as
\begin{equation}
  \label{eq:av}
  \Nhshield = 6 \times 10^{20} \left(\nh / 10 \pcc\right)^{0.3} \pcs,
\end{equation}
but with $\sim 0.5$ dex of scatter between particles of the same density. This is close to the $\Nhshield \propto \nh^{0.5}$ scaling that would result from shielding by a Jeans length of material at the same volume density (as adopted in \citealt{priestley2023b}); the slightly lower exponent likely represents the fact that the average density within a Jeans length of a point tends to be lower than the density {\it at} that point.

Figure \ref{fig:chemtest} shows how the post-processed abundances of H$_2$ and CO compare to the internal {\sc arepo} abundances. The H$_2$ abundances predicted by NEATH are consistent with the {\sc arepo} values over the entire $10-10^7 \pcc$ density range; while the values for individual particles are not always identical, the differences are comparable to or smaller than the scatter between different particles of the same density, and the overall trend of H$_2$ abundance with density is reproduced. The CO abundances are similarly consistent between NEATH and {\sc arepo} between $10^2-10^4 \pcc$. At lower densities, NEATH appears to underpredict the amount of CO, likely because our self-shielding column densities from Equation \ref{eq:sscol} are underestimating the true `effective' value. At higher densities, NEATH again predicts lower CO abundances than {\sc arepo}, with increasing severity as the density increases, but in this regime the difference is due to the depletion of CO onto grain surfaces. Freeze-out is not modelled in the \citet{gong2017} network, so this behaviour is expected, and the post-processed NEATH abundances are overall in good agreement with those in the underlying MHD simulation.

\subsection{Effects of time resolution}
\label{sec:time}

We investigate the sensitivity of the chemistry to unresolved changes in the gas properties with a simplified model of {a transient density enhancement (hereafter `shock', although we note that this is different from genuine shocks such as in protostellar outflows)}, treated as a temporary and instantaneous change in the gas density. The temperature and column density are assumed to follow the relationships given by Equations \ref{eq:temp} and \ref{eq:av} respectively. Column densities for self-shielding are obtained by multiplying Equation \ref{eq:av} by the local abundance of the relevant molecule. The initial density is $10^3 \pcc$, which is enhanced to $10^5 \pcc$ after $0.515 \myr$ of evolution (sufficient to establish chemical near-equilbrium) for a duration of $40 \kyr$, after which it reverts to the original value and evolution continues up to an end point of $1 \myr$. We perform the chemical modelling with timesteps of $10 \kyr$ and $50 \kyr$; in the latter case, the shock is missed completely, and the chemical model assumes all physical properties remain constant.

Figure \ref{fig:shocktest} shows the evolution of the gas density and the abundances of CO and NH$_3$, chosen for their different chemical behaviour (see Section \ref{sec:chemdiff}). With a time resolution of $50 \kyr$, the transient density enhancement is not captured, and the abundances remain nearly constant (there is a slight increase for both molecules as they have not quite reached their equilibrium values). CO is barely affected by the shock; the increased density causes a slight decrease (around a factor of two) due to depletion, but the duration is not long enough for this to have much effect, and once the initial density is restored the CO abundance returns to its original value within a few $100 \kyr$.

The shock has a much greater impact on the NH$_3$ abundance, which increases abruptly by several orders of magnitude during the density enhancement. However, it declines similarly abruptly once the density enhancement ends. While the abundance does not return to its pre-shock value over the $\myr$ duration of the model, the difference is only a factor of a few compared to the case where the shock was not captured at all. These differences are negligible compared to those caused by different tracer particle histories and/or the current physical conditions. Temporary changes in the gas properties on timescales too short to be captured are unlikely to have any significant impact on our results.

While unresolved transient changes do not appear to be an issue, poor resolution of rapid evolution is more problematic. Figure \ref{fig:restest} shows the median CO abundance versus density for a smaller sample of $10^4$ tracer particles, with the time resolution ranging from $8.8$ to $88 \kyr$. The chemical model is well-converged up to densities of $\sim 10^4 \pcc$, but at higher densities the gas-phase CO abundance decreases as the time resolution increases. Much of the material in this density regime is gravitationally collapsing, and so evolves on approximately the freefall timescale $t_{\rm ff} \sim 0.8 \, (\nh / 10^4 \pcc)^{-0.5} \myr$. If the interval between updates of physical properties does not fully resolve the freefall time, the chemistry spends too long at too low a density, and so undergoes less freeze-out than it should, leading to the higher CO abundances in the lower resolution models.

This effect becomes increasingly significant as the density increases and the freefall timescales decreases; while our fiducial $44 \kyr$ timestep is converged up to a density of $10^5 \pcc$, this is clearly not the case at $10^7 \pcc$. {\citet{ferrada2021} suggest a chemical timestep below $0.01 \, t_{\rm ff}$ is required for percent-level accuracy in the molecular abundances, which corresponds to $0.3 \kyr$ at a number density of $10^7 \pcc$.} However, increasing the time resolution by a factor of $5$ {from our fiducial value} only results in a factor of a few decrease in the CO abundance at densities up to $10^6 \pcc$, suggesting that we are reasonably close to convergence in this regime, and the mass fraction of gas with higher densities is extremely small (about $10^{-4}$). {Reaction rates in the chemical network are rarely, if ever, known to better precision than a factor of a few \citep[e.g.][]{rocha2023}, so a percent-level numerical accuracy in abundance would be dwarfed by systematic uncertainties. Based on the results in Figure \ref{fig:restest}, we suggest that a time resolution of $\sim 0.3 \, t_{\rm ff}$ is sufficient to obtain abundances accurate to within a factor of two.} The fiducial time resolution of $44 \kyr$ is likely to be adequate for investigating the overall chemical composition of the clouds, {being effectively converged between number densities of $10^2-10^6 \pcc$}.

\citet{panessa2023} deal with the issue of poorly-resolved changes in physical properties by `subcycling'; using multiple shorter timesteps for the chemical evolution and interpolating between the initial and final physical properties if the difference between these is above a certain threshold. With this approach, it is not clear how exactly the interpolation should be done. \citet{panessa2023} use a linear interpolation, but the gas density of a gravitationally collapsing region grows exponentially \citep{larson1969}, so a linear interpolation will tend to overestimate the amount of time spent at high density and thus the extent to which molecules are depleted. This is unlikely to be an issue in their case, as their simulations do not extent far beyond a density of $10^4 \pcc$, but the method is not easily applicable to our smaller-scale, higher-density simulations.

\begin{figure*}
  \centering
  \includegraphics[width=0.31\textwidth]{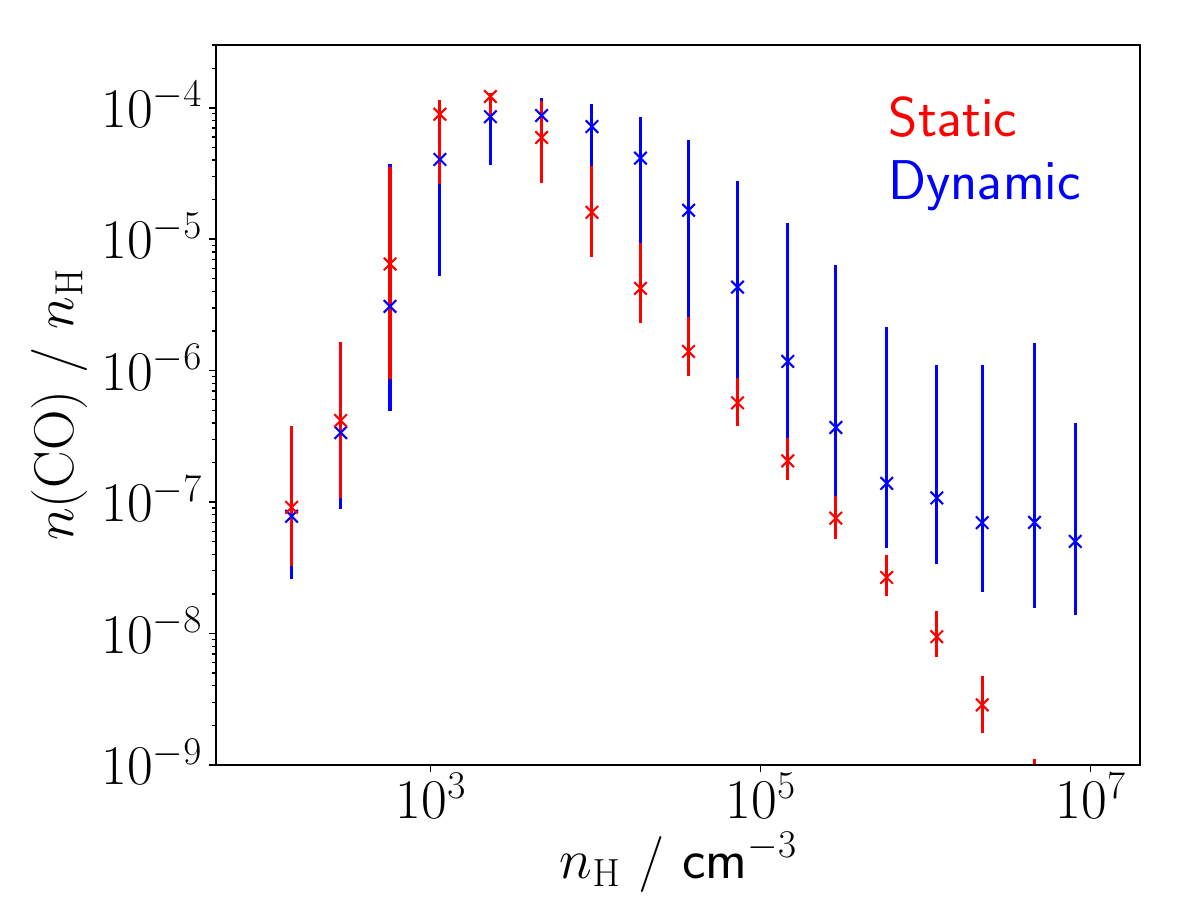}\quad
  \includegraphics[width=0.31\textwidth]{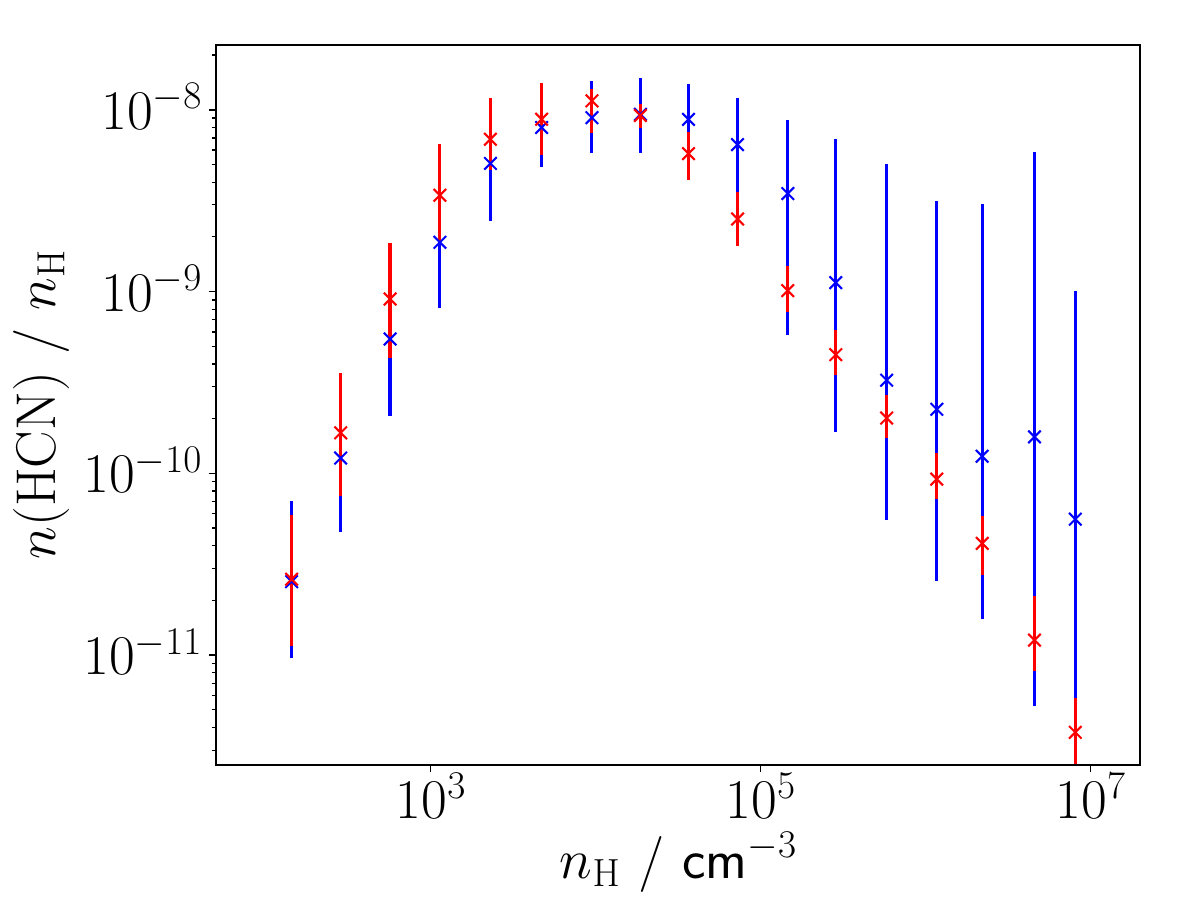}\quad
  \includegraphics[width=0.31\textwidth]{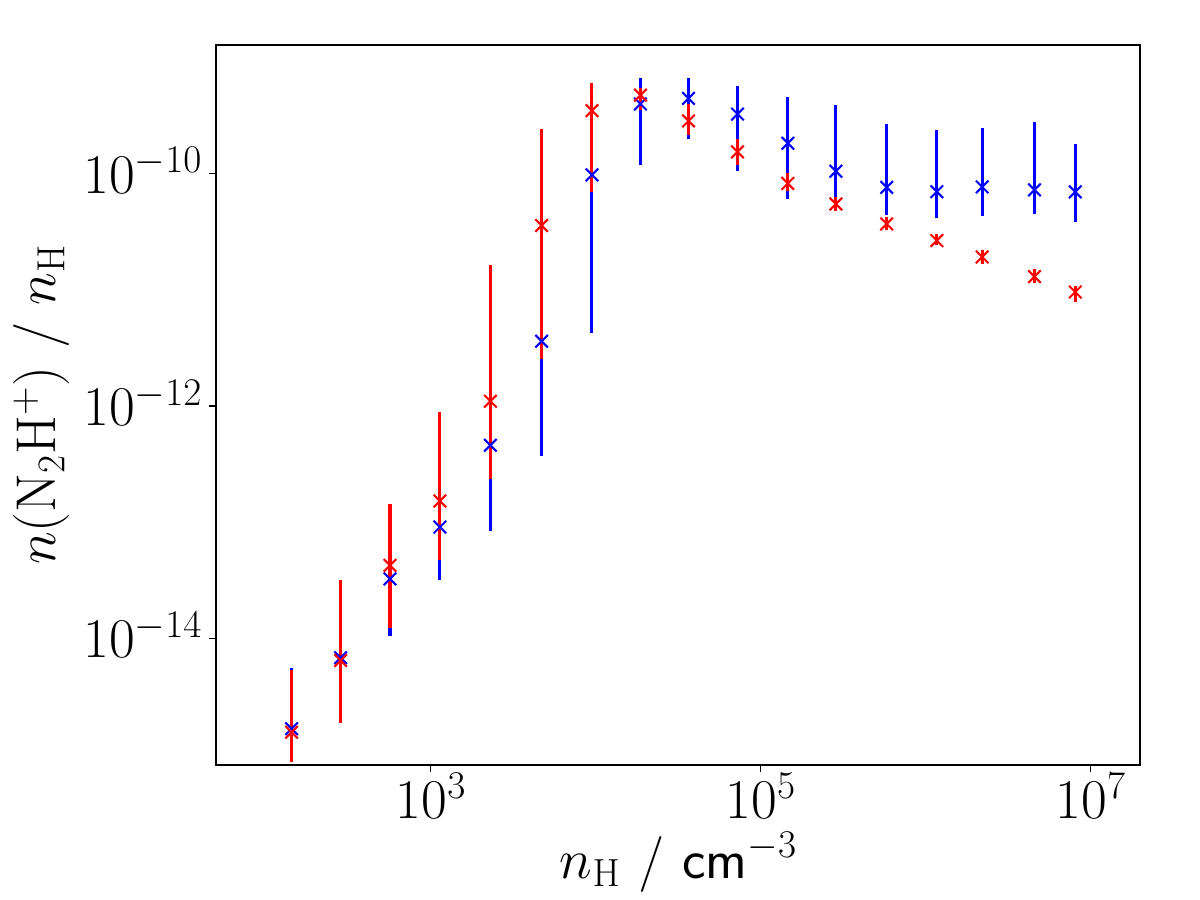}\quad
  \caption{Average abundances versus gas density of CO (left), HCN (centre) and N$_2$H$^+$ (right), for {dynamic} (blue) and {static} (red) models. Crosses show the median values, with the 16th and 84th percentiles as error bars.}
  \label{fig:equibtest}
\end{figure*}

\begin{figure*}
  \centering
  \includegraphics[width=0.31\textwidth]{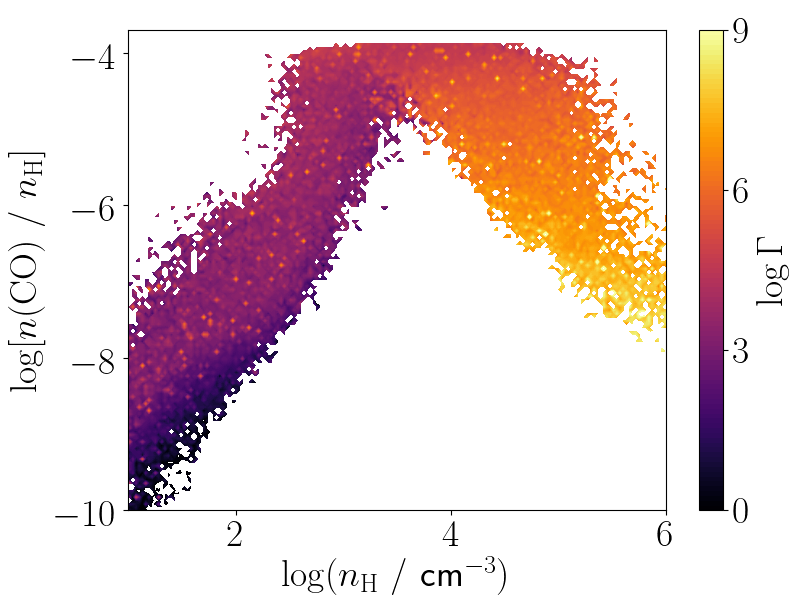}\quad
  \includegraphics[width=0.31\textwidth]{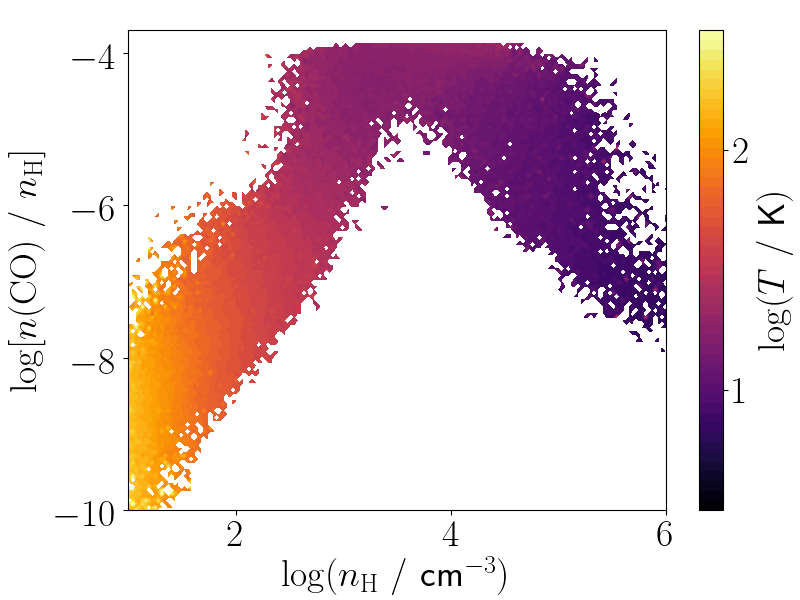}\quad
  \includegraphics[width=0.31\textwidth]{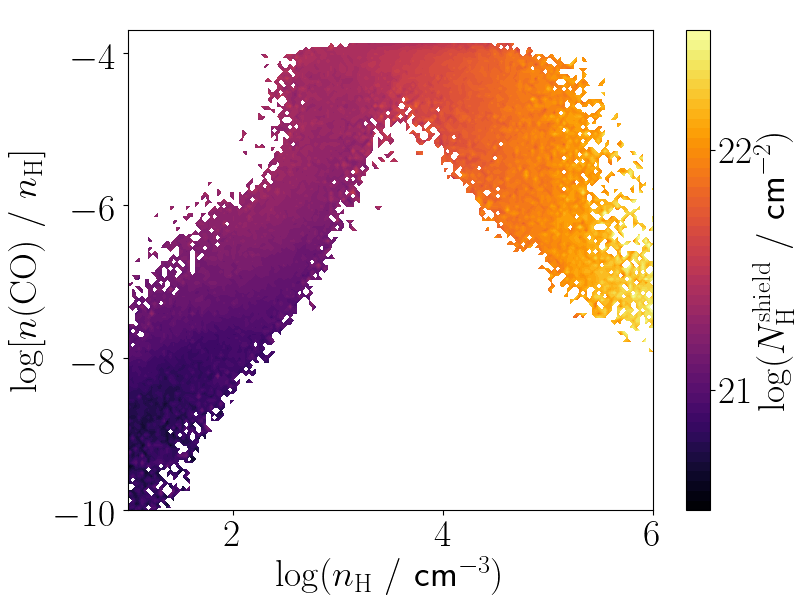}\quad
  \caption{Average value of the overdensity parameter (left), gas temperature (centre) and shielding column density (right), as a function of gas density and CO abundance.}
  \label{fig:overdense}
\end{figure*}

\begin{figure*}
  \centering
  \includegraphics[width=\textwidth]{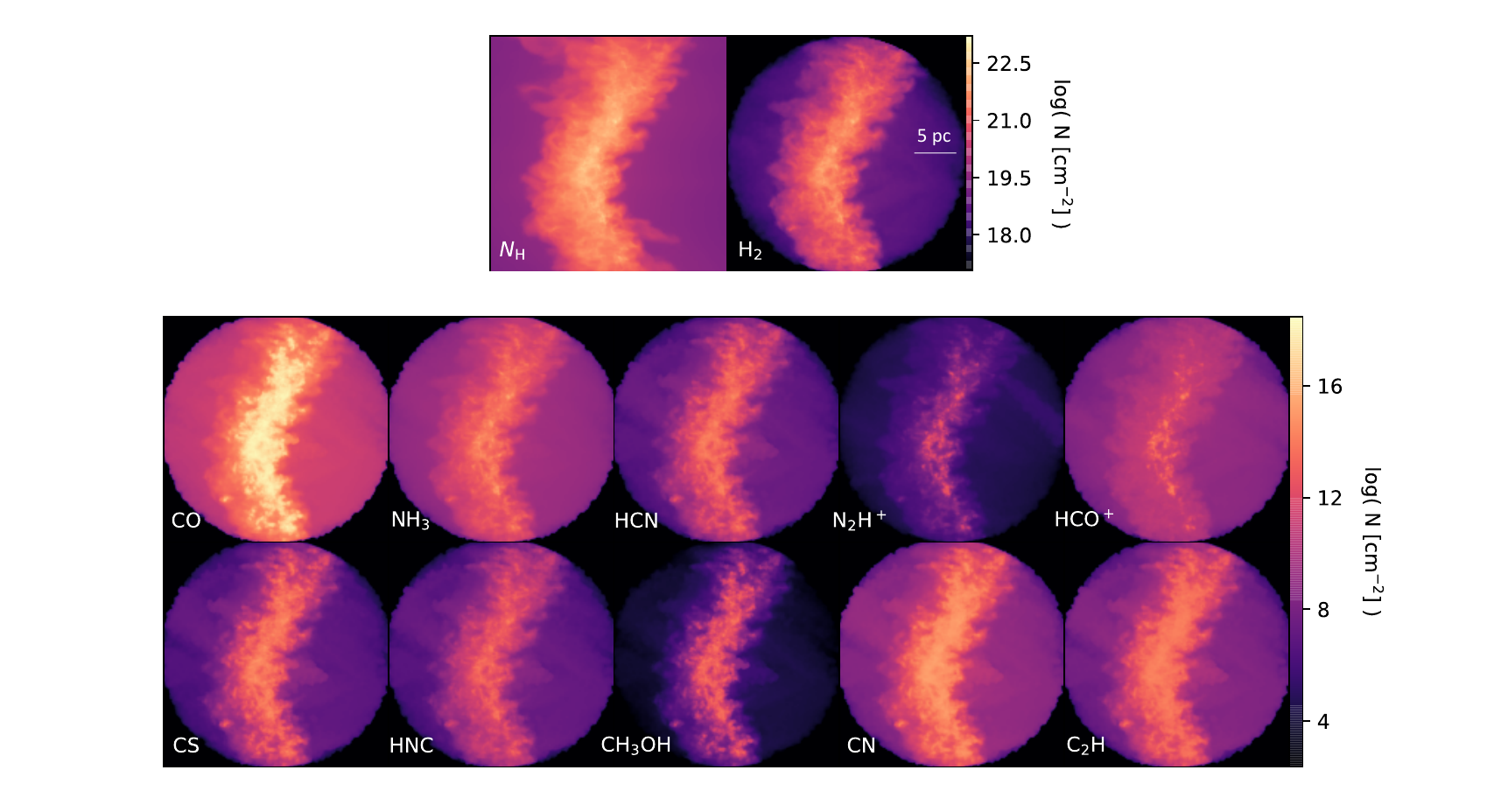}
  \caption{Column densities in the $x-y$ plane of total hydrogen nuclei and H$_2$ (top panel), and of CO, NH$_3$, HCN, N$_2$H$^+$, HCO$^+$, CS, HNC, CH$_3$OH, CN and C$_2$H (bottom panel). {Note that all post-processed tracer particles are located within a sphere of radius $16.2 \pc$, so the abundances and column densities of all molecules are set to zero outside of this region.}}
  \label{fig:col}
\end{figure*}

\begin{figure}
  \centering
  \includegraphics[width=\columnwidth]{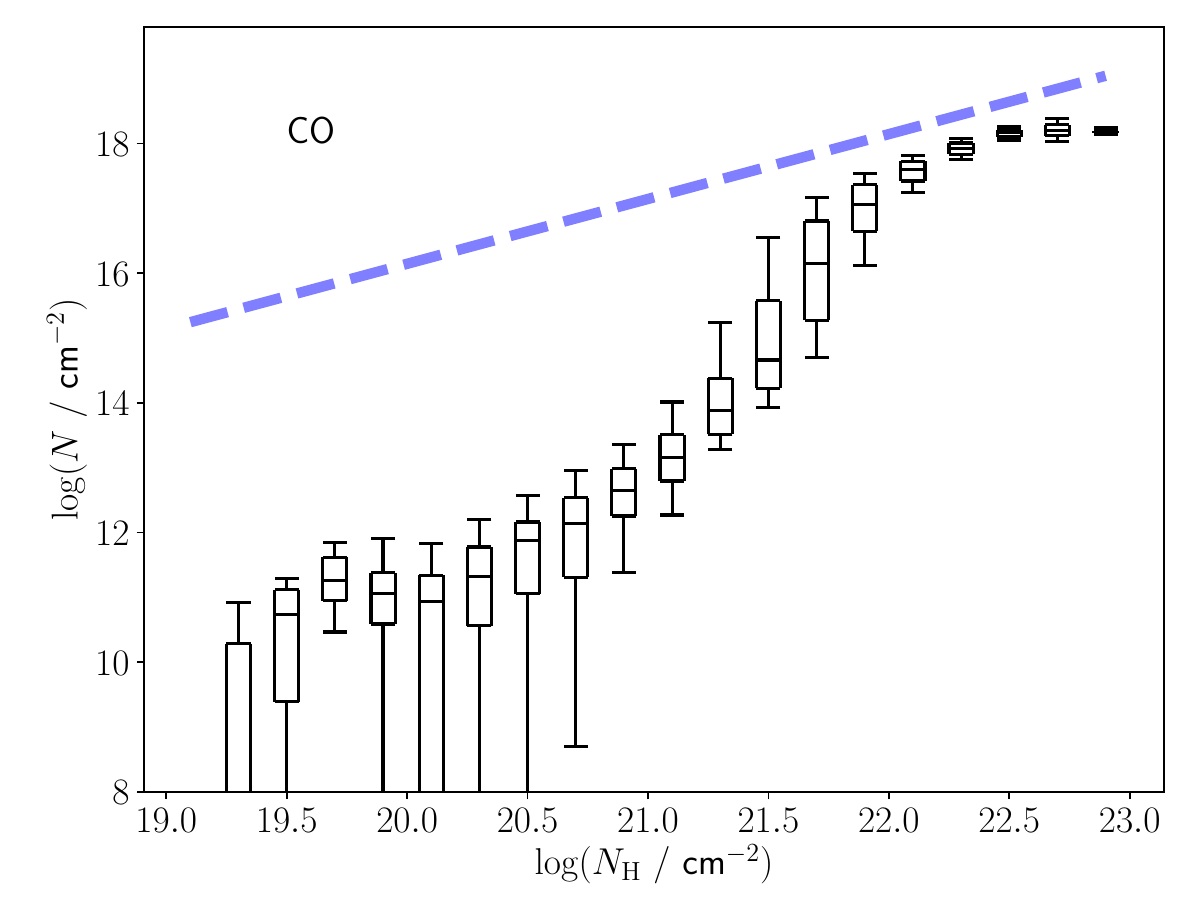}
  \caption{Box plot of CO column density versus $\Nhcol$. Boxes show the median and the 25th/75th percentiles; whiskers show the 10th and 90th percentiles. The dashed blue line shows the column density expected if all carbon is in the form of CO.}
  \label{fig:boxcolco}
\end{figure}

\begin{figure*}
  \centering
  \includegraphics[width=0.31\textwidth]{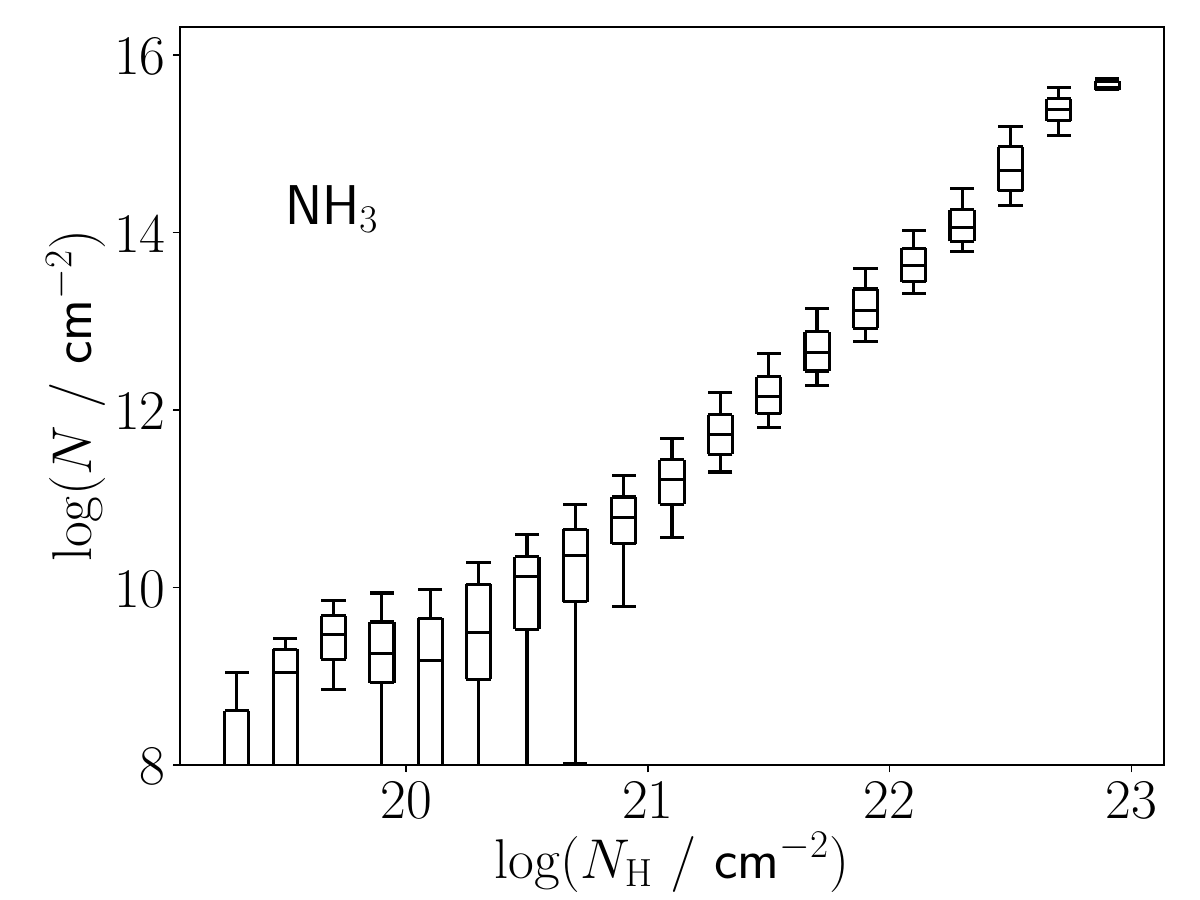}\quad
  \includegraphics[width=0.31\textwidth]{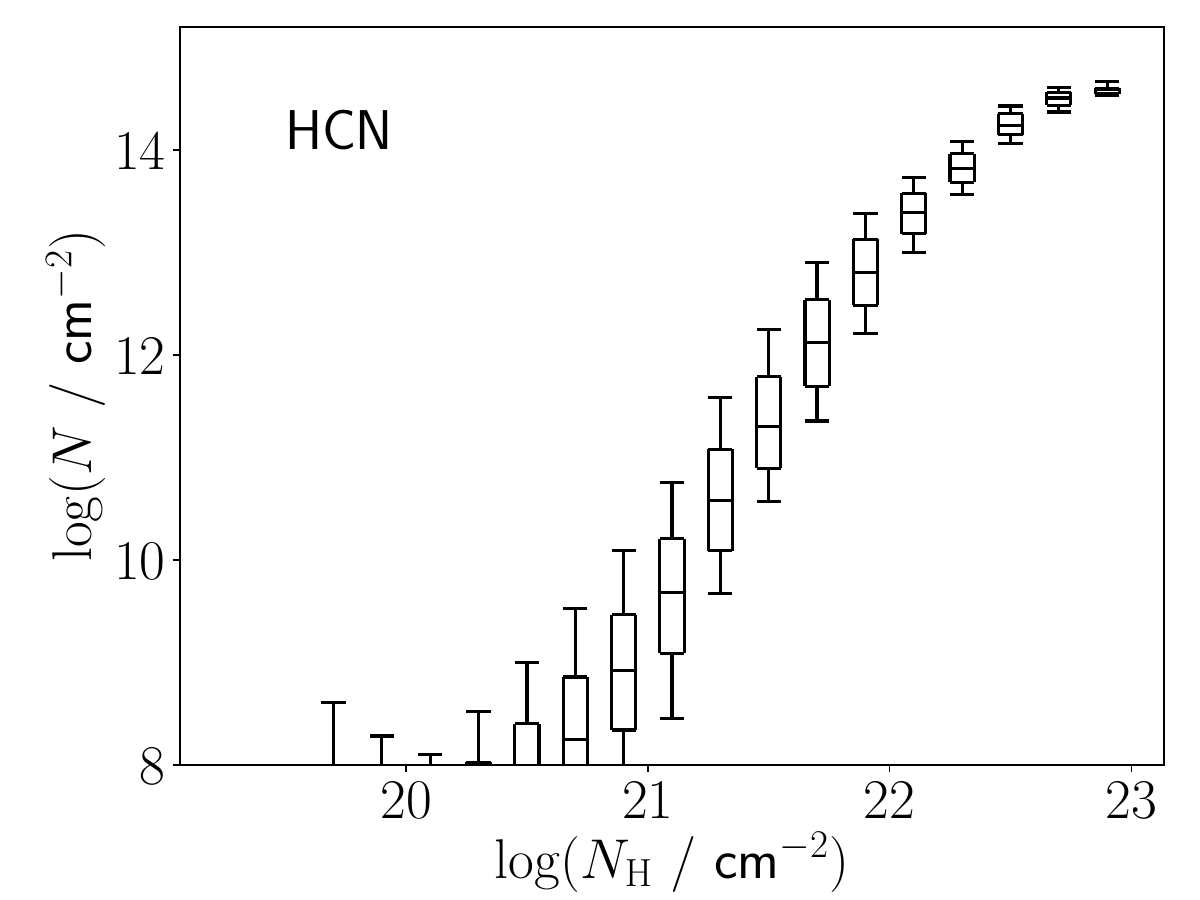}\quad
  \includegraphics[width=0.31\textwidth]{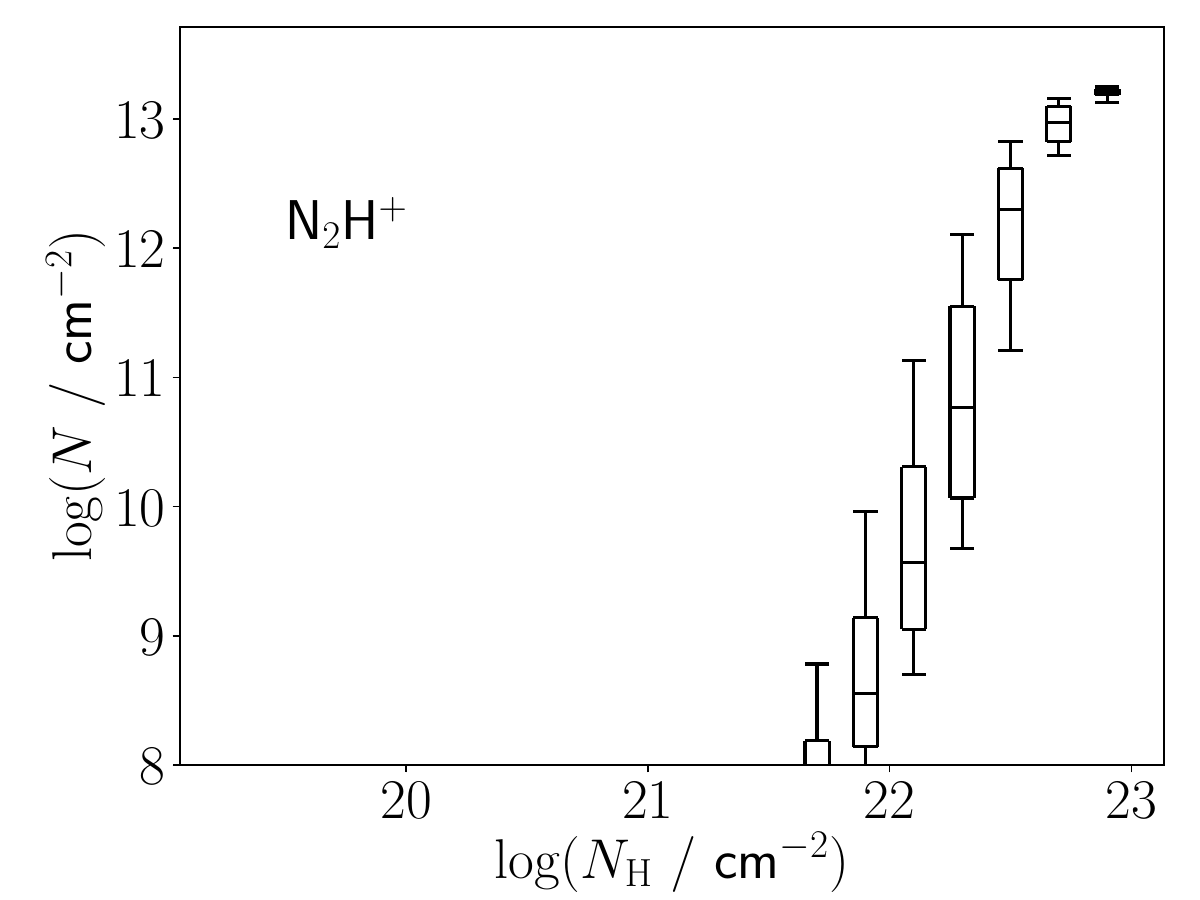}\\
  \includegraphics[width=0.31\textwidth]{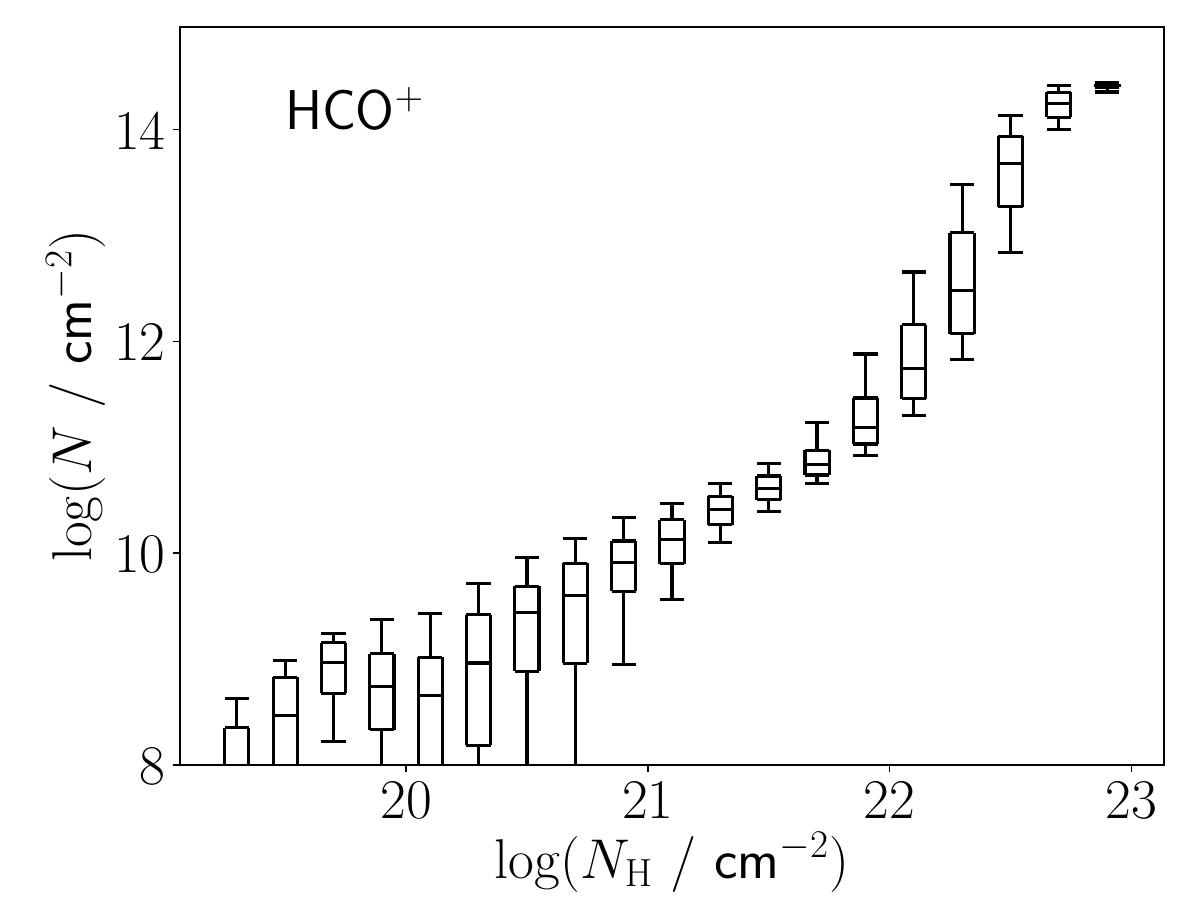}\quad
  \includegraphics[width=0.31\textwidth]{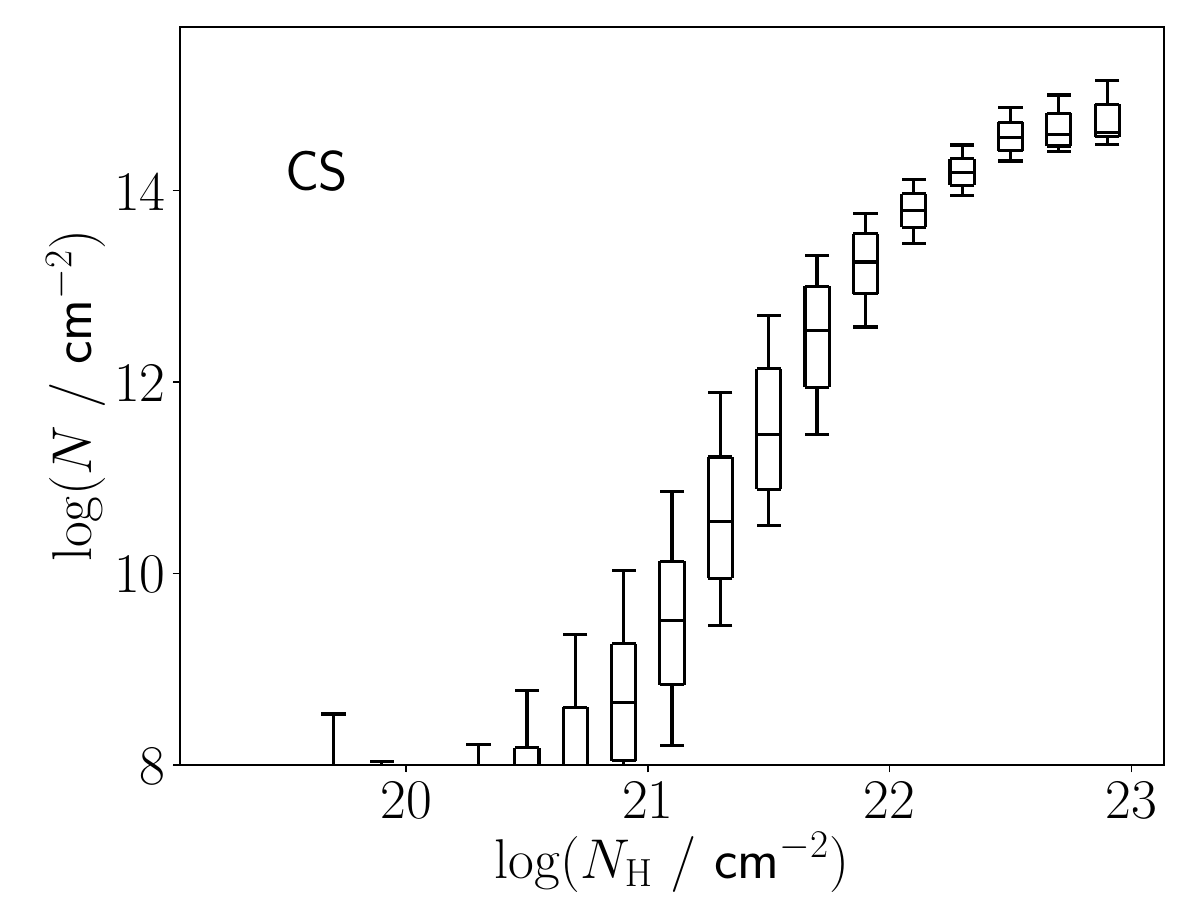}\quad
  \includegraphics[width=0.31\textwidth]{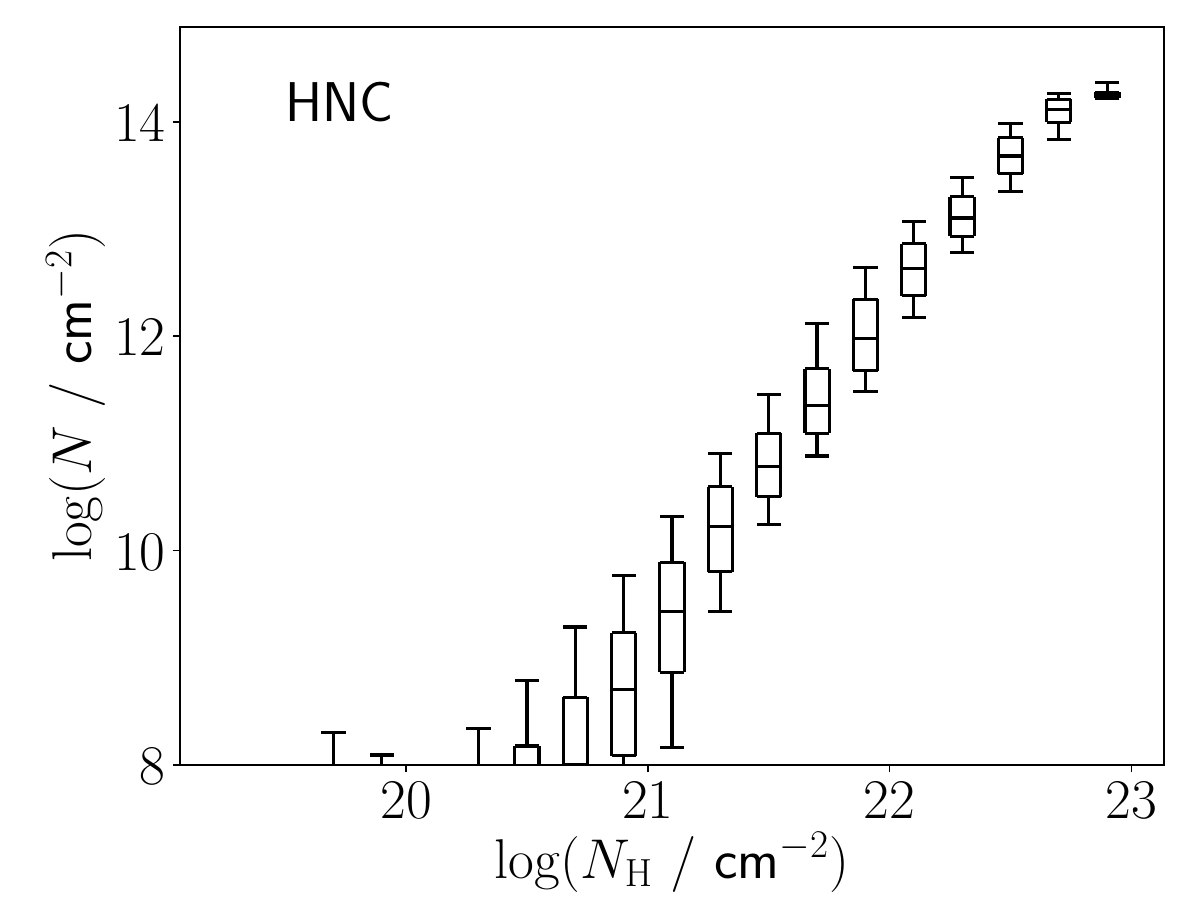}\\
  \includegraphics[width=0.31\textwidth]{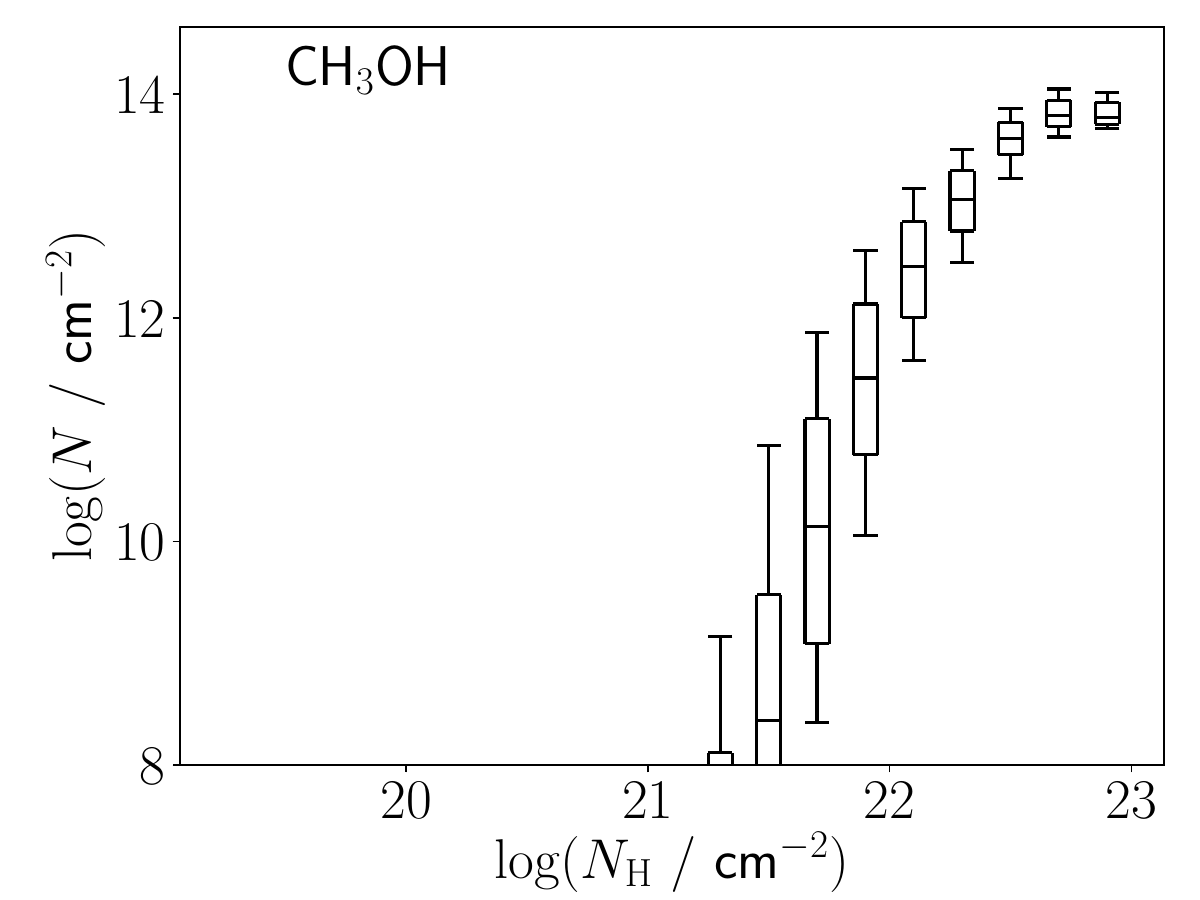}\quad
  \includegraphics[width=0.31\textwidth]{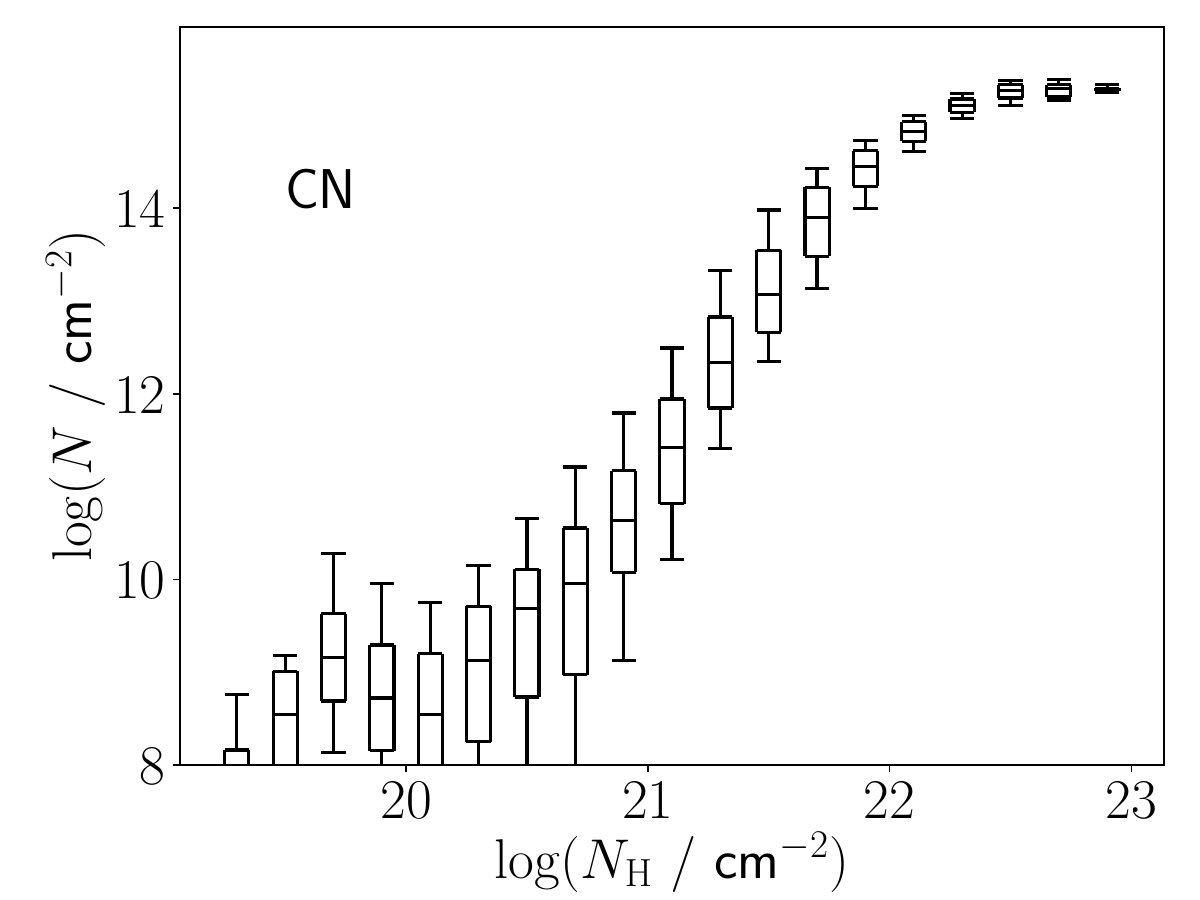}\quad
  \includegraphics[width=0.31\textwidth]{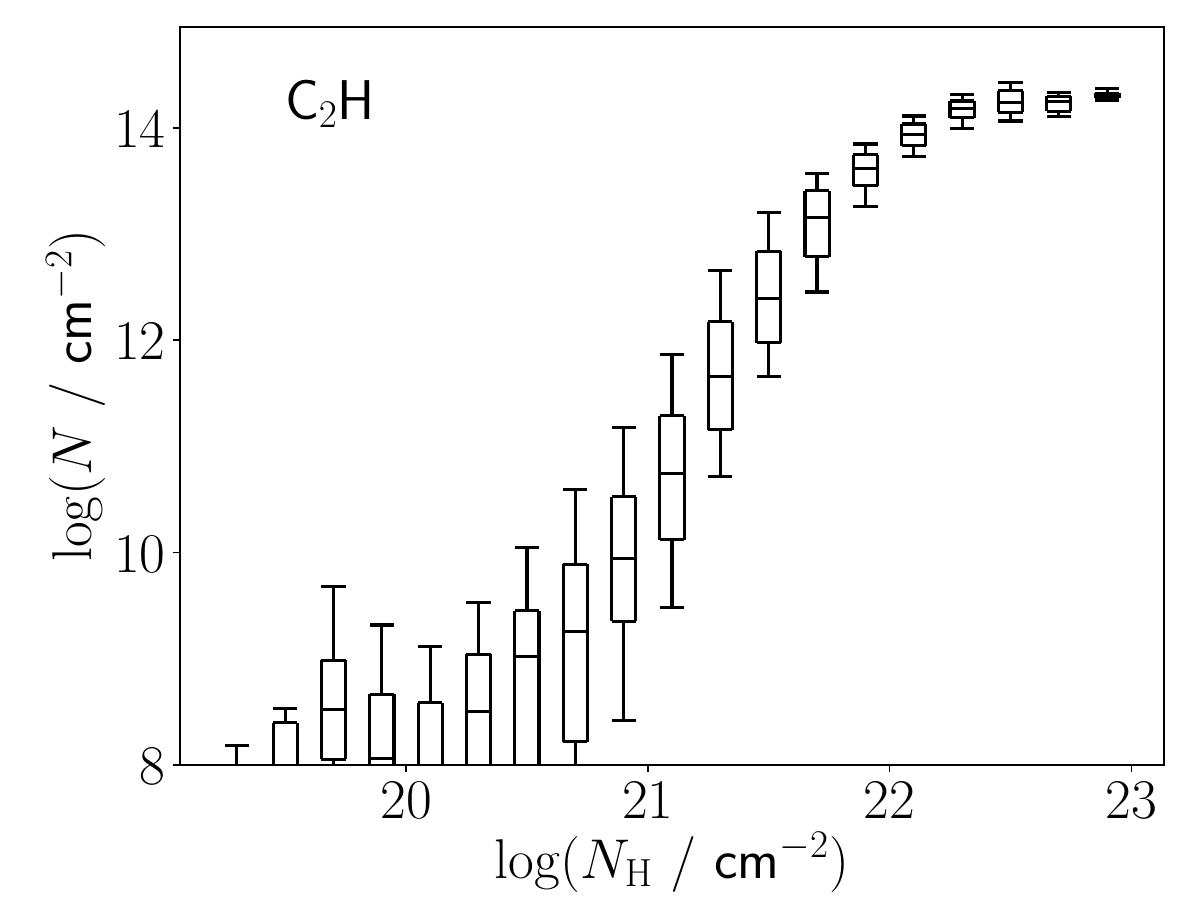}\\
  \caption{Box plots of molecular column density versus $\Nhcol$ for; NH$_3$, HCN and N$_2$H$^+$ (top row); HCO$^+$, CS and HNC (middle row); CH$_3$OH, CN and C$_2$H (bottom row). Boxes show the median and the 25th/75th percentiles; whiskers show the 10th and 90th percentiles.}
  \label{fig:boxcol}
\end{figure*}

\begin{figure}
  \centering
  \includegraphics[width=\columnwidth]{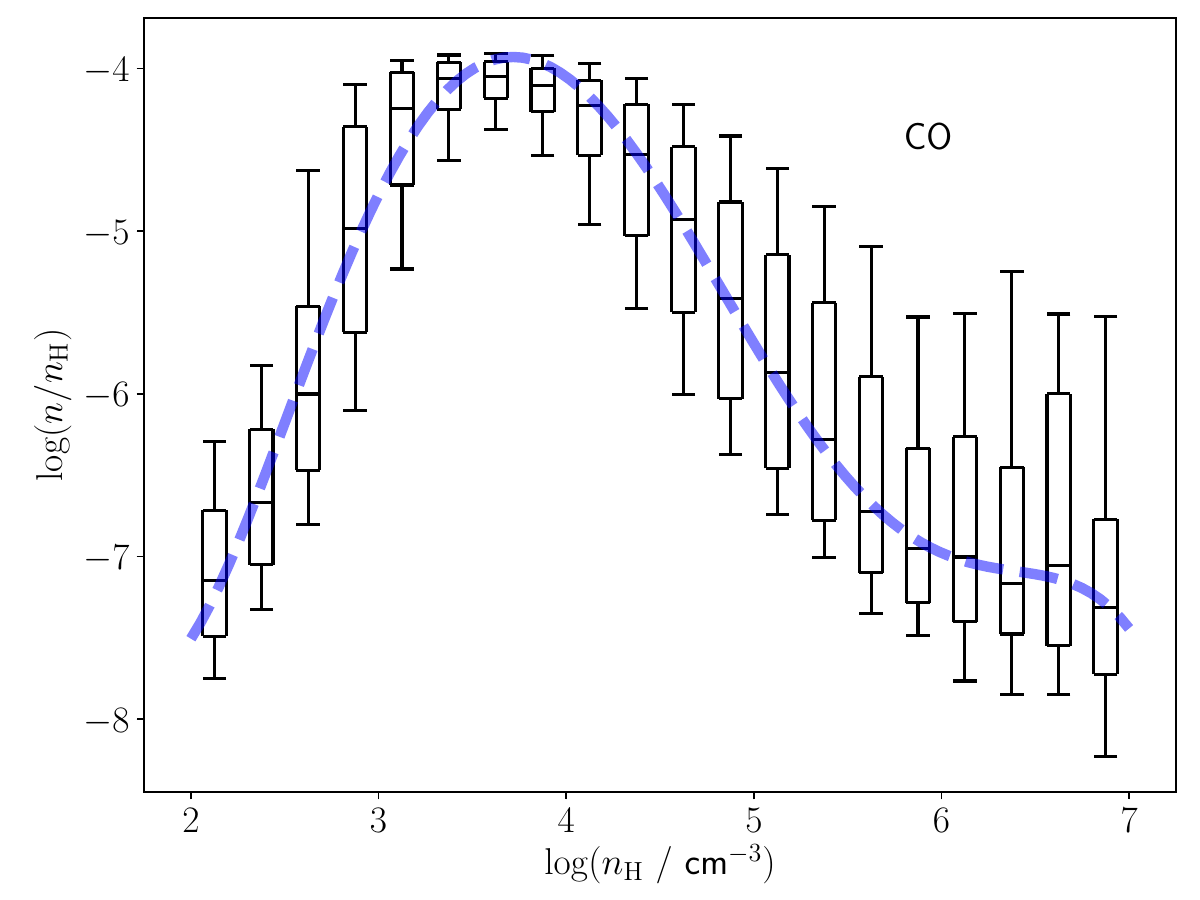}
  \caption{Box plot of CO abundance versus gas density. Boxes show the median and the 25th/75th percentiles; whiskers show the 10th and 90th percentiles. {The dashed blue line shows a fifth-order polynomial fit to the median abundances.}}
  \label{fig:boxco}
\end{figure}

\begin{figure*}
  \centering
  \includegraphics[width=0.31\textwidth]{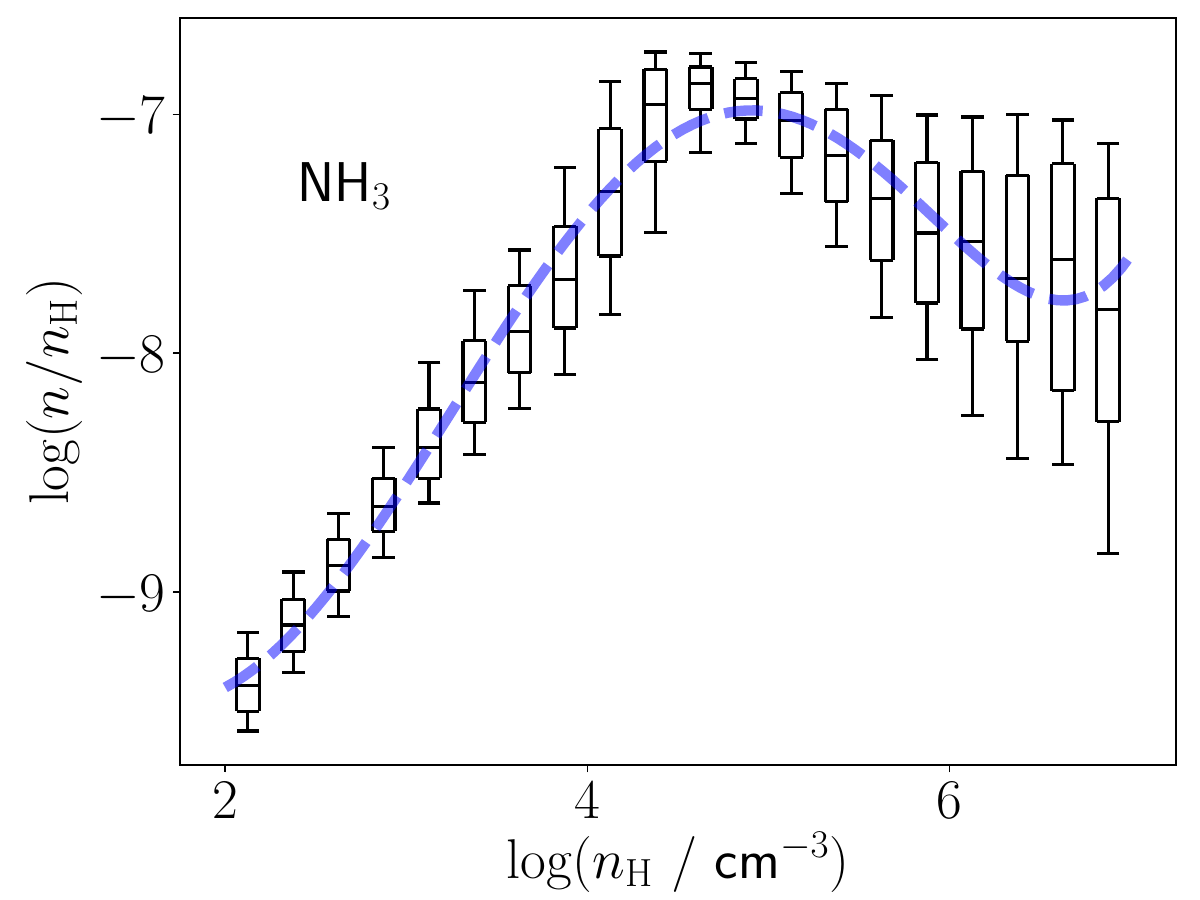}\quad
  \includegraphics[width=0.31\textwidth]{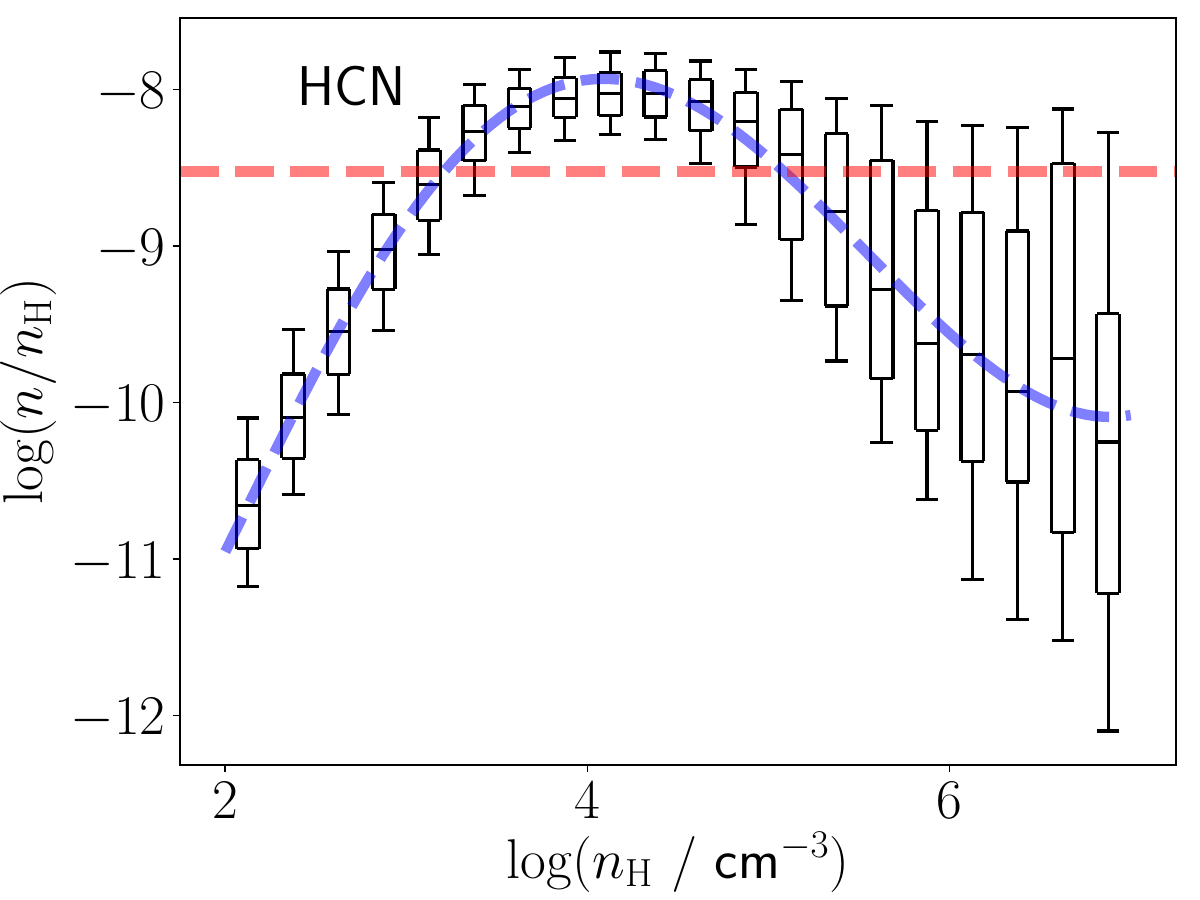}\quad
  \includegraphics[width=0.31\textwidth]{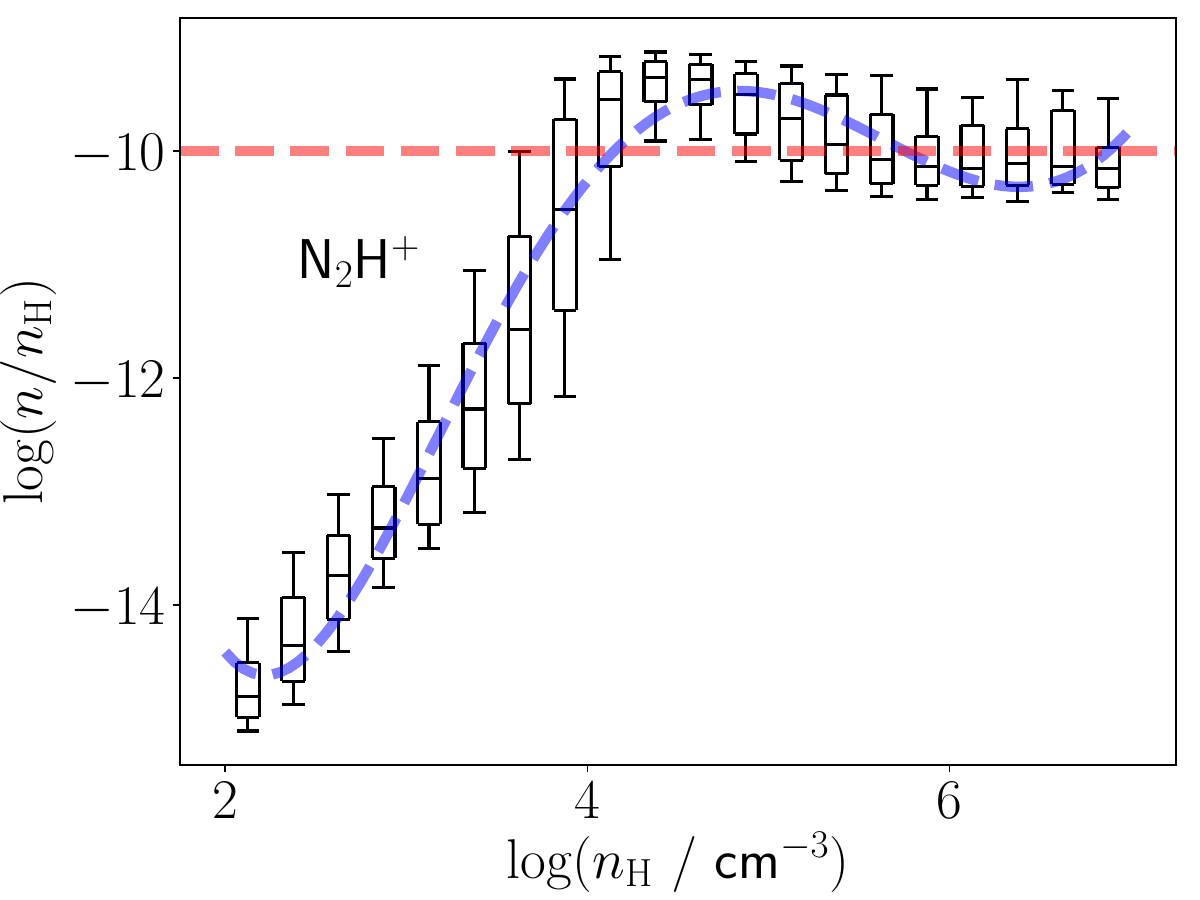}\\
  \includegraphics[width=0.31\textwidth]{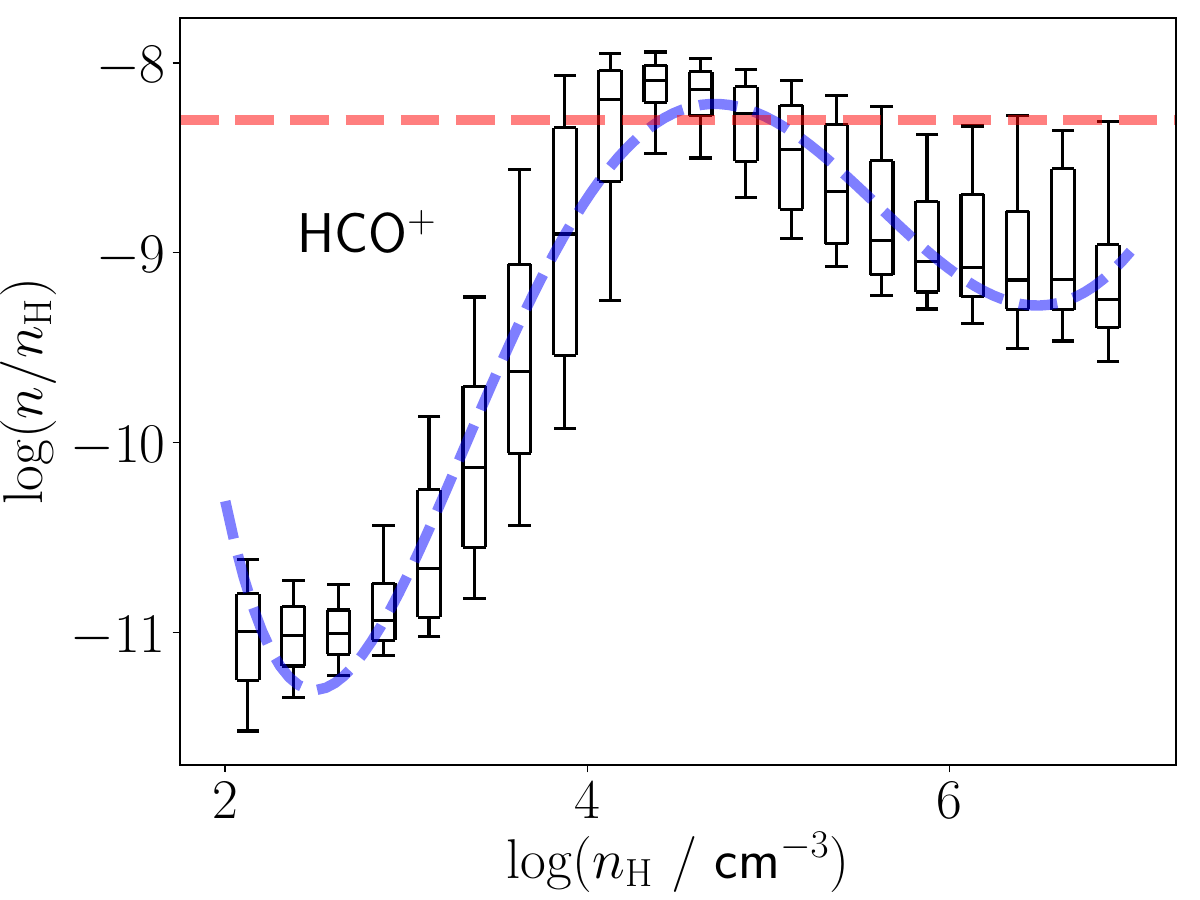}\quad
  \includegraphics[width=0.31\textwidth]{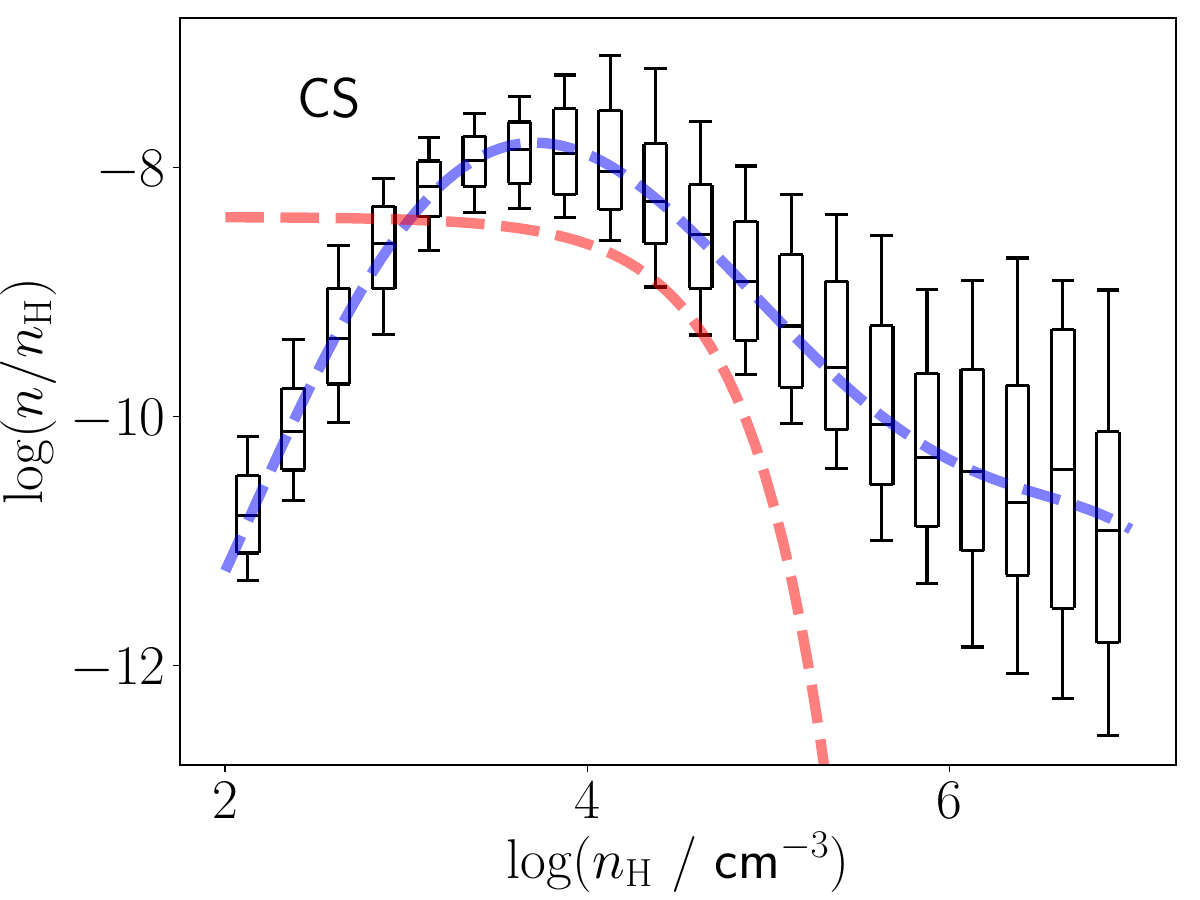}\quad
  \includegraphics[width=0.31\textwidth]{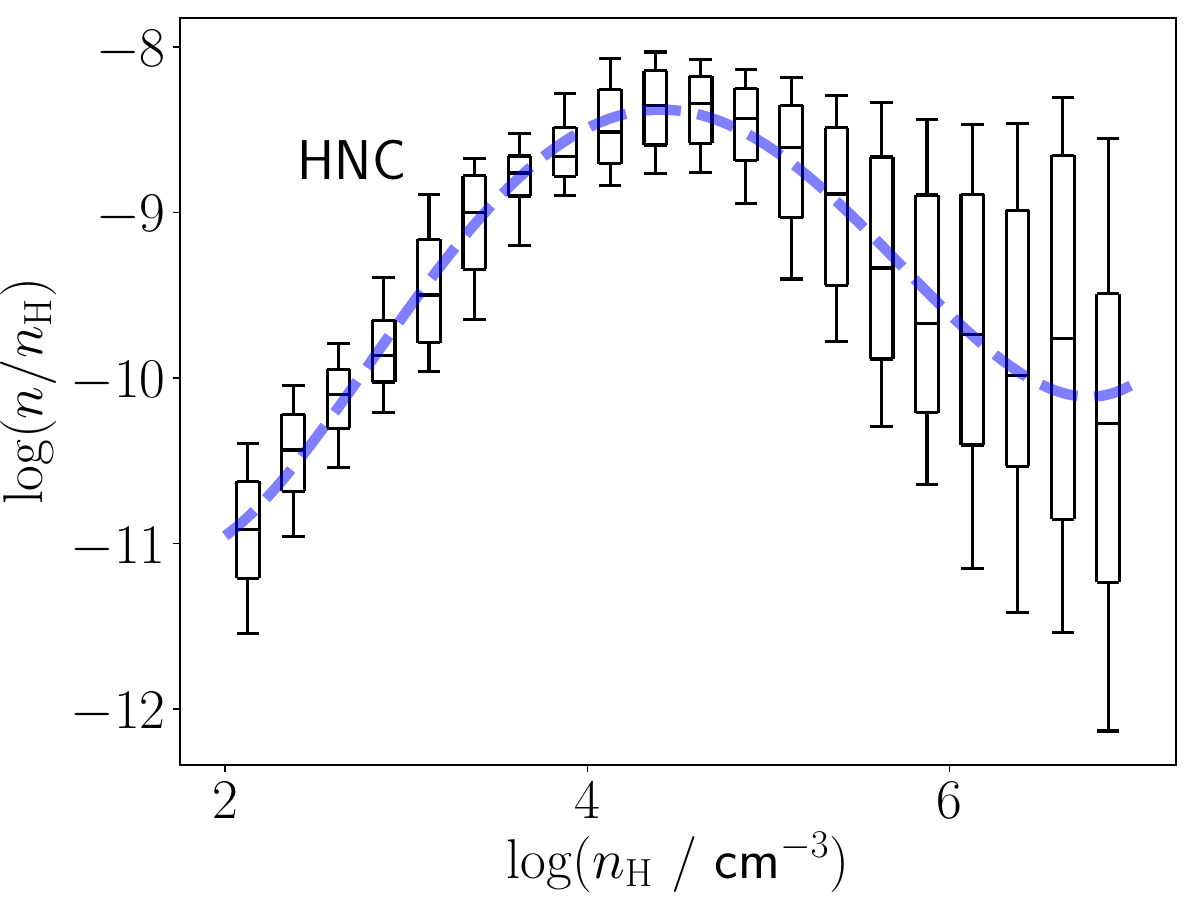}\\
  \includegraphics[width=0.31\textwidth]{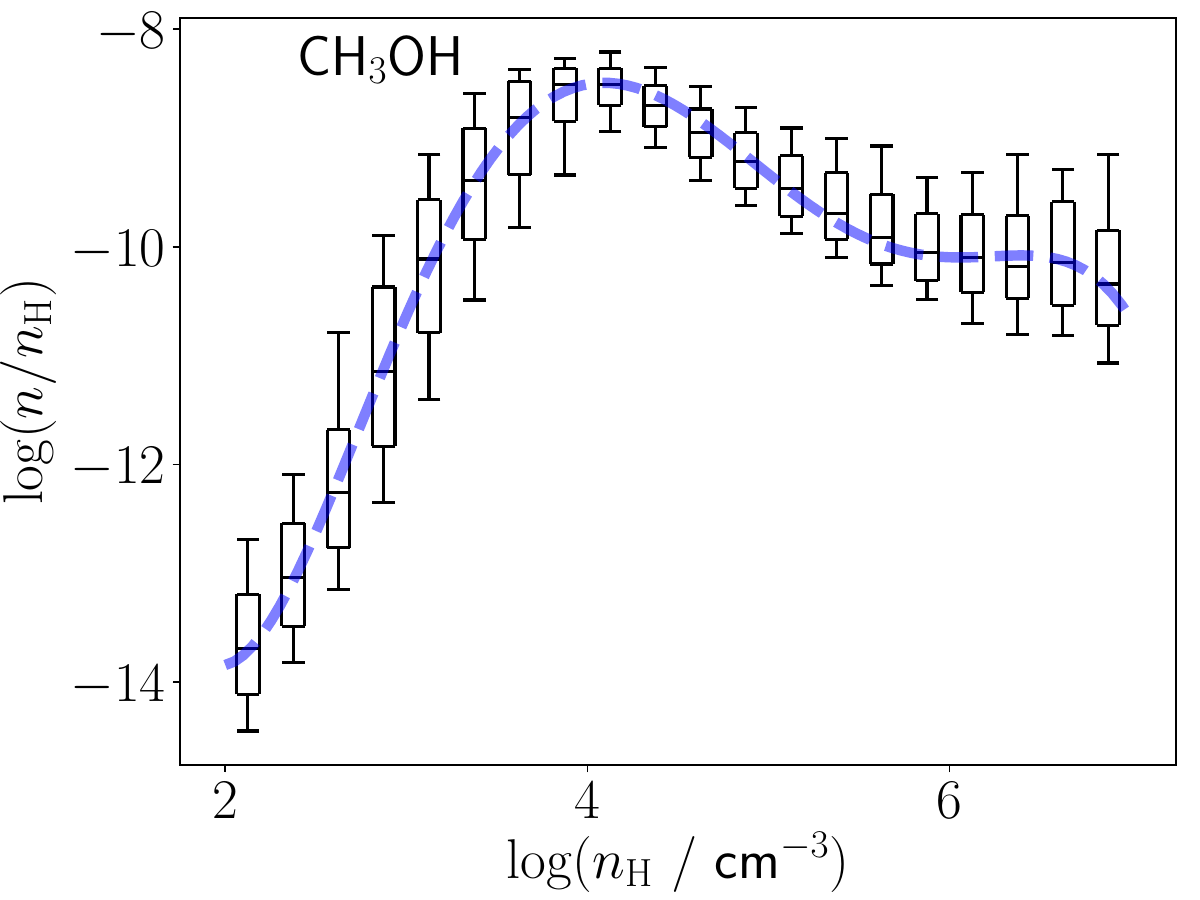}\quad
  \includegraphics[width=0.31\textwidth]{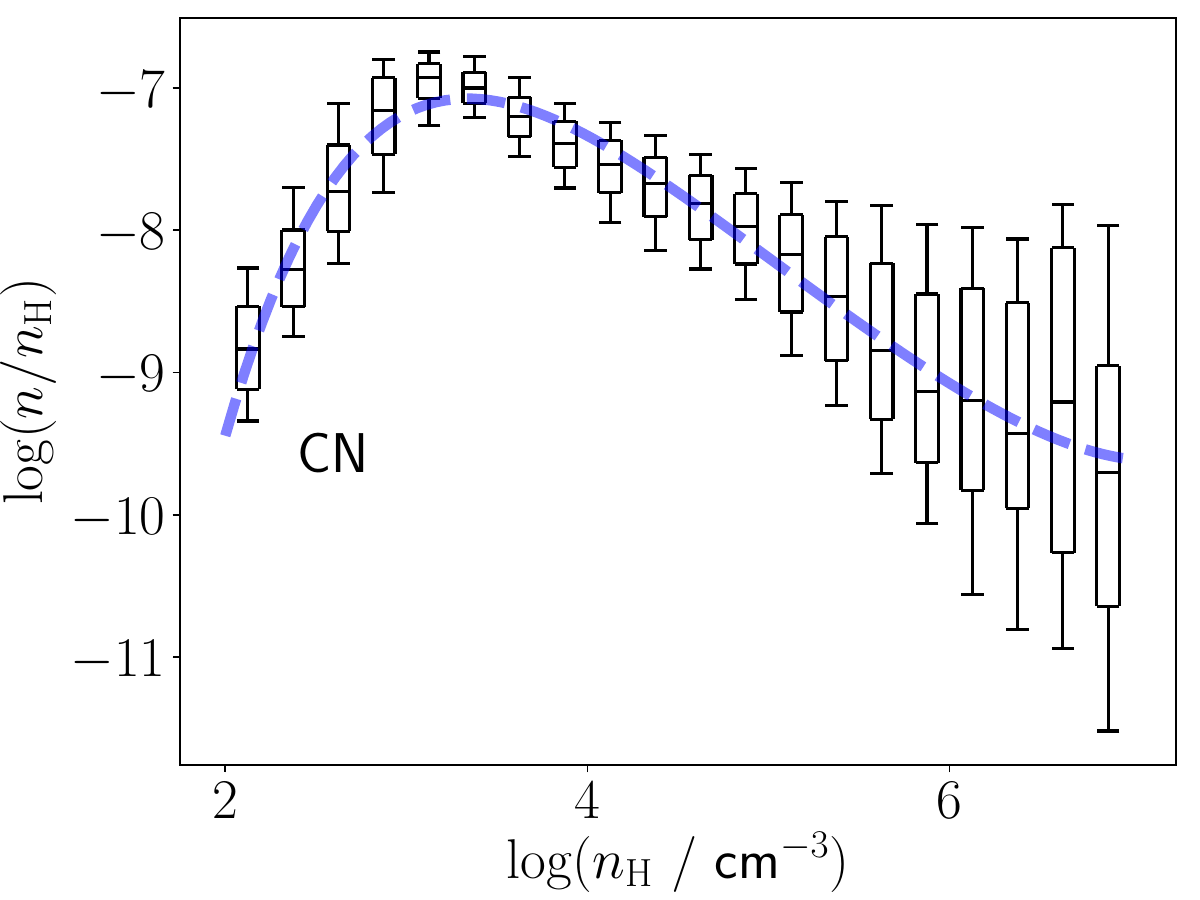}\quad
  \includegraphics[width=0.31\textwidth]{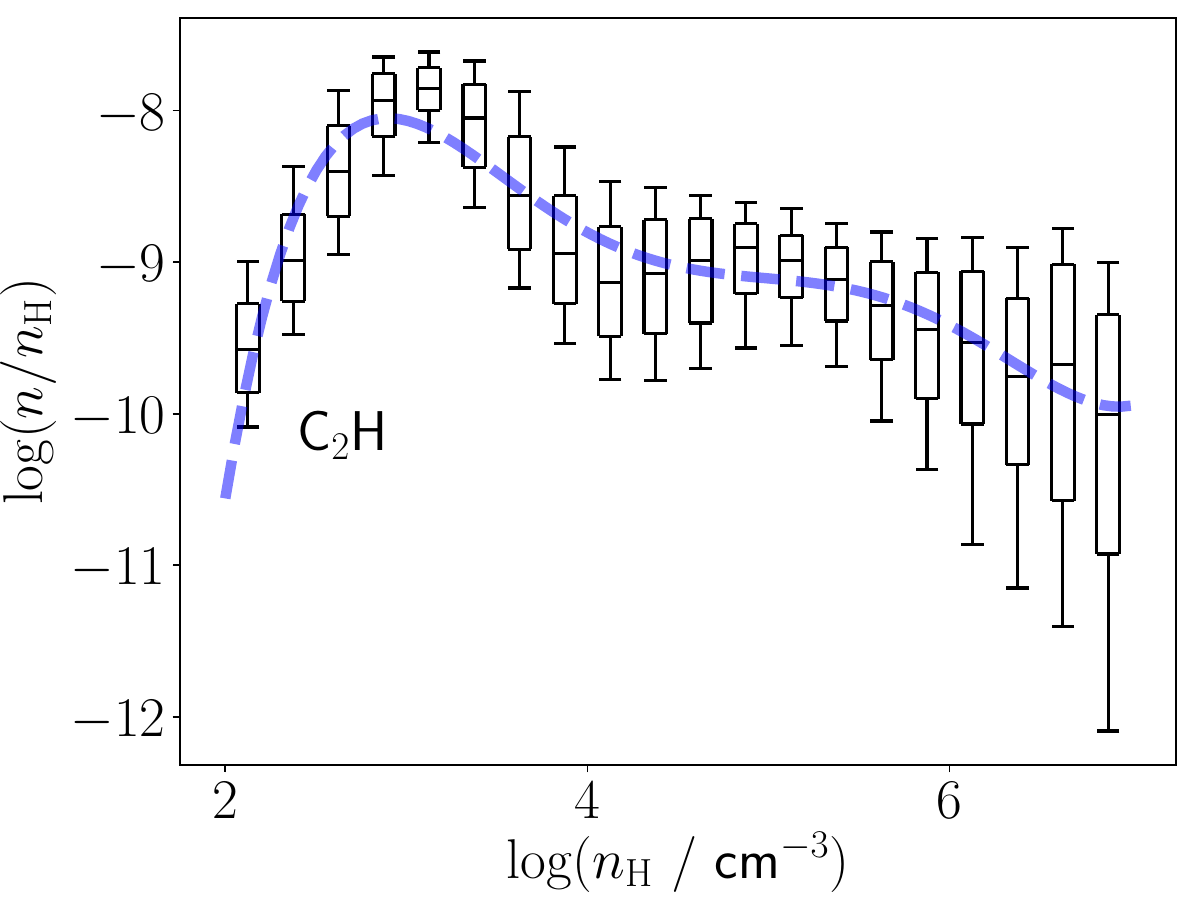}\\
  \caption{Box plots of molecular abundance versus gas density for; NH$_3$, HCN and N$_2$H$^+$ (top row); HCO$^+$, CS and HNC (middle row); CH$_3$OH, CN and C$_2$H (bottom row). Boxes show the median and the 25th/75th percentiles; whiskers show the 10th and 90th percentiles. {The dashed blue lines show fifth-order polynomial fits to the median abundances.} For HCN, N$_2$H$^+$, HCO$^+$ and CS, the abundance-density relationships assumed by \citet{smith2012,smith2013} are shown as dashed red lines.}
  \label{fig:box}
\end{figure*}

\begin{figure*}
  \centering
  \includegraphics[width=0.31\textwidth]{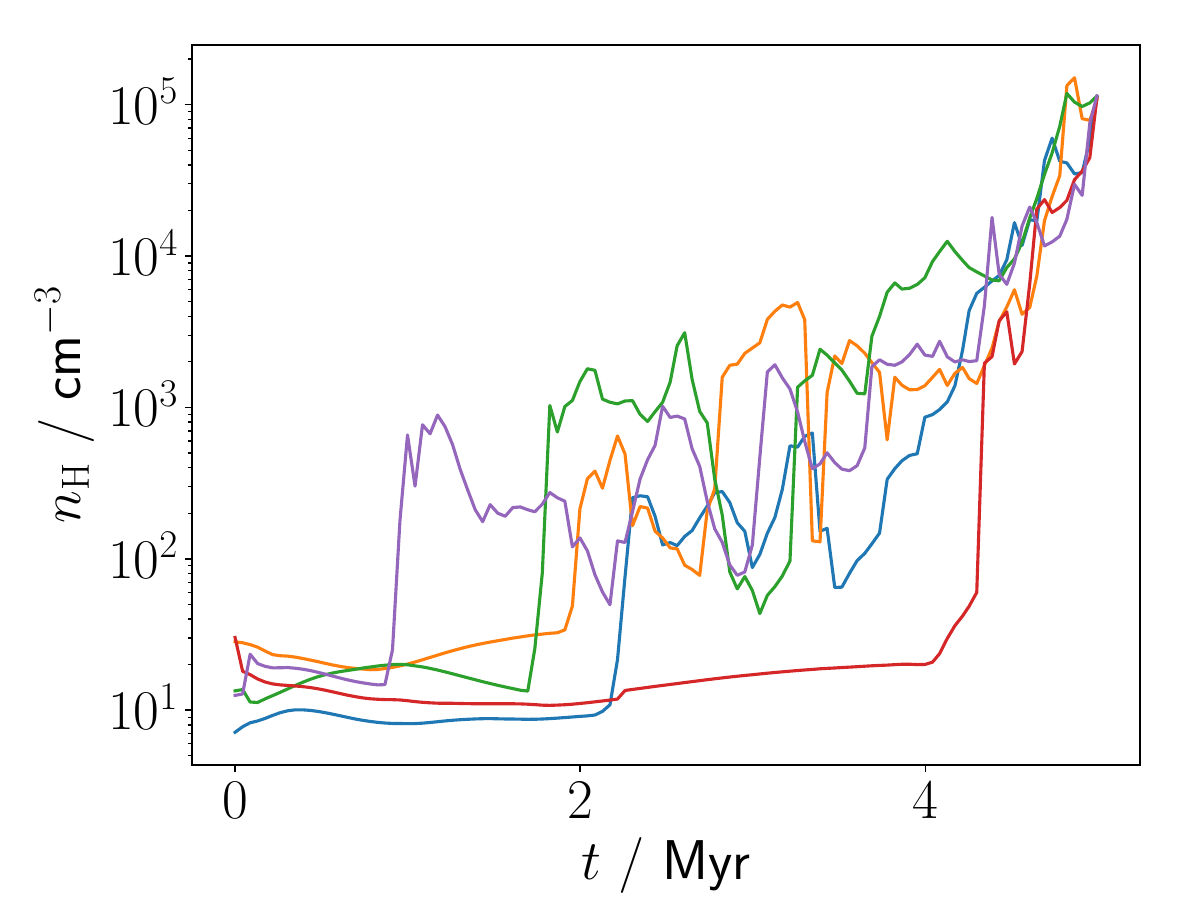}\quad
  \includegraphics[width=0.31\textwidth]{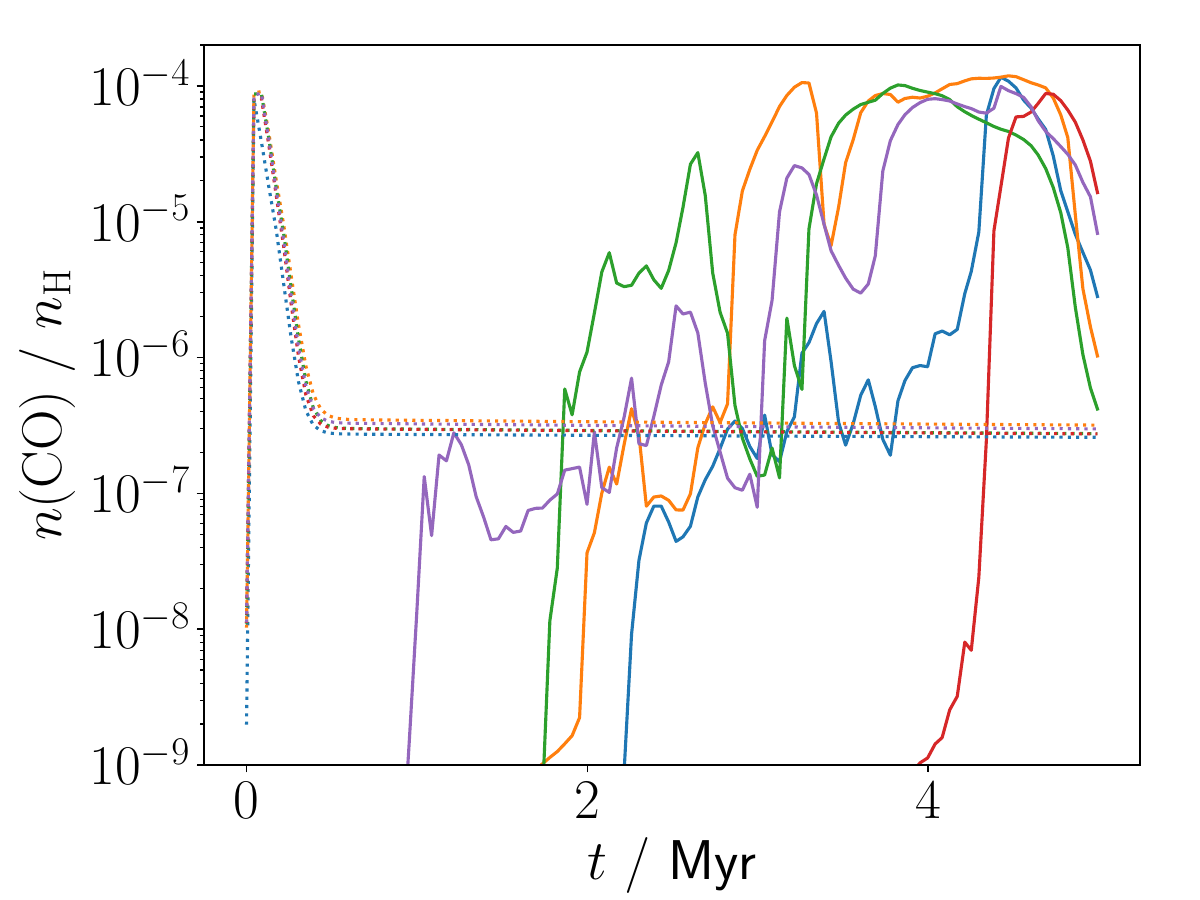}\quad
  \includegraphics[width=0.31\textwidth]{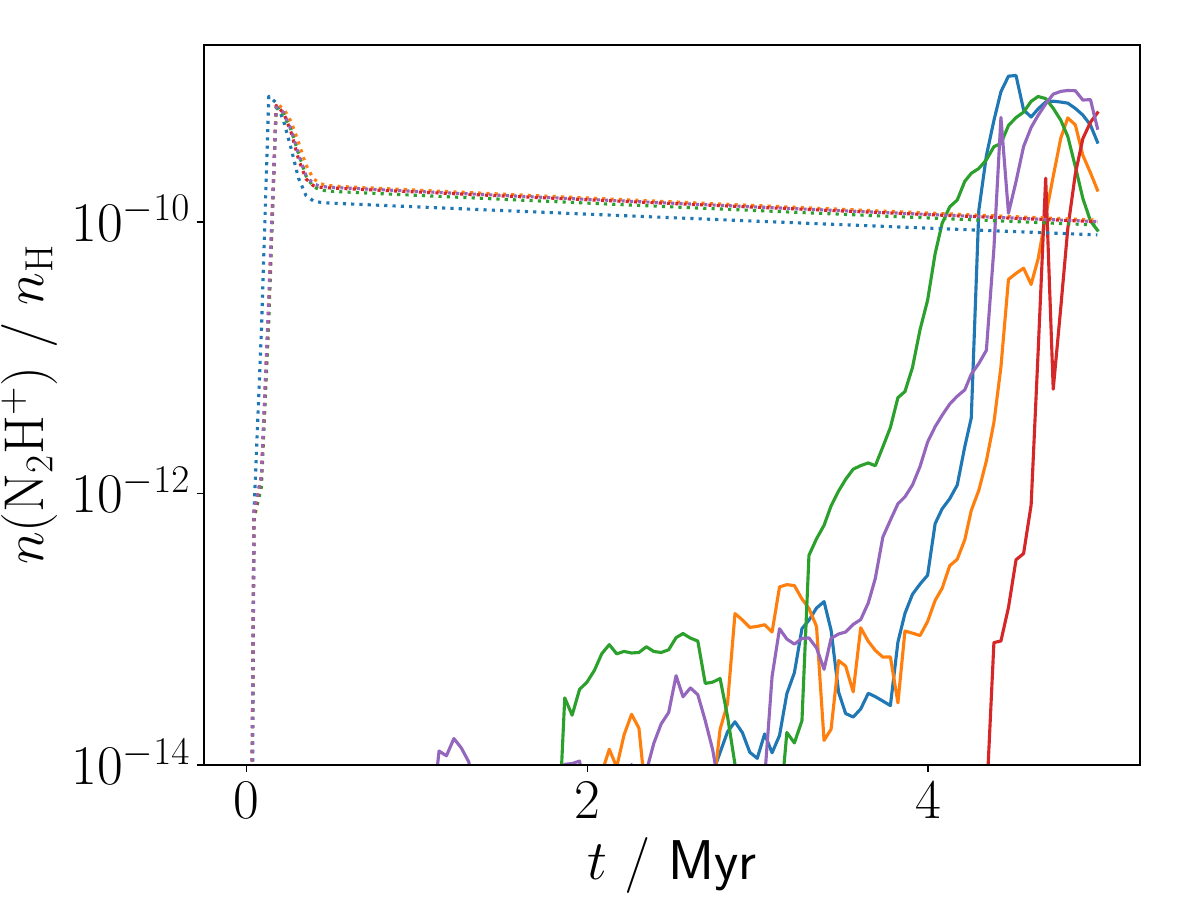}\quad
  \caption{Time evolution of the gas density (left) and the CO (middle) and N$_2$H$^+$ (right) abundances for five tracer particles with a final density of $\sim 10^5 \pcc$. {Solid lines show the abundances for a dynamic physical model, dotted lines for a static model with physical properties fixed at their final values.}}
  \label{fig:timeevol}
\end{figure*}

\begin{figure*}
  \centering
  \includegraphics[width=0.31\textwidth]{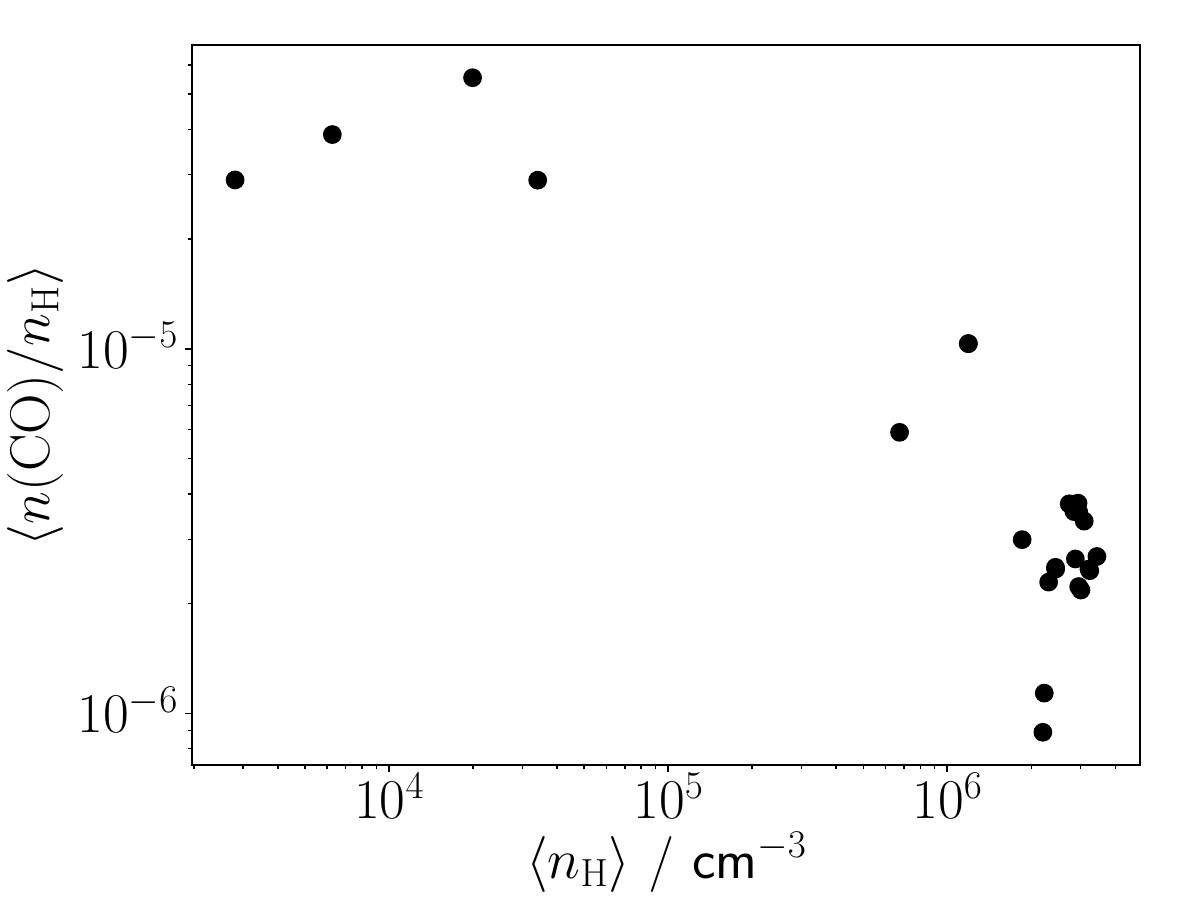}\quad
  \includegraphics[width=0.31\textwidth]{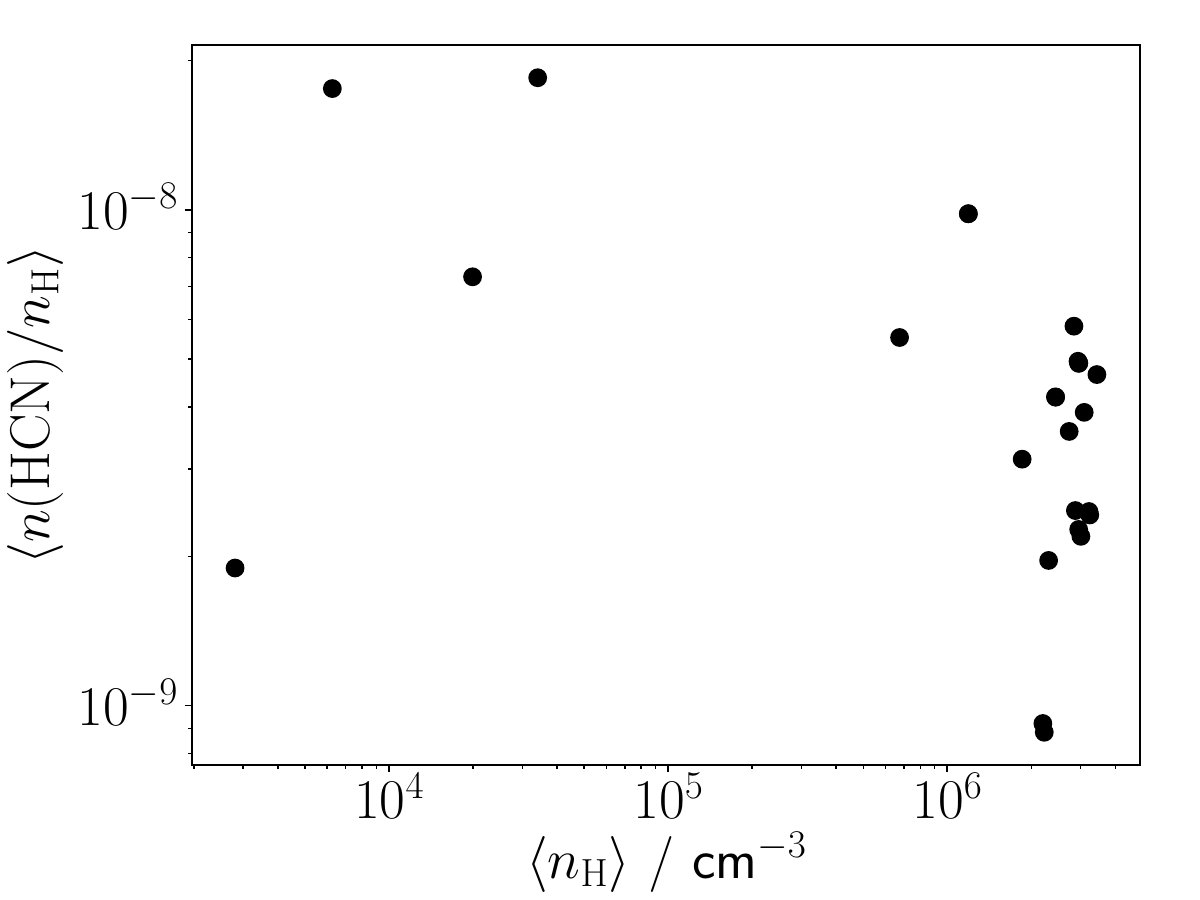}\quad
  \includegraphics[width=0.31\textwidth]{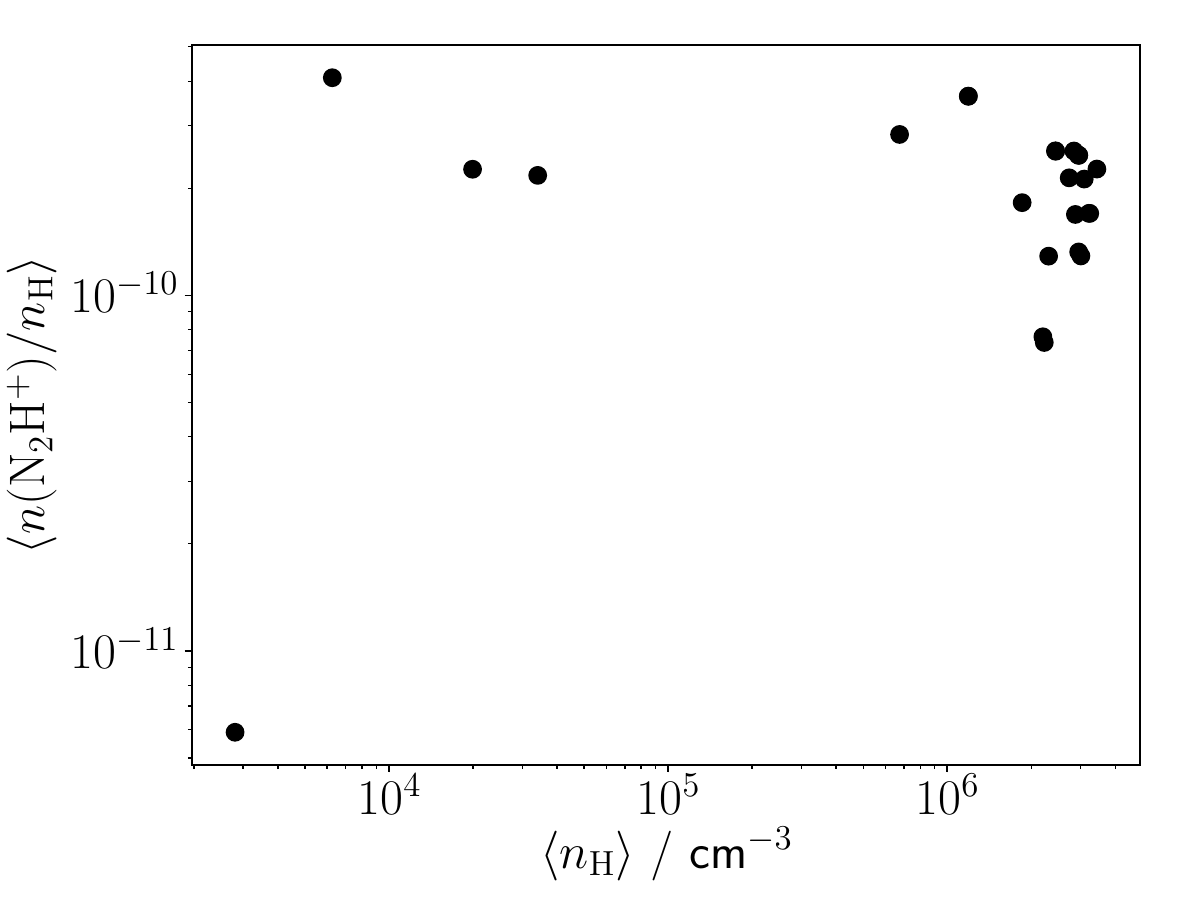}\quad
  \caption{Average abundances of CO (left), HCN (middle) and N$_2$H$^+$ (right) versus average density, within a radius of $0.05 \pc$ of each of the $28$ sink particles at $5.0 \myr$.}
  \label{fig:coreabun}
\end{figure*}

\section{Results}

\subsection{Dynamic versus static chemical evolution}

The impact of {the dynamical evolution of the gas on the time-dependent chemical evolution} is shown in Figure \ref{fig:equibtest} for CO, HCN and N$_2$H$^+$, comparing the molecular abundances {calculated with evolving gas properties (hereafter `dynamic') to values} calculated by evolving the chemistry for the same amount of time ($5.0 \myr$), but holding all physical properties constant at their final values {(hereafter `static')}. At low to moderate densities ($\lesssim 10^4 \pcc$), the differences are minor, with static chemistry tending to slightly overestimate the abundances of all species. Under these conditions, the chemistry can react on short enough timescales to effectively adopt the `appropriate' abundances for the relevant physical conditions \citep{holdship2022}, although there is more scatter in abundance at fixed density in the {dynamic} case due to the different tracer particle histories.

At higher densities ($\gtrsim 10^4 \pcc$), the {static} and {dynamic} abundances diverge significantly. Holding the physical properties constant for several $\myr$ allows the gas-phase abundances to reach equilibrium between freeze-out onto grain surfaces and desorption of molecules from ice mantles back into the gas-phase. As such, the abundances of all molecules decrease monotonically with density, because the rate of depletion increases as $\nh^2$ compared to $\nh$ for desorption processes. There is essentially no scatter in the abundances at fixed density in this regime {for the static model}; differences in the other gas properties have a negligible impact on the chemistry.

The {dynamic} abundances show similar qualitative behaviour {to the static ones}, but with a shallower decline which begins to plateau towards a constant average abundance at densities above $\sim 10^5 \pcc$ (although to some extent this may be due to insufficient time resolution; Figure \ref{fig:restest}). Particles spend only a very small fraction of their $5.0 \myr$ evolution at these high densities, so there is insufficient time for an equilibrium between freeze-out and desorption to be established. This also means the abundances of a given tracer particle are strongly dependent on its history, {rather than being mostly established by its current properties as at lower densities}. At fixed density, the HCN abundance can vary by up to three orders of magnitude, compared to almost no scatter predicted by the {static} model. While the {static} abundances for HCN are within the scatter of the {dynamic} values, they are located towards the extreme lower end of the distribution, so are still poorly representative of the typical chemical conditions. For CO and N$_2$H$^+$, the {static} abundances are significantly discrepant with the range of {dynamic} values, even at densities where the chemical evolution is well-converged.

{The scatter in abundance at fixed density in the dynamic results is not random; it is a deterministic result of the different pathways taken by different tracer particles to reach the same final state.} To investigate the impact of evolutionary history on the distribution of abundances at fixed density, we introduce the `overdensity' parameter,
\begin{equation}
  \label{eq:neq}
  \Gamma = \frac{1}{\nh^2(0) \, t_{\rm end}} \int^{t_{\rm end}}_{0} \nh^2(t) \, {\rm d}t,
\end{equation}
quantifying how long each particle has spent at a higher density than the original $10 \pcc$ (the $\nh^2$ weighting corresponds to the density dependence of two-body reactions). Figure \ref{fig:overdense} shows how the scatter in CO abundance at fixed density is affected by $\Gamma$, $T$ and $\Nhshield$. Below a density of $10^4 \pcc$, more overdense particles tend to have higher abundances, as the higher densities are more favourable for forming CO. The trend reverses above $10^4 \pcc$, with more overdense particles having spent longer under conditions favourable to freeze-out, and so having lower gas-phase CO abundance.

{In contrast to $\Gamma$, the CO abundance has almost no dependence on the gas temperature at fixed density. There is a mild dependence on shielding column density at lower densities, but by the point where the abundance peaks ($\sim 10^3 \pcc$), the gas is typically well-shielded enough to make any variations in the attenuated UV flux unimportant. The prior history of the gas is much more important for determining its chemical composition than its present-day physical properties (other than density).} The several orders of magnitude of scatter in abundance at fixed density, {which is not present in the static model}, demonstrates that the assumption of {unchanging physical properties} is completely inappropriate, particularly at the highest densities. The physical and chemical evolution of the gas have to be considered simultaneously.

\subsection{The molecular structure of clouds}
\label{sec:chemdiff}

Figure \ref{fig:col} shows the column densities of total hydrogen nuclei and of various commonly-observed molecular species, viewed in the $x-y$ plane (i.e. perpendicular to the collision axis). Except in the lowest density regions, the H$_2$ column density generally traces the total gas density well; hydrogen is mostly in molecular form. CO, by contrast, is noticeably deficient in the low-density parts of the cloud (the so-called `CO dark' gas; \citealt{wolfire2010,glover2011}). Of the other molecules we consider, a distinction can be drawn between those which essentially follow the distribution of CO (HCN, CS, HNC, CN, C$_2$H), and those which are biased towards higher density regions (NH$_3$, N$_2$H$^+$, HCO$^+$, CH$_3$OH), in that there is a higher column density contrast when compared to that of the total gas column density. This is particularly apparent for N$_2$H$^+$ (Figure \ref{fig:col}), one of only a few species which is found to genuinely trace dense gas preferentially \citep{pety2017,kauffmann2017,tafalla2021}. The latter category of molecules either include nitrogen (NH$_3$, N$_2$H$^+$), are charged (N$_2$H$^+$, HCO$^+$), or are formed by a relatively complex series of grain-surface reactions and subsequently returned to the gas phase (CH$_3$OH).

{More quantitatively, Figure \ref{fig:boxcolco} shows the column density of CO versus the total column density {(note that this is the column {\it along the line of sight}, which is distinct from the effective shielding column $\Nhshield$ discussed in Section \ref{sec:chem}; \citealt{clark2014})}. The CO column remains fairly low below $\Nhcol \sim 10^{21} \pcs$, but rises quite rapidly beyond this point until seemingly reaching a constant value of $\sim 10^{18} \pcs$ for the highest density sightlines. The gas-phase CO column never reaches the value expected if all elemental carbon was in this form ($1.4 \times 10^{-4} \, \Nhcol$), suggesting that photodissociation, freeze-out, or some combination of both are always in effect to some extent. The column densities of most other species, shown in Figure \ref{fig:boxcol}, are qualitatively similar to CO, but the value of $\Nhcol$ at which the rapid increase in molecular column begins varies significantly; N$_2$H$^+$, for example, has a negligible column density for all sightlines with $\Nhcol < 10^{22} \pcs$, {consistent with the observed threshold column density at which this molecule becomes detectable in nearby\footnote{{\citet{barnes2020} find different behaviour in the distant high-mass star forming region W49B, with N$_2$H$^+$ becoming detectable at comparable column densities to HCN ($\sim 10^{21} \pcs$). While this could indicate chemical differences compared to local Gould Belt clouds, a full investigation of the underlying reasons is beyond the scope of this work.}} clouds \citep{pety2017,kauffmann2017,tafalla2021}.}

The onset of this rise in molecular column effectively replicates the distinction between CO-like species and those tracing higher densities outlined above, with the latter category only existing in detectable quantities {(very roughly, molecular columns greater than $\sim 10^{10} \pcs$; see e.g. \citealt{ragan2006})} at higher values of $\Nhcol$. The exception is NH$_3$, which has a fairly smooth increase in molecular column density over the entire range covered. In this case, the bias towards denser gas apparent in Figure \ref{fig:col} is due to the fact that the NH$_3$ column density continues to increase with total column up to $10^{23} \pcs$, whereas CO-like species plateau before this point due to depletion onto grain surfaces, reducing the contrast between high- and low-density sightlines.}

Figure \ref{fig:boxco} shows the distribution of the CO abundance at fixed gas density. Scatter in the abundance reaches a minimum at a density of a few $10^3 \pcc$; in this regime, variations in other gas properties are not large enough to significantly affect the chemistry, while freeze-out is still too inefficient for the tracer history to have an impact. Even then, there is still $\sim 0.5$ dex of scatter. While the median CO abundance declines quite sharply above densities of $10^4 \pcc$, the amount of scatter increases substantially, with the interquartile range of our sample of tracer particles spanning an order of magnitude or more.

The variation of molecular abundances with volume density for molecules other than CO are shown in Figure \ref{fig:box}; the median abundances and interquartile ranges at various densities are given for all species in Table \ref{tab:molabun}. As with the column densities, there are two distinct classes of molecules: CO and molecules which behave similarly (HCN, CS, HNC, CN), peaking in abundance at $\nh \sim 10^3 \pcc$ and then declining quite rapidly due to freeze-out; and those which peak at $\nh \sim 10^4 \pcc$ (NH$_3$, N$_2$H$^+$, HCO$^+$, CH$_3$OH) and either decline more slowly toward higher densities, or maintain a roughly constant abundance. The latter category also show significantly less scatter in abundance at fixed density, typically less than $0.5$ dex, whereas the abundances of CO-like molecules can vary by several orders of magnitude beyond $10^4 \pcc$. Of note is C$_2$H, which shows unique behaviour among the molecules investigated here. Its abundance peaks at $10^3 \pcc$, like CO, but then declines sharply and levels off to a nearly-constant value above $10^4 \pcc$, more comparable to nitrogen-bearing species such as NH$_3$ and N$_2$H$^+$. This is likely due to its sensitivity to photodissociation, as discussed by \citet{tafalla2021}.

The abundances for HCN, N$_2$H$^+$, HCO$^+$ and CS assumed by \citet{smith2012,smith2013} are overplotted on the relevant panels in Figure \ref{fig:box}. For all four species, molecular abundances are seriously overestimated at low densities, and the density-independent abundances assumed for HCN and HCO$^+$ are much higher than the NEATH values at high densities. The exponential depletion factor assumed for CS, by contrast, significantly overestimates the actual extent of gas-phase depletion for this molecule. \citet{smith2012,smith2013} used these abundances to model low-$J$ line emission, so the poor agreement at high densities may be of limited concern; these lines are typically optically thick, and so the emission is dominated by relatively low-density ($\sim 10^3 \pcc$) gas \citep{jones2023}. However, the peak abundances in our simulations are higher than the adopted values, and the constant N$_2$H$^+$ and HCO$^+$ abundances are many orders of magnitude higher than the true values at $10^3 \pcc$, where significant line emission is expected to be generated \citep{shirley2015}. These differences are likely to dramatically alter the resulting line emission properties.

{The trends in median abundance versus density can be more accurately reproduced by fifth-order polynomial fits, also overplotted on Figures \ref{fig:boxco} and \ref{fig:box}, the coefficients for which are given in Appendix \ref{sec:polyfit}. However, these relationships are still insufficient to fully capture the behaviour of the models. At high densities, abundances may differ by orders of magnitude from the median values, and we emphasise again that this scatter is not random, but is related to each particle's individual evolutionary history (Figure \ref{fig:overdense}).} Extracting consistent observational predictions from models requires a full treatment of their time-dependent chemical evolution, {something that we plan to investigate in future work}.

\subsection{Stochasticity in the chemical composition of dense gas}

Figure \ref{fig:timeevol} shows the histories of five tracer particles with nearly identical final densities of $\sim 10^5 \pcc$. Despite the similar end point, the density evolution varies dramatically between tracers. Some particles undergo a relatively gradual rise from the initial to final density over the entire $5 \myr$ of the simulation, while others remain at low density until the last $\myr$. All particles undergo at least one period of shock compression. Some are shocked multiple times, and the density evolution is not monotonic. Tracers can oscillate between $10^2-10^3 \pcc$ for long periods of time, although those which exceed the upper end of this range seem to undergo irreversible collapse or compression. The five tracers in Figure \ref{fig:timeevol} were selected randomly, rather than being chosen as extreme examples; this complex evolution is an inevitable consequence of the dynamics of star-forming clouds.

The highly variable tracer histories result in similarly large variations in their chemical evolution. The formation of significant quantities of CO seems to be tied to the first instance of shock compression causing the gas density to rise above $100 \pcc$. The abundance then roughly follows the evolution of the gas density, with rarefaction episodes tied to decreasing CO abundance and vice versa, up until the density approaches $10^5 \pcc$ and freeze-out results in a rapid decline for all five tracers. The onset and extent of this decline vary significantly, however, leading to a difference of nearly two orders of magnitude in final CO abundance despite the near-identical final densities. N$_2$H$^+$ behaves somewhat differently, forming relatively late for all tracers, and roughly at the point where the density exceeds $10^3 \pcc$ for the first time. Compared to CO, the abundance evolution is closer to a monotonic increase, but there are still sharp, temporary declines (related to the CO abundance reaching a maximum, as N$_2$H$^+$ is destroyed by reactions with CO), and for some tracers evidence of late-time freeze-out.

{Figure \ref{fig:timeevol} also shows the evolution of the tracer abundances under a static model, where the physical conditions are held constant at their final values. This completely fails to reproduce the actual behaviour of the time-dependent chemistry. Both CO and N$_2$H$^+$ effectively reach chemical equilibrium within $\sim 1 \myr$, at abundances noticeably lower than their true final values, and there are almost no chemical differences between the different tracers at any point. The complete lack of chemical evolution over most of the $5 \myr$ duration of the static model, compared to the highly complex variations seen when physical evolution is taken into account, demonstrates that any concept of chemical age based on static models will have little, if any, correspondance to the true physical age of an object.}

This variation in history affects the chemical composition of the gas in the immediate surroundings of the sink particles, corresponding to the material in prestellar or protostellar cores. Figure \ref{fig:coreabun} shows the average abundances of CO, HCN and N$_2$H$^+$ versus average gas density for the tracers within $0.05 \pc$ of each of the $28$ sink particles that have formed by the end of the simulation.\footnote{Strictly speaking, these averages should be mass-weighted, but as we do not have any information on the gas mass represented by each tracer, we perform an unweighted average over all tracers.} Even for cores with the same (average) density, there are order-of-magnitude differences in the amount of CO and HCN present in the gas-phase. Although N$_2$H$^+$ is more consistent between sinks, it is clear that the chemical composition of the material being delivered to pre- or protostellar cores can vary significantly from object to object.

\begin{table*}
  \centering
  \caption{Median molecular abundances and interquartile ranges at different gas densities.}
  \begin{tabular}{ccccccccccc}
\hline
 & \multicolumn{5}{c}{$\nh$ / $\pcc$} \\
 & \multicolumn{1}{c}{$10^3$} & \multicolumn{1}{c}{$10^4$} & \multicolumn{1}{c}{$10^5$} & \multicolumn{1}{c}{$10^6$} & \multicolumn{1}{c}{$10^7$} \\
\hline
Molecule & \multicolumn{5}{c}{Median} \\
\hline
CO & $2.9 \times 10^{-5}$ & $7.0 \times 10^{-5}$ & $2.4 \times 10^{-6}$ & $1.0 \times 10^{-7}$ & $5.0 \times 10^{-8}$ \\
NH$_3$ & $3.0 \times 10^{-9}$ & $2.9 \times 10^{-8}$ & $1.1 \times 10^{-7}$ & $3.1 \times 10^{-8}$ & $1.6 \times 10^{-8}$ \\
HCN & $1.5 \times 10^{-9}$ & $9.0 \times 10^{-9}$ & $4.9 \times 10^{-9}$ & $2.2 \times 10^{-10}$ & $6.7 \times 10^{-11}$ \\
N$_2$H$^+$ & $7.5 \times 10^{-14}$ & $1.1 \times 10^{-10}$ & $2.6 \times 10^{-10}$ & $7.3 \times 10^{-11}$ & $7.0 \times 10^{-11}$ \\
HCO$^+$ & $1.5 \times 10^{-11}$ & $3.2 \times 10^{-9}$ & $4.4 \times 10^{-9}$ & $8.6 \times 10^{-10}$ & $5.9 \times 10^{-10}$ \\
CS & $4.4 \times 10^{-9}$ & $1.1 \times 10^{-8}$ & $8.2 \times 10^{-10}$ & $4.2 \times 10^{-11}$ & $1.2 \times 10^{-11}$ \\
HNC & $1.9 \times 10^{-10}$ & $2.5 \times 10^{-9}$ & $3.0 \times 10^{-9}$ & $2.0 \times 10^{-10}$ & $6.3 \times 10^{-11}$ \\
CH$_3$OH & $2.7 \times 10^{-11}$ & $3.0 \times 10^{-9}$ & $4.6 \times 10^{-10}$ & $8.6 \times 10^{-11}$ & $4.7 \times 10^{-11}$ \\
CN & $9.6 \times 10^{-8}$ & $3.5 \times 10^{-8}$ & $8.6 \times 10^{-9}$ & $6.9 \times 10^{-10}$ & $2.3 \times 10^{-10}$ \\
C$_2$H & $1.3 \times 10^{-8}$ & $9.6 \times 10^{-10}$ & $1.1 \times 10^{-9}$ & $3.2 \times 10^{-10}$ & $1.1 \times 10^{-10}$ \\
\hline
Molecule & \multicolumn{5}{c}{Range [$x \times 10^y$ is represented as $x(y)$]} \\
\hline
CO & $5.1 (-6)$ - $7.7 (-5)$ & $4.0 (-5)$ - $9.4 (-5)$ & $5.4 (-7)$ - $1.1 (-5)$ & $4.4 (-8)$ - $5.2 (-7)$ & $2.0 (-8)$ - $1.7 (-7)$ \\
NH$_3$ & $2.1 (-9)$ - $4.6 (-9)$ & $1.5 (-8)$ - $6.1 (-8)$ & $7.7 (-8)$ - $1.3 (-7)$ & $1.3 (-8)$ - $6.0 (-8)$ & $5.2 (-9)$ - $4.4 (-8)$ \\
HCN & $7.4 (-10)$ - $3.1 (-9)$ & $6.7 (-9)$ - $1.2 (-8)$ & $1.8 (-9)$ - $8.5 (-9)$ & $5.0 (-11)$ - $1.7 (-9)$ & $8.8 (-12)$ - $4.3 (-10)$ \\
N$_2$H$^+$ & $3.4 (-14)$ - $2.4 (-13)$ & $9.9 (-12)$ - $3.9 (-10)$ & $1.0 (-10)$ - $4.4 (-10)$ & $4.9 (-11)$ - $1.6 (-10)$ & $4.9 (-11)$ - $1.3 (-10)$ \\
HCO$^+$ & $1.0 (-11)$ - $3.5 (-11)$ & $5.4 (-10)$ - $7.8 (-9)$ & $2.2 (-9)$ - $6.9 (-9)$ & $5.9 (-10)$ - $2.0 (-9)$ & $4.2 (-10)$ - $1.1 (-9)$ \\
CS & $1.8 (-9)$ - $8.6 (-9)$ & $5.2 (-9)$ - $2.8 (-8)$ & $2.6 (-10)$ - $2.9 (-9)$ & $9.8 (-12)$ - $2.3 (-10)$ & $1.8 (-12)$ - $8.6 (-11)$ \\
HNC & $1.1 (-10)$ - $4.5 (-10)$ & $1.7 (-9)$ - $4.2 (-9)$ & $1.4 (-9)$ - $5.0 (-9)$ & $4.7 (-11)$ - $1.3 (-9)$ & $8.4 (-12)$ - $3.8 (-10)$ \\
CH$_3$OH & $3.4 (-12)$ - $1.4 (-10)$ & $1.6 (-9)$ - $4.3 (-9)$ & $2.5 (-10)$ - $9.2 (-10)$ & $4.2 (-11)$ - $2.0 (-10)$ & $2.0 (-11)$ - $1.4 (-10)$ \\
CN & $5.0 (-8)$ - $1.3 (-7)$ & $2.3 (-8)$ - $5.1 (-8)$ & $3.8 (-9)$ - $1.6 (-8)$ & $1.8 (-10)$ - $3.9 (-9)$ & $3.3 (-11)$ - $1.3 (-9)$ \\
C$_2$H & $7.6 (-9)$ - $1.8 (-8)$ & $4.4 (-10)$ - $2.3 (-9)$ & $5.7 (-10)$ - $1.7 (-9)$ & $8.9 (-11)$ - $8.5 (-10)$ & $1.6 (-11)$ - $5.3 (-10)$ \\
\hline
\end{tabular}

  \label{tab:molabun}
\end{table*}

\section{Discussion}

\subsection{Chemical tracers in molecular clouds}

Using MHD simulations with an essentially complete treatment of gas dynamics and the relevant thermochemistry, we are able to investigate how molecular abundances evolve under realistic conditions, as opposed to the idealised models commonly adopted in previous studies. Of the species we investigate here, there appears to be a distinction between CO-like molecules, which are effectively depleted from the gas-phase beyond densities of $10^4 \pcc$, and those where some competing process mitigates freeze-out onto grain surfaces. The latter (NH$_3$-like) molecules may show this behaviour due to reduced destruction by gas-phase CO (e.g. N$_2$H$^+$), their requirement to be formed via depletion of said CO (CH$_3$OH), or some other reason. In all cases they present nearly constant abundances at high density, and relatively little scatter compared to CO-like species. They are thus relatively insensitive to the prior evolution of the gas, and may represent good targets for tracing the present-day conditions of denser structures within molecular clouds. The abundances of CO-like molecules, on the other hand, are strongly dependent on the evolutionary history (Figure \ref{fig:overdense}), and so would make better tracers of, for example, core ages. However, the complex and non-monotonic chemical evolution of all species (Figure \ref{fig:timeevol}) makes either of these uses challenging. Diagnostics based on simplified evolutionary models should be considered speculative at best.

\subsection{Dynamic physical evolution in prestellar cores}

Chemical models of prestellar cores often assume {a static} density profile, and subsequently evolve the chemistry at different points along this profile for several $\myr$ in order to calculate molecular abundances \citep[e.g.][]{caselli2022,megias2022}. Figure \ref{fig:equibtest} shows that this will inevitably overestimate the amount of depletion in the high-density central regions of cores, even for species which are supposedly resistant to freeze-out such as N$_2$H$^+$, because it is unrealistic for gas to remain at these densities for this length of time (orders of magnitude longer than the free-fall timescale). The material at high density will likely have spent only a very small fraction of time {\it at} this density, and so will retain much of its gas-phase molecular content. The chemical makeup also depends critically on the core's prior evolutionary history; cores with identical density structures can (and most likely will) have completely different molecular compositions, which depend on the details of their formation. {While we have focused on the gas-phase abundances, \citet{clement2023} find similar history-dependent effects for the composition of ices in prestellar cores.} These effects may have some bearing on otherwise-unexpected phenomena such as the lack of depletion of nitrogen-bearing molecules \citep{sipila2019} and the ubiquitous detection of complex organic molecules \citep{jimenez2016,jimenez2021,scibelli2020,scibelli2021,ambrose2021} in cold prestellar environments.

\subsection{Comparison with previous work}

Most previous studies {using large chemical networks} on the scales of molecular clouds adopt a simplified treatment of the thermal physics \citep[e.g.][]{ruaud2018,wakelam2019,wakelam2020,bovino2021,jensen2021,ferrada2021,gomez2022,priestley2023b}. {To our knowledge, the only other peer-reviewed study on these scales}\footnote{{\citet{ilee2017} also post-process hydrodynamical models which include a detailed treatment of the thermal evolution, but on the much smaller scales of protoplanetary discs. Preliminary results from a comparable study to ours were presented in a conference proceedings by \citet{szucs2015}.}} which uses a hydrodynamical simulation with self-consistent thermodynamical evolution as a base for chemical post-processing is that of \citet{panessa2023}, who primarily focus on the abundance of HCO$^+$. They find a similar peak HCO$^+$ column density to ours ($\sim 10^{15} \pcs$), and a similar {threshold value of $\Nhcol$ beyond which the abundance sharply rises ($\sim 10^{22} \pcs$)}. While this agrees with our classification of HCO$^+$ as an NH$_3$-like molecule, we note that this behaviour is in conflict with its observed line emission properties \citep{tafalla2021}, suggesting that some part of the reaction network for this molecule is incorrect \citep{priestley2023b}.

The key difference between our simulations and those of \citet{panessa2023} is that we use an idealised model of colliding spherical clouds, whereas they study zoomed-in regions from within a galactic disc-scale simulation \citep{walch2015,seifried2017b}. Our approach allows more flexibility in choosing cloud parameters, rather than being restricted to whatever happens to form in the parent simulations; theirs has the benefit of forming clouds self-consistently from the larger-scale structure, rather than beginning with arbitrary (and possibly unrealistic) initial conditions. We note that idealised cloud models of the sort considered here have been shown to accurately reproduce the line emission properties of observed molecular clouds \citep{priestley2023b}, suggesting that they are successfully capturing the important details of the chemical evolution.

{\citet{lupi2021} take the alternative approach to post-processing, and expand the internal chemical network in their underlying MHD simulations to include more species and reactions. While this allows for a completely self-consistent treatment of the coupled physical, thermal, and chemical evolution, the increased computational expense puts a limit on the size of network that can be employed; \citet{lupi2021} do not include any nitrogen chemistry, for example, a key focus of this work due to the observational importance of many nitrogen-bearing species. Given that the additional chemical complexity is unlikely to greatly affect the gas or dust temperatures \citep{goldsmith2001,gong2017}, post-processing remains a viable alternative for modelling time-dependent chemistry in molecular clouds.}

\section{Conclusions}

We present a framework for chemically post-processing MHD simulations of molecular clouds. Our post-processed molecular abundances {are consistent} with those of the underlying MHD simulation {for the small number of species required to follow the thermal evolution of the gas, and which are therefore included in the on-the-fly chemical evolution. Our post-processing framework allows us to investigate the behaviour of many additional species at low computational cost, such as nitrogen-bearing molecules which are observationally important as dense gas tracers.}

We find that most molecules fall into one of two families: CO-like molecules, which reach peak abundances at gas densities of $10^3 \pcc$, decline toward higher densities, and have large scatter in abundance at fixed density due to the varying evolutionary history of the gas; and NH$_3$-like molecules, which reach their peak abundance at higher ($\sim 10^4 \pcc$) densities, decline either slowly or not at all beyond this peak, and have {a tighter ({although still not one-to-one}) relationship between density and abundance}.

{Evolving the chemistry with the physical properties held constant produces results accurate} to within a factor of a few up to densities of $\sim 10^3 \pcc$, but diverges from the {true behaviour} at higher densities. {Ignoring the dynamical evolution of the gas tends to} severely overestimate the amount of depletion of molecules onto grain surfaces. {Differences in the prior evolution of tracer particles additionally result in significant} scatter in the molecular abundances at fixed gas density, in contrast to {static-density} models. Chemical models which assume a static density structure will inevitably produce erroneous (and potentially misleading) results; the objects being modelled are neither static nor in equilibrium, {and their simultaneous physical and chemical evolution must be taken into account}.

\section*{Acknowledgements}
FDP, PCC, SER and OF acknowledge the support of a consolidated grant (ST/K00926/1) from the UK Science and Technology Facilities Council (STFC). SCOG and RSK acknowledge funding from the European Research Council (ERC) via the ERC Synergy Grant “ECOGAL-Understanding our Galactic ecosystem: From the disk of the Milky Way to the formation sites of stars and planets” (project ID 855130, from the Heidelberg Cluster of Excellence (EXC 2181 - 390900948) “STRUCTURES: A unifying approach to emergent phenomena in the physical world, mathematics, and complex data”, funded by the German Excellence Strategy, and from the German Ministry for Economic Affairs and Climate Action in project ``MAINN'' (funding ID 50OO2206). The team in Heidelberg also thanks for computing resources provided by {\em The L\"{a}nd} through bwHPC and DFG through grant INST 35/1134-1 FUGG and for data storage at SDS@hd through grant INST 35/1314-1 FUGG. This research was undertaken using the supercomputing facilities at Cardiff University operated by Advanced Research Computing at Cardiff (ARCCA) on behalf of the Cardiff Supercomputing Facility and the Supercomputing Wales (SCW) project. We acknowledge the support of the latter, which is part-funded by the European Regional Development Fund (ERDF) via the Welsh Government.

\section*{Data Availability}
The data underlying this article will be shared on request.

\bibliographystyle{mnras}
\bibliography{arepochem}

\begin{thebibliography}{}
\makeatletter
\relax
\def\mn@urlcharsother{\let\do\@makeother \do\$\do\&\do\#\do\^\do\_\do\%\do\~}
\def\mn@doi{\begingroup\mn@urlcharsother \@ifnextchar [ {\mn@doi@}
  {\mn@doi@[]}}
\def\mn@doi@[#1]#2{\def\@tempa{#1}\ifx\@tempa\@empty \href
  {http://dx.doi.org/#2} {doi:#2}\else \href {http://dx.doi.org/#2} {#1}\fi
  \endgroup}
\def\mn@eprint#1#2{\mn@eprint@#1:#2::\@nil}
\def\mn@eprint@arXiv#1{\href {http://arxiv.org/abs/#1} {{\tt arXiv:#1}}}
\def\mn@eprint@dblp#1{\href {http://dblp.uni-trier.de/rec/bibtex/#1.xml}
  {dblp:#1}}
\def\mn@eprint@#1:#2:#3:#4\@nil{\def\@tempa {#1}\def\@tempb {#2}\def\@tempc
  {#3}\ifx \@tempc \@empty \let \@tempc \@tempb \let \@tempb \@tempa \fi \ifx
  \@tempb \@empty \def\@tempb {arXiv}\fi \@ifundefined
  {mn@eprint@\@tempb}{\@tempb:\@tempc}{\expandafter \expandafter \csname
  mn@eprint@\@tempb\endcsname \expandafter{\@tempc}}}

\bibitem[\protect\citeauthoryear{{Aikawa}, {Herbst}, {Roberts}  \&
  {Caselli}}{{Aikawa} et~al.}{2005}]{aikawa2005}
{Aikawa} Y.,  {Herbst} E.,  {Roberts} H.,   {Caselli} P.,  2005, \mn@doi [\apj]
  {10.1086/427017}, \href {http://adsabs.harvard.edu/abs/2005ApJ...620..330A}
  {620, 330}

\bibitem[\protect\citeauthoryear{{Ambrose}, {Shirley}  \& {Scibelli}}{{Ambrose}
  et~al.}{2021}]{ambrose2021}
{Ambrose} H.~E.,  {Shirley} Y.~L.,   {Scibelli} S.,  2021, \mn@doi [\mnras]
  {10.1093/mnras/staa3649}, \href
  {https://ui.adsabs.harvard.edu/abs/2021MNRAS.501..347A} {501, 347}

\bibitem[\protect\citeauthoryear{{Banerji}, {Viti}, {Williams}  \&
  {Rawlings}}{{Banerji} et~al.}{2009}]{banerji2009}
{Banerji} M.,  {Viti} S.,  {Williams} D.~A.,   {Rawlings} J.~M.~C.,  2009,
  \mn@doi [\apj] {10.1088/0004-637X/692/1/283}, \href
  {https://ui.adsabs.harvard.edu/abs/2009ApJ...692..283B} {692, 283}

\bibitem[\protect\citeauthoryear{{Barnes} et~al.,}{{Barnes}
  et~al.}{2020}]{barnes2020}
{Barnes} A.~T.,  et~al., 2020, \mn@doi [\mnras] {10.1093/mnras/staa1814}, \href
  {https://ui.adsabs.harvard.edu/abs/2020MNRAS.497.1972B} {497, 1972}

\bibitem[\protect\citeauthoryear{{Bergin} \& {Tafalla}}{{Bergin} \&
  {Tafalla}}{2007}]{bergin2007}
{Bergin} E.~A.,  {Tafalla} M.,  2007, \mn@doi [\araa]
  {10.1146/annurev.astro.45.071206.100404}, \href
  {http://adsabs.harvard.edu/abs/2007ARA%26A..45..339B} {45, 339}

\bibitem[\protect\citeauthoryear{{Bohlin}, {Savage}  \& {Drake}}{{Bohlin}
  et~al.}{1978}]{bohlin1978}
{Bohlin} R.~C.,  {Savage} B.~D.,   {Drake} J.~F.,  1978, \mn@doi [\apj]
  {10.1086/156357}, \href
  {https://ui.adsabs.harvard.edu/abs/1978ApJ...224..132B} {224, 132}

\bibitem[\protect\citeauthoryear{{Boogert}, {Gerakines}  \&
  {Whittet}}{{Boogert} et~al.}{2015}]{boogert2015}
{Boogert} A.~C.~A.,  {Gerakines} P.~A.,   {Whittet} D. C.~B.,  2015, \mn@doi
  [\araa] {10.1146/annurev-astro-082214-122348}, \href
  {https://ui.adsabs.harvard.edu/abs/2015ARA&A..53..541B} {53, 541}

\bibitem[\protect\citeauthoryear{{Bovino}, {Lupi}, {Giannetti}, {Sabatini},
  {Schleicher}, {Wyrowski}  \& {Menten}}{{Bovino} et~al.}{2021}]{bovino2021}
{Bovino} S.,  {Lupi} A.,  {Giannetti} A.,  {Sabatini} G.,  {Schleicher} D.
  R.~G.,  {Wyrowski} F.,   {Menten} K.~M.,  2021, \mn@doi [\aap]
  {10.1051/0004-6361/202141252}, \href
  {https://ui.adsabs.harvard.edu/abs/2021A&A...654A..34B} {654, A34}

\bibitem[\protect\citeauthoryear{{Caselli} et~al.,}{{Caselli}
  et~al.}{2022}]{caselli2022}
{Caselli} P.,  et~al., 2022, \mn@doi [\apj] {10.3847/1538-4357/ac5913}, \href
  {https://ui.adsabs.harvard.edu/abs/2022ApJ...929...13C} {929, 13}

\bibitem[\protect\citeauthoryear{{Clark} \& {Glover}}{{Clark} \&
  {Glover}}{2014}]{clark2014}
{Clark} P.~C.,  {Glover} S. C.~O.,  2014, \mn@doi [\mnras]
  {10.1093/mnras/stu1589}, \href
  {https://ui.adsabs.harvard.edu/abs/2014MNRAS.444.2396C} {444, 2396}

\bibitem[\protect\citeauthoryear{{Clark}, {Glover}  \& {Klessen}}{{Clark}
  et~al.}{2012a}]{clark2012}
{Clark} P.~C.,  {Glover} S. C.~O.,   {Klessen} R.~S.,  2012a, \mn@doi [\mnras]
  {10.1111/j.1365-2966.2011.20087.x}, \href
  {https://ui.adsabs.harvard.edu/abs/2012MNRAS.420..745C} {420, 745}

\bibitem[\protect\citeauthoryear{{Clark}, {Glover}, {Klessen}  \&
  {Bonnell}}{{Clark} et~al.}{2012b}]{clark2012b}
{Clark} P.~C.,  {Glover} S. C.~O.,  {Klessen} R.~S.,   {Bonnell} I.~A.,  2012b,
  \mn@doi [\mnras] {10.1111/j.1365-2966.2012.21259.x}, \href
  {https://ui.adsabs.harvard.edu/abs/2012MNRAS.424.2599C} {424, 2599}

\bibitem[\protect\citeauthoryear{{Clark}, {Glover}, {Ragan}  \&
  {Duarte-Cabral}}{{Clark} et~al.}{2019}]{clark2019}
{Clark} P.~C.,  {Glover} S. C.~O.,  {Ragan} S.~E.,   {Duarte-Cabral} A.,  2019,
  \mn@doi [\mnras] {10.1093/mnras/stz1119}, \href
  {https://ui.adsabs.harvard.edu/abs/2019MNRAS.486.4622C} {486, 4622}

\bibitem[\protect\citeauthoryear{{Cl{\'e}ment} et~al.,}{{Cl{\'e}ment}
  et~al.}{2023}]{clement2023}
{Cl{\'e}ment} A.,  et~al., 2023, \mn@doi [arXiv e-prints]
  {10.48550/arXiv.2306.08346}, \href
  {https://ui.adsabs.harvard.edu/abs/2023arXiv230608346C} {p. arXiv:2306.08346}

\bibitem[\protect\citeauthoryear{{Federman}, {Glassgold}  \& {Kwan}}{{Federman}
  et~al.}{1979}]{federman1979}
{Federman} S.~R.,  {Glassgold} A.~E.,   {Kwan} J.,  1979, \mn@doi [\apj]
  {10.1086/156753}, \href
  {https://ui.adsabs.harvard.edu/abs/1979ApJ...227..466F} {227, 466}

\bibitem[\protect\citeauthoryear{{Ferrada-Chamorro}, {Lupi}  \&
  {Bovino}}{{Ferrada-Chamorro} et~al.}{2021}]{ferrada2021}
{Ferrada-Chamorro} S.,  {Lupi} A.,   {Bovino} S.,  2021, \mn@doi [\mnras]
  {10.1093/mnras/stab1525}, \href
  {https://ui.adsabs.harvard.edu/abs/2021MNRAS.505.3442F} {505, 3442}

\bibitem[\protect\citeauthoryear{{Genel}, {Vogelsberger}, {Nelson}, {Sijacki},
  {Springel}  \& {Hernquist}}{{Genel} et~al.}{2013}]{genel2013}
{Genel} S.,  {Vogelsberger} M.,  {Nelson} D.,  {Sijacki} D.,  {Springel} V.,
  {Hernquist} L.,  2013, \mn@doi [\mnras] {10.1093/mnras/stt1383}, \href
  {https://ui.adsabs.harvard.edu/abs/2013MNRAS.435.1426G} {435, 1426}

\bibitem[\protect\citeauthoryear{{Glover} \& {Clark}}{{Glover} \&
  {Clark}}{2012}]{glover2012}
{Glover} S. C.~O.,  {Clark} P.~C.,  2012, \mn@doi [\mnras]
  {10.1111/j.1365-2966.2011.20260.x}, \href
  {https://ui.adsabs.harvard.edu/abs/2012MNRAS.421..116G} {421, 116}

\bibitem[\protect\citeauthoryear{{Glover} \& {Mac Low}}{{Glover} \& {Mac
  Low}}{2007}]{glover2007}
{Glover} S. C.~O.,  {Mac Low} M.-M.,  2007, \mn@doi [\apjs] {10.1086/512238},
  \href {https://ui.adsabs.harvard.edu/abs/2007ApJS..169..239G} {169, 239}

\bibitem[\protect\citeauthoryear{{Glover} \& {Mac Low}}{{Glover} \& {Mac
  Low}}{2011}]{glover2011}
{Glover} S.~C.~O.,  {Mac Low} M.~M.,  2011, \mn@doi [\mnras]
  {10.1111/j.1365-2966.2010.17907.x}, \href
  {https://ui.adsabs.harvard.edu/abs/2011MNRAS.412..337G} {412, 337}

\bibitem[\protect\citeauthoryear{{Goldsmith}}{{Goldsmith}}{2001}]{goldsmith2001}
{Goldsmith} P.~F.,  2001, \mn@doi [\apj] {10.1086/322255}, \href
  {https://ui.adsabs.harvard.edu/abs/2001ApJ...557..736G} {557, 736}

\bibitem[\protect\citeauthoryear{{G{\'o}mez}, {Walsh}  \& {Palau}}{{G{\'o}mez}
  et~al.}{2022}]{gomez2022}
{G{\'o}mez} G.~C.,  {Walsh} C.,   {Palau} A.,  2022, \mn@doi [\mnras]
  {10.1093/mnras/stac912}, \href
  {https://ui.adsabs.harvard.edu/abs/2022MNRAS.513.1244G} {513, 1244}

\bibitem[\protect\citeauthoryear{{Gong}, {Ostriker}  \& {Wolfire}}{{Gong}
  et~al.}{2017}]{gong2017}
{Gong} M.,  {Ostriker} E.~C.,   {Wolfire} M.~G.,  2017, \mn@doi [\apj]
  {10.3847/1538-4357/aa7561}, \href
  {https://ui.adsabs.harvard.edu/abs/2017ApJ...843...38G} {843, 38}

\bibitem[\protect\citeauthoryear{{Habing}}{{Habing}}{1968}]{habing1968}
{Habing} H.~J.,  1968, \bain, \href
  {https://ui.adsabs.harvard.edu/abs/1968BAN....19..421H} {19, 421}

\bibitem[\protect\citeauthoryear{{Hily-Blant}, {Pineau des For{\^e}ts}, {Faure}
   \& {Lique}}{{Hily-Blant} et~al.}{2022}]{hily2022}
{Hily-Blant} P.,  {Pineau des For{\^e}ts} G.,  {Faure} A.,   {Lique} F.,  2022,
  \mn@doi [\aap] {10.1051/0004-6361/201936498}, \href
  {https://ui.adsabs.harvard.edu/abs/2022A&A...658A.168H} {658, A168}

\bibitem[\protect\citeauthoryear{{Holdship} \& {Viti}}{{Holdship} \&
  {Viti}}{2022}]{holdship2022}
{Holdship} J.,  {Viti} S.,  2022, \mn@doi [\aap] {10.1051/0004-6361/202142398},
  \href {https://ui.adsabs.harvard.edu/abs/2022A&A...658A.103H} {658, A103}

\bibitem[\protect\citeauthoryear{{Holdship}, {Viti}, {Jim{\'e}nez-Serra},
  {Makrymallis}  \& {Priestley}}{{Holdship} et~al.}{2017}]{holdship2017}
{Holdship} J.,  {Viti} S.,  {Jim{\'e}nez-Serra} I.,  {Makrymallis} A.,
  {Priestley} F.,  2017, \mn@doi [\aj] {10.3847/1538-3881/aa773f}, \href
  {https://ui.adsabs.harvard.edu/abs/2017AJ....154...38H} {154, 38}

\bibitem[\protect\citeauthoryear{{Hunter}, {Clark}, {Glover}  \&
  {Klessen}}{{Hunter} et~al.}{2023}]{hunter2023}
{Hunter} G.~H.,  {Clark} P.~C.,  {Glover} S. C.~O.,   {Klessen} R.~S.,  2023,
  \mn@doi [\mnras] {10.1093/mnras/stac3751}, \href
  {https://ui.adsabs.harvard.edu/abs/2023MNRAS.519.4152H} {519, 4152}

\bibitem[\protect\citeauthoryear{{Ilee} et~al.,}{{Ilee}
  et~al.}{2017}]{ilee2017}
{Ilee} J.~D.,  et~al., 2017, \mn@doi [\mnras] {10.1093/mnras/stx1966}, \href
  {https://ui.adsabs.harvard.edu/abs/2017MNRAS.472..189I} {472, 189}

\bibitem[\protect\citeauthoryear{{Jenkins}}{{Jenkins}}{2009}]{jenkins2009}
{Jenkins} E.~B.,  2009, \mn@doi [\apj] {10.1088/0004-637X/700/2/1299}, \href
  {https://ui.adsabs.harvard.edu/abs/2009ApJ...700.1299J} {700, 1299}

\bibitem[\protect\citeauthoryear{{Jensen}, {J{\o}rgensen}, {Furuya},
  {Haugb{\o}lle}  \& {Aikawa}}{{Jensen} et~al.}{2021}]{jensen2021}
{Jensen} S.~S.,  {J{\o}rgensen} J.~K.,  {Furuya} K.,  {Haugb{\o}lle} T.,
  {Aikawa} Y.,  2021, \mn@doi [\aap] {10.1051/0004-6361/202040196}, \href
  {https://ui.adsabs.harvard.edu/abs/2021A&A...649A..66J} {649, A66}

\bibitem[\protect\citeauthoryear{{Jensen}, {Spezzano}, {Caselli}, {Grassi}  \&
  {Haugb{\o}lle}}{{Jensen} et~al.}{2023}]{jensen2023}
{Jensen} S.~S.,  {Spezzano} S.,  {Caselli} P.,  {Grassi} T.,   {Haugb{\o}lle}
  T.,  2023, \mn@doi [\aap] {10.1051/0004-6361/202245466}, \href
  {https://ui.adsabs.harvard.edu/abs/2023A&A...675A..34J} {675, A34}

\bibitem[\protect\citeauthoryear{{Jim{\'e}nez-Serra}
  et~al.,}{{Jim{\'e}nez-Serra} et~al.}{2016}]{jimenez2016}
{Jim{\'e}nez-Serra} I.,  et~al., 2016, \mn@doi [\apjl]
  {10.3847/2041-8205/830/1/L6}, \href
  {https://ui.adsabs.harvard.edu/abs/2016ApJ...830L...6J} {830, L6}

\bibitem[\protect\citeauthoryear{{Jim{\'e}nez-Serra}, {Vasyunin}, {Spezzano},
  {Caselli}, {Cosentino}  \& {Viti}}{{Jim{\'e}nez-Serra}
  et~al.}{2021}]{jimenez2021}
{Jim{\'e}nez-Serra} I.,  {Vasyunin} A.~I.,  {Spezzano} S.,  {Caselli} P.,
  {Cosentino} G.,   {Viti} S.,  2021, \mn@doi [\apj]
  {10.3847/1538-4357/ac024c}, \href
  {https://ui.adsabs.harvard.edu/abs/2021ApJ...917...44J} {917, 44}

\bibitem[\protect\citeauthoryear{{Jones}, {Clark}, {Glover}  \&
  {Hacar}}{{Jones} et~al.}{2023}]{jones2023}
{Jones} G.~H.,  {Clark} P.~C.,  {Glover} S. C.~O.,   {Hacar} A.,  2023, \mn@doi
  [\mnras] {10.1093/mnras/stad202}, \href
  {https://ui.adsabs.harvard.edu/abs/2023MNRAS.520.1005J} {520, 1005}

\bibitem[\protect\citeauthoryear{{Kaminski}, {Frank}, {Carroll}  \&
  {Myers}}{{Kaminski} et~al.}{2014}]{kaminski2014}
{Kaminski} E.,  {Frank} A.,  {Carroll} J.,   {Myers} P.,  2014, \mn@doi [\apj]
  {10.1088/0004-637X/790/1/70}, \href
  {https://ui.adsabs.harvard.edu/abs/2014ApJ...790...70K} {790, 70}

\bibitem[\protect\citeauthoryear{{Kauffmann}, {Goldsmith}, {Melnick}, {Tolls},
  {Guzman}  \& {Menten}}{{Kauffmann} et~al.}{2017}]{kauffmann2017}
{Kauffmann} J.,  {Goldsmith} P.~F.,  {Melnick} G.,  {Tolls} V.,  {Guzman} A.,
  {Menten} K.~M.,  2017, \mn@doi [\aap] {10.1051/0004-6361/201731123}, \href
  {https://ui.adsabs.harvard.edu/abs/2017A&A...605L...5K} {605, L5}

\bibitem[\protect\citeauthoryear{{Larson}}{{Larson}}{1969}]{larson1969}
{Larson} R.~B.,  1969, \mn@doi [\mnras] {10.1093/mnras/145.3.271}, \href
  {http://adsabs.harvard.edu/abs/1969MNRAS.145..271L} {145, 271}

\bibitem[\protect\citeauthoryear{{Lupi}, {Bovino}  \& {Grassi}}{{Lupi}
  et~al.}{2021}]{lupi2021}
{Lupi} A.,  {Bovino} S.,   {Grassi} T.,  2021, \mn@doi [\aap]
  {10.1051/0004-6361/202142145}, \href
  {https://ui.adsabs.harvard.edu/abs/2021A&A...654L...6L} {654, L6}

\bibitem[\protect\citeauthoryear{{Mathis}, {Mezger}  \& {Panagia}}{{Mathis}
  et~al.}{1983}]{mathis1983}
{Mathis} J.~S.,  {Mezger} P.~G.,   {Panagia} N.,  1983, \aap, \href
  {https://ui.adsabs.harvard.edu/abs/1983A&A...128..212M} {128, 212}

\bibitem[\protect\citeauthoryear{{McElroy}, {Walsh}, {Markwick}, {Cordiner},
  {Smith}  \& {Millar}}{{McElroy} et~al.}{2013}]{mcelroy2013}
{McElroy} D.,  {Walsh} C.,  {Markwick} A.~J.,  {Cordiner} M.~A.,  {Smith} K.,
  {Millar} T.~J.,  2013, \mn@doi [\aap] {10.1051/0004-6361/201220465}, \href
  {http://adsabs.harvard.edu/abs/2013A%26A...550A..36M} {550, A36}

\bibitem[\protect\citeauthoryear{{Meg{\'\i}as}, {Jim{\'e}nez-Serra},
  {Mart{\'\i}n-Pintado}, {Vasyunin}, {Spezzano}, {Caselli}, {Cosentino}  \&
  {Viti}}{{Meg{\'\i}as} et~al.}{2022}]{megias2022}
{Meg{\'\i}as} A.,  {Jim{\'e}nez-Serra} I.,  {Mart{\'\i}n-Pintado} J.,
  {Vasyunin} A.~I.,  {Spezzano} S.,  {Caselli} P.,  {Cosentino} G.,   {Viti}
  S.,  2022, \mn@doi [\mnras] {10.1093/mnras/stac3449}, \href
  {https://ui.adsabs.harvard.edu/abs/2022MNRAS.tmp.3255M} {}

\bibitem[\protect\citeauthoryear{{Navarro-Almaida} et~al.,}{{Navarro-Almaida}
  et~al.}{2020}]{navarro2020}
{Navarro-Almaida} D.,  et~al., 2020, \mn@doi [\aap]
  {10.1051/0004-6361/201937180}, \href
  {https://ui.adsabs.harvard.edu/abs/2020A&A...637A..39N} {637, A39}

\bibitem[\protect\citeauthoryear{{Panessa}, {Seifried}, {Walch}, {Gaches},
  {Barnes}, {Bigiel}  \& {Neumann}}{{Panessa} et~al.}{2023}]{panessa2023}
{Panessa} M.,  {Seifried} D.,  {Walch} S.,  {Gaches} B.,  {Barnes} A.~T.,
  {Bigiel} F.,   {Neumann} L.,  2023, \mn@doi [\mnras]
  {10.1093/mnras/stad1741}, \href
  {https://ui.adsabs.harvard.edu/abs/2023MNRAS.523.6138P} {523, 6138}

\bibitem[\protect\citeauthoryear{{Pety} et~al.,}{{Pety}
  et~al.}{2017}]{pety2017}
{Pety} J.,  et~al., 2017, \mn@doi [\aap] {10.1051/0004-6361/201629862}, \href
  {https://ui.adsabs.harvard.edu/abs/2017A&A...599A..98P} {599, A98}

\bibitem[\protect\citeauthoryear{{Priestley}, {Whitworth}  \&
  {Fogerty}}{{Priestley} et~al.}{2023a}]{priestley2023}
{Priestley} F.~D.,  {Whitworth} A.~P.,   {Fogerty} E.,  2023a, \mn@doi [\mnras]
  {10.1093/mnras/stac3444}, \href
  {https://ui.adsabs.harvard.edu/abs/2023MNRAS.518.4839P} {518, 4839}

\bibitem[\protect\citeauthoryear{{Priestley}, {Clark}  \&
  {Whitworth}}{{Priestley} et~al.}{2023b}]{priestley2023b}
{Priestley} F.~D.,  {Clark} P.~C.,   {Whitworth} A.~P.,  2023b, \mn@doi
  [\mnras] {10.1093/mnras/stad150}, \href
  {https://ui.adsabs.harvard.edu/abs/2023MNRAS.519.6392P} {519, 6392}

\bibitem[\protect\citeauthoryear{{Prole}, {Clark}, {Klessen}  \&
  {Glover}}{{Prole} et~al.}{2022}]{prole2022}
{Prole} L.~R.,  {Clark} P.~C.,  {Klessen} R.~S.,   {Glover} S. C.~O.,  2022,
  \mn@doi [\mnras] {10.1093/mnras/stab3697}, \href
  {https://ui.adsabs.harvard.edu/abs/2022MNRAS.510.4019P} {510, 4019}

\bibitem[\protect\citeauthoryear{{Ragan}, {Bergin}, {Plume}, {Gibson},
  {Wilner}, {O'Brien}  \& {Hails}}{{Ragan} et~al.}{2006}]{ragan2006}
{Ragan} S.~E.,  {Bergin} E.~A.,  {Plume} R.,  {Gibson} D.~L.,  {Wilner} D.~J.,
  {O'Brien} S.,   {Hails} E.,  2006, \mn@doi [\apjs] {10.1086/506594}, \href
  {https://ui.adsabs.harvard.edu/abs/2006ApJS..166..567R} {166, 567}

\bibitem[\protect\citeauthoryear{{Rocha} et~al.,}{{Rocha}
  et~al.}{2023}]{rocha2023}
{Rocha} C. M.~R.,  et~al., 2023, \mn@doi [arXiv e-prints]
  {10.48550/arXiv.2307.00311}, \href
  {https://ui.adsabs.harvard.edu/abs/2023arXiv230700311R} {p. arXiv:2307.00311}

\bibitem[\protect\citeauthoryear{{Ruaud}, {Wakelam}, {Gratier}  \&
  {Bonnell}}{{Ruaud} et~al.}{2018}]{ruaud2018}
{Ruaud} M.,  {Wakelam} V.,  {Gratier} P.,   {Bonnell} I.~A.,  2018, \mn@doi
  [\aap] {10.1051/0004-6361/201731693}, \href
  {https://ui.adsabs.harvard.edu/abs/2018A&A...611A..96R} {611, A96}

\bibitem[\protect\citeauthoryear{{Scibelli} \& {Shirley}}{{Scibelli} \&
  {Shirley}}{2020}]{scibelli2020}
{Scibelli} S.,  {Shirley} Y.,  2020, \mn@doi [\apj] {10.3847/1538-4357/ab7375},
  \href {https://ui.adsabs.harvard.edu/abs/2020ApJ...891...73S} {891, 73}

\bibitem[\protect\citeauthoryear{{Scibelli}, {Shirley}, {Vasyunin}  \&
  {Launhardt}}{{Scibelli} et~al.}{2021}]{scibelli2021}
{Scibelli} S.,  {Shirley} Y.,  {Vasyunin} A.,   {Launhardt} R.,  2021, \mn@doi
  [\mnras] {10.1093/mnras/stab1151}, \href
  {https://ui.adsabs.harvard.edu/abs/2021MNRAS.504.5754S} {504, 5754}

\bibitem[\protect\citeauthoryear{{Seifried} et~al.,}{{Seifried}
  et~al.}{2017}]{seifried2017b}
{Seifried} D.,  et~al., 2017, \mn@doi [\mnras] {10.1093/mnras/stx2343}, \href
  {https://ui.adsabs.harvard.edu/abs/2017MNRAS.472.4797S} {472, 4797}

\bibitem[\protect\citeauthoryear{{Sembach}, {Howk}, {Ryans}  \&
  {Keenan}}{{Sembach} et~al.}{2000}]{sembach2000}
{Sembach} K.~R.,  {Howk} J.~C.,  {Ryans} R. S.~I.,   {Keenan} F.~P.,  2000,
  \mn@doi [\apj] {10.1086/308173}, \href
  {https://ui.adsabs.harvard.edu/abs/2000ApJ...528..310S} {528, 310}

\bibitem[\protect\citeauthoryear{{Shirley}}{{Shirley}}{2015}]{shirley2015}
{Shirley} Y.~L.,  2015, \mn@doi [\pasp] {10.1086/680342}, \href
  {https://ui.adsabs.harvard.edu/abs/2015PASP..127..299S} {127, 299}

\bibitem[\protect\citeauthoryear{{Sipil{\"a}} \& {Caselli}}{{Sipil{\"a}} \&
  {Caselli}}{2018}]{sipila2018}
{Sipil{\"a}} O.,  {Caselli} P.,  2018, \mn@doi [\aap]
  {10.1051/0004-6361/201732326}, \href
  {https://ui.adsabs.harvard.edu/abs/2018A&A...615A..15S} {615, A15}

\bibitem[\protect\citeauthoryear{{Sipil{\"a}}, {Caselli}, {Redaelli}, {Juvela}
  \& {Bizzocchi}}{{Sipil{\"a}} et~al.}{2019}]{sipila2019}
{Sipil{\"a}} O.,  {Caselli} P.,  {Redaelli} E.,  {Juvela} M.,   {Bizzocchi} L.,
   2019, \mn@doi [\mnras] {10.1093/mnras/stz1344}, \href
  {https://ui.adsabs.harvard.edu/abs/2019MNRAS.487.1269S} {487, 1269}

\bibitem[\protect\citeauthoryear{{Smith}, {Shetty}, {Stutz}  \&
  {Klessen}}{{Smith} et~al.}{2012}]{smith2012}
{Smith} R.~J.,  {Shetty} R.,  {Stutz} A.~M.,   {Klessen} R.~S.,  2012, \mn@doi
  [\apj] {10.1088/0004-637X/750/1/64}, \href
  {https://ui.adsabs.harvard.edu/abs/2012ApJ...750...64S} {750, 64}

\bibitem[\protect\citeauthoryear{{Smith}, {Shetty}, {Beuther}, {Klessen}  \&
  {Bonnell}}{{Smith} et~al.}{2013}]{smith2013}
{Smith} R.~J.,  {Shetty} R.,  {Beuther} H.,  {Klessen} R.~S.,   {Bonnell}
  I.~A.,  2013, \mn@doi [\apj] {10.1088/0004-637X/771/1/24}, \href
  {https://ui.adsabs.harvard.edu/abs/2013ApJ...771...24S} {771, 24}

\bibitem[\protect\citeauthoryear{{Smith} et~al.,}{{Smith}
  et~al.}{2020}]{smith2020}
{Smith} R.~J.,  et~al., 2020, \mn@doi [\mnras] {10.1093/mnras/stz3328}, \href
  {https://ui.adsabs.harvard.edu/abs/2020MNRAS.492.1594S} {492, 1594}

\bibitem[\protect\citeauthoryear{{Springel}}{{Springel}}{2010}]{springel2010}
{Springel} V.,  2010, \mn@doi [\mnras] {10.1111/j.1365-2966.2009.15715.x},
  \href {https://ui.adsabs.harvard.edu/abs/2010MNRAS.401..791S} {401, 791}

\bibitem[\protect\citeauthoryear{{Szucs}, {Glover}  \& {Caselli}}{{Szucs}
  et~al.}{2015}]{szucs2015}
{Szucs} L.,  {Glover} S.,   {Caselli} P.,  2015, in EAS Publications Series. pp
  391--392, \mn@doi{10.1051/eas/1575079}

\bibitem[\protect\citeauthoryear{{Tafalla}, {Myers}, {Caselli}, {Walmsley}  \&
  {Comito}}{{Tafalla} et~al.}{2002}]{tafalla2002}
{Tafalla} M.,  {Myers} P.~C.,  {Caselli} P.,  {Walmsley} C.~M.,   {Comito} C.,
  2002, \mn@doi [\apj] {10.1086/339321}, \href
  {http://adsabs.harvard.edu/abs/2002ApJ...569..815T} {569, 815}

\bibitem[\protect\citeauthoryear{{Tafalla}, {Usero}  \& {Hacar}}{{Tafalla}
  et~al.}{2021}]{tafalla2021}
{Tafalla} M.,  {Usero} A.,   {Hacar} A.,  2021, \mn@doi [\aap]
  {10.1051/0004-6361/202038727}, \href
  {https://ui.adsabs.harvard.edu/abs/2021A&A...646A..97T} {646, A97}

\bibitem[\protect\citeauthoryear{{Tress}, {Smith}, {Sormani}, {Glover},
  {Klessen}, {Mac Low}  \& {Clark}}{{Tress} et~al.}{2020}]{tress2020}
{Tress} R.~G.,  {Smith} R.~J.,  {Sormani} M.~C.,  {Glover} S. C.~O.,  {Klessen}
  R.~S.,  {Mac Low} M.-M.,   {Clark} P.~C.,  2020, \mn@doi [\mnras]
  {10.1093/mnras/stz3600}, \href
  {https://ui.adsabs.harvard.edu/abs/2020MNRAS.492.2973T} {492, 2973}

\bibitem[\protect\citeauthoryear{{Wakelam}, {Ruaud}, {Gratier}  \&
  {Bonnell}}{{Wakelam} et~al.}{2019}]{wakelam2019}
{Wakelam} V.,  {Ruaud} M.,  {Gratier} P.,   {Bonnell} I.~A.,  2019, \mn@doi
  [\mnras] {10.1093/mnras/stz1122}, \href
  {https://ui.adsabs.harvard.edu/abs/2019MNRAS.486.4198W} {486, 4198}

\bibitem[\protect\citeauthoryear{{Wakelam}, {Iqbal}, {Melisse}, {Gratier},
  {Ruaud}  \& {Bonnell}}{{Wakelam} et~al.}{2020}]{wakelam2020}
{Wakelam} V.,  {Iqbal} W.,  {Melisse} J.~P.,  {Gratier} P.,  {Ruaud} M.,
  {Bonnell} I.,  2020, \mn@doi [\mnras] {10.1093/mnras/staa2016}, \href
  {https://ui.adsabs.harvard.edu/abs/2020MNRAS.497.2309W} {497, 2309}

\bibitem[\protect\citeauthoryear{{Walch} et~al.,}{{Walch}
  et~al.}{2015}]{walch2015}
{Walch} S.,  et~al., 2015, \mn@doi [\mnras] {10.1093/mnras/stv1975}, \href
  {https://ui.adsabs.harvard.edu/abs/2015MNRAS.454..238W} {454, 238}

\bibitem[\protect\citeauthoryear{{Wolfire}, {Hollenbach}  \& {McKee}}{{Wolfire}
  et~al.}{2010}]{wolfire2010}
{Wolfire} M.~G.,  {Hollenbach} D.,   {McKee} C.~F.,  2010, \mn@doi [\apj]
  {10.1088/0004-637X/716/2/1191}, \href
  {https://ui.adsabs.harvard.edu/abs/2010ApJ...716.1191W} {716, 1191}

\bibitem[\protect\citeauthoryear{{Wurster}}{{Wurster}}{2021}]{wurster2021}
{Wurster} J.,  2021, \mn@doi [\mnras] {10.1093/mnras/staa3943}, \href
  {https://ui.adsabs.harvard.edu/abs/2021MNRAS.501.5873W} {501, 5873}

\bibitem[\protect\citeauthoryear{{de Jong}}{{de Jong}}{1977}]{dejong1977}
{de Jong} T.,  1977, \aap, \href
  {https://ui.adsabs.harvard.edu/abs/1977A&A....55..137D} {55, 137}

\bibitem[\protect\citeauthoryear{{van Dishoeck} \& {Black}}{{van Dishoeck} \&
  {Black}}{1988}]{vandishoeck1988}
{van Dishoeck} E.~F.,  {Black} J.~H.,  1988, \mn@doi [\apj] {10.1086/166877},
  \href {https://ui.adsabs.harvard.edu/abs/1988ApJ...334..771V} {334, 771}

\makeatother
\end{thebibliography}

\appendix

\section{Analytical fits to molecular abundances}
\label{sec:polyfit}

{Table \ref{tab:polyfit} lists the coefficients of the polynomial fits to median molecular abundance shown in Figures \ref{fig:boxco} and \ref{fig:box}, in logarithmic space so that}
\begin{equation}
  \log_{10} (n_X / \nh) = \ssum_{i=0}^5 a_i \left(\log_{10} \nh\right)^i
\end{equation}
{with $n_X$ and $\nh$ in units of $\pcc$. While these values are reasonably accurate in reproducing the {\it median} abundance at a given density, there is no one-to-one connection between density and abundance for any species, with variations particularly significant for CO-like molecules at densities above $10^3 \pcc$. {In addition, it remains to be seen how well these fits will reproduce the median abundances in molecular clouds formed in conditions very different to those simulated here, e.g. in the Milky Way's Central Molecular Zone.} As such, the polynomial fits should be seen as rough approximations, suitable when order-of-magnitude level errors can be tolerated, and not as a genuine alternative to full chemical modelling.}

\newpage

\begin{table}
  \centering
  \caption{Coefficients of polynomial fits to median molecular abundances as a function of density.}
  \begin{tabular}{ccccccc}
    \hline
    Molecule & $a_0$ & $a_1$ & $a_2$ & $a_3$ & $a_4$ & $a_5$ \\
    \hline
    CO & $-0.044$ & $1$ & $-8.7$ & $34$ & $-60$ & $30$ \\
    NH$_3$ & $0.0073$ & $-0.12$ & $0.55$ & $-0.55$ & $-0.92$ & $-8.2$ \\
    HCN & $-0.0041$ & $0.12$ & $-1.2$ & $4.8$ & $-6$ & $-10$ \\
    N$_2$H$^+$ & $-0.018$ & $0.52$ & $-5.7$ & $28$ & $-62$ & $34$ \\
    HCO$^+$ & $-0.022$ & $0.62$ & $-6.6$ & $33$ & $-74$ & $51$ \\
    CS & $-0.022$ & $0.5$ & $-4.2$ & $15$ & $-22$ & $-2.2$ \\
    HNC & $-0.0011$ & $0.077$ & $-1.1$ & $5.9$ & $-12$ & $-3.1$ \\
    CH$_3$OH & $-0.065$ & $1.5$ & $-13$ & $54$ & $-100$ & $55$ \\
    CN & $0.0064$ & $-0.17$ & $1.9$ & $-11$ & $29$ & $-38$ \\
    C$_2$H & $0.036$ & $-0.89$ & $8.6$ & $-40$ & $90$ & $-86$ \\
    \hline
  \end{tabular}
  \label{tab:polyfit}
\end{table}

\bsp	
\label{lastpage}
\end{document}